\documentclass[twocolumn]{aastex62}
\usepackage{natbib}
\usepackage{amsmath}
\usepackage{graphicx}
\usepackage{makecell}
\usepackage{kbordermatrix}

\usepackage[algoruled]{algorithm2e}
\SetAlgoSkip{medskip}
\SetAlgoInsideSkip{medskip}
\SetAlgoCaptionSeparator{.}
\SetAlgorithmName{ALGORITHM}{}
\DontPrintSemicolon

\graphicspath{{figures/}}

\newcommand{\Ss}{\S\S}  
\newcommand{\PapI}{Pap~I} 

\newcommand{\sci}[2]{\mbox{$#1 \times 10^{#2}$}}

\newcommand{\rmscriptsize}[1]{\textrm{\scriptsize{#1}}}

\newcommand{\SUB}[3][]{
  \ifx\relax#1\relax
  \ensuremath{#2_{\rmscriptsize{#3}}}
  \else
  \ensuremath{#2_{#3,\,\rmscriptsize{#1}}}
  \fi}

\newcommand{\GJ}[1]{\SUB{#1}{\textsc{gj}}}
\newcommand{\PC}[1]{\SUB{#1}{pc}}    
\newcommand{\RNS}{\SUB{R}{\textsc{ns}}}
\newcommand{\rhoC}{\SUB{\rho}{c}}

\newcommand{\jm}{\SUB{j}{m}}

\newcommand{\lambdaC}{\SUB{\lambdabar{}}{\textsc{c}}}
\newcommand{\alphaF}{\ensuremath{\alpha_f}}    

\newcommand{\chiA}{\ensuremath{\chi_{a}}}

\newcommand{\epsilonG}{\ensuremath{\epsilon_{\gamma}}}    
\newcommand{\epsilonP}{\ensuremath{\epsilon_{\pm}}}    

\newcommand{\ESC}[1]{\SUB{#1}{esc}}
\newcommand{\epsilonGesc}{\ensuremath{\epsilon_{\gamma,\,\rmscriptsize{esc}}}}
\newcommand{\chiAesc}{\ensuremath{\chi_{a,\,\rmscriptsize{esc}}}}    

\newcommand{\epsilonPacc}{\SUB[acc]{\epsilon}{\pm}}
\newcommand{\chiAacc}{\ensuremath{\chi_{a,\,\rmscriptsize{acc}}}}

\newcommand{\SYN}[1]{\SUB{#1}{syn}}
\newcommand{\CR}[1]{\SUB{#1}{\textsc{cr}}}
\newcommand{\RICS}[1]{\SUB{#1}{\textsc{rics}}}

\newcommand{\epsilonGRICS}{\SUB[\textsc{RICS}]{\epsilon}{\gamma}}

\newcommand{\Pair}[1]{\ensuremath{#1_{\gamma\rightarrow\pm}}}
\newcommand{\Split}[1]{\ensuremath{#1_{\gamma\rightarrow\gamma\gamma}}}
\newcommand{\MFP}{\Pair{\lambda}}
\newcommand{\MFPSplit}{\Split{\lambda}}

\newcommand{\sigmaT}{\SUB{\sigma}{T}}

\newcommand{\kappaArr}{\ensuremath{\left<\kappa_x\right>}}
\newcommand{\orgn}{\ensuremath{\texttt{orgn}}}
\newcommand{\igen}{\SUB{i}{gen}}
\newcommand{\iproc}{\SUB{i}{proc}}

\shorttitle{maximum multiplicity of pulsar cascades}

\shortauthors{Timokhin \& Harding}

\begin{document}

\title{On the maximum pair multiplicity of pulsar cascades}

\author{A.~N.~Timokhin$^{1,2}$ and A.~K.~Harding}

\affil{%
  Astrophysics Science Division, NASA/Goddard Space Flight Center,
  Greenbelt, MD 20771, USA\\
  $^2$University of Maryland, College Park (UMDCP/CRESSTII), College Park, MD 20742, USA}

\email{andrey.timokhin@nasa.gov}

\date{Received ; accepted ; published }

\begin{abstract}

  We study electron-positron pair production in polar caps of energetic pulsars
  to determine the maximum multiplicity of pair plasma a pulsar can produce
  under the most favorable conditions.  This paper complements and updates our
  study of pair cascades presented in \citet{TimokhinHarding2015} with more
  accurate treatment of the effects of ultra strong $B\gtrsim\sci{3}{12}$G
  magnetic fields and emission processes of primary and secondary particles. We
  include pairs produced by curvature and synchrotron radiation photons as well
  as resonant Compton scattered photons.  We develop a semi-analytical model of
  electron-positrons cascades which can efficiently simulate pair cascades with
  an arbitrary number of microphysical processes and use it to explore cascade
  properties for a wide range of pulsar parameters.  We argue that the maximum
  cascade multiplicity can not exceed $\sim\sci{\mbox{a few}}{5}$ and the
  multiplicity has a rather weak dependence on pulsar period on pulsar
  period. The highest multiplicity is achieved in pulsars with magnetic field
  $\sci{4}{12}\lesssim{}B\lesssim10^{13}$G and hot surfaces, with
  $T\gtrsim10^{6}$K.  We also derive analytical expressions for several physical
  quantities relevant for electromagnetic cascade in pulsars which may be useful
  in future works on pulsar cascades, including the upper limit on cascade
  multiplicity and various approximations for the parameter $\chi$, the
  exponential factor in the expression for photon attenuation in strong magnetic
  field.
  
\end{abstract}


\section{Introduction}
\label{sec:introduction}

Dense electron-positron pair plasmas are an integral part of the standard model
for rotation-powered pulsars which was initially proposed by \citet{GJ} and
\citet{Sturrock71}.  According to the standard model, a pulsar magnetosphere is
filled with dense pair plasma which screens the electric field along magnetic
field lines everywhere, except in some small zones responsible for particle
acceleration and emission.  The sharpness of peaks in pulsar light curves is a
strong argument in favor of thin acceleration zones and screened electric field
in most parts of the pulsar magnetosphere.  There is also direct observational
evidence for plasma creation in pulsars: the most energetic ones are surrounded
by ``cocoons'' of dense relativistic plasma -- Pulsar Wind Nebulae (PWNe) --
powered by plasma outflow from their pulsars. Understanding pair creation is
important for unraveling the mystery of the pulsar emission mechanism(s) and
understanding pulsar surroundings on both small and large scales.  Pair plasma
flows out of the magnetosphere providing the radiating particles for PWNe and
could make a make significant contribution to the lepton component of cosmic rays.

The regions responsible for production of most of the pair plasma are believed
to be pulsar polar caps (PCs) -- small regions near the magnetic poles
\citep{Sturrock71,RudermanSutherland1975,AronsScharlemann1979}.  Without dense
plasma produced in the PCs, at the base of open magnetic field lines, the
magnetosphere would have large volumes with unscreened electric field, as pair
creation in e.g. outer gaps \citep{Cheng/Ruderman76} cannot screen the electric
field over the rest of the magnetosphere.  The physical process responsible for
pair production in the PCs is conversion of high energy $\gamma$-rays into
electron-positron pairs in strong ($\gtrsim10^{11}$G) magnetic field. According
to the recent self-consistent PC models, specific regions of the PC
intermittently become charge starved, when the number density of charged
particles is not enough to support both the change and the current density
required by the global structure of the pulsar magnetosphere
\citep{Timokhin2010::TDC_MNRAS_I,TimokhinArons2013}.  This gives rise to a
strong accelerating electric field and formation of (intermittent) accelerating
zone(s).  Some charged particles enter these zones, are accelerated to very high
energies and emit $\gamma$-rays, creating electron-positron pairs. The pairs can
also emit pair producing photons and so the avalanche develops until photons
emitted by the last generation of pairs can no longer produce pairs and escape
the magnetosphere.

The cascade process in pulsar polar caps has been the subject of extensive
studies \citep[e.g.][]{Daugherty/Harding82,GurevichIstomin1985,Zhang2000,
  Hibschman/Arons:pair_multipl::2001,Medin2010}. Those works considered pair
creation together with particle acceleration and so the results were dependent
on the acceleration model used. The most popular acceleration model assumed
steady, time-independent acceleration of the primary particles in a flow with a
relatively weak accelerating electric field
\citep{AronsScharlemann1979,Muslimov/Tsygan92}, which recently was shown to be
incorrect by means of direct self-consistent numerical simulations
\citep{TimokhinArons2013}. The necessity of bringing PC pair creation models up
to date with the self-consistent description of pair acceleration motivated us
to develop a simple semi-analytical model for pair cascades in pulsar polar caps
that can be easily decoupled from the details of the particle acceleration model
and that allows easy exploration of the parameter space --
\citet{TimokhinHarding2015}, hereafter \PapI.  In \PapI{} we considered cascades
at PCs of young pulsars with moderate magnetic fields, when the dominant process
of high energy photon emission are Curvature Radiation (CR) of primary particles
and Synchrotron Radiation (SR) of secondary particles. This model agrees very
well with the results of elaborate numerical simulations of pair cascade.  We
have shown that the maximum pair multiplicity achievable in pulsars does not
exceed $\sim\text{few}\times10^5$ which sets stringent limits on PWNe models.

In this paper we present a significant improvement to the semi-analytical
model from \PapI.  The new model allows inclusion of additional emission
processes, provides detailed information about the spatial distribution of pair
creation, and is applicable for pulsars with strong, up to $\sim10^{13}$G,
magnetic fields.  More specifically, the new model (i) can include an arbitrary
number of emission processes and collect detailed information about where and in
what cascade branch pairs are created, (ii) incorporates strong field
corrections to the expression for the attenuation coefficient for one-photon pair 
creation, and (iii) takes into account the effect of photons splitting on
the cascade multiplicity.

The main question we try to answer in this paper is: What is the maximum number
density of pair plasma a pulsar can generate?  It has been shown in numerous
studies that the highest cascade multiplicity is expected in young energetic
pulsars with high accelerating electric fields, where cascades are initiated by
CR of primary particles.  In such pulsars primary particles are accelerated to
higher energies over a short distance, emit photons via CR, which is the most
efficient radiation mechanism in such physical conditions, and those photons
propagate a short distance before being absorbed in a still strong magnetic
field.  In \PapI, we limited ourselves to CR-synchrotron cascades, without
considering Resonant Inverse Compton Scattering (RICS) of thermal photons from
the neutron star (NS) surface by particles in the cascade. As we argued in that
paper it was an adequate approximation for most young energetic
pulsars. However, for $B\gtrsim10^{12}$G, right where CR-synchrotron cascades
reach their highest multiplicity, RICS becomes an important emission mechanism,
while Inverse Compton Scattering in the non-resonant regime remain irrelevant
for polar cap cascades. In order to get an accurate limit on the maximum cascade
multiplicity RICS must be taken into account.

In this paper we apply our new semi-analytical model to cascades where pairs are
created by photons emitted via CR of primary particles, and by photons emitted
by secondary particle via SR as well as RICS of soft X-rays from the NS surface%
\footnote{Pairs may also be created via RICS of primary particles.  However, it
  was shown \citep{Harding/Muslimov:heating_2::2002} that pair cascades from
  primary RICS have very low multiplicity.  We therefore neglect this channel of
  pair production here.}.
Similar to \PapI{} we consider the physical processes
in pair cascades and particle acceleration models separately to clearly set
apart different factors influencing the efficiency of pair cascades. Results
presented in this paper supersede results of \PapI{} for high (around
$\sim10^{13}$G) field pulsars and improve multiplicity estimates for pulsars
with medium magnetic fields ($\sim10^{12}$G) covering all ranges of parameters
for pulsars capable of generating high multiplicity pair plasma.

We do not attempt to model how the entire magnetosphere is filled with plasma
\citep[like e.g.][]{PhilippovSpitkovsky_2015,Brambilla:2018}.  We adopt the standard pulsar
model and concentrate on microphysics of the polar cap cascade zone -- the most
important supplier of pair plasma in the magnetosphere -- to determine the
upper limit on pair plasma density that a pulsar can generate.

The plan of the paper is as follows.  In \S\ref{sec:cascades-overview} we first
discuss general properties of electron-positron cascades and then give an
overview of the microphysical processes in polar cap cascades.  In
\S\ref{sec:photon-absorption} we consider photon absorption in strong magnetic
fields: single photon pair creation in \S\ref{sec:phot_pair_creation}, photon
splitting in \S\ref{sec:phot_split}, and the energy of photons escaping from the
cascade in \S\ref{sec:E_esc}.  We discuss particle acceleration in
\S\ref{sec:particle-acceleration}. In \S\ref{sec:simple-multiplicity} we give a
simple estimate for the upper limit of the cascade multiplicity from first
principles.  \S\ref{sec:cascade-devel} gives an overview of our semi-analytical
cascade model (with more technical details described in
Appendix~\ref{sec:algorithms}). The main results are described in
\S\ref{sec:results}.  We summarize our findings and discuss limitations of our
model in \S\ref{sec:discussion}.

\section{Physics of polar cap cascades: an overview}
\label{sec:cascades-overview}

In pulsar magnetospheres electron-positron pairs can be created by single-photon 
absorption in a strong magnetic field ($\gamma{}B$), which can happen only close
to the NS where the magnetic field is strong enough, and in two-photon collisions
($\gamma\gamma$) which are relevant mostly in the outer magnetosphere.  In an
electron-positron cascade primary particles lose energy by some emission
mechanism, creating high energy photons which are absorbed in a pair creation
process and produce electron-positron pairs.  Pairs can also emit high energy
photons which then create the next generation of pairs.  As the cascade develops it
``alternates'' between electron/positron and photon states.  At each step in the
cascade the energy of the parent particle is divided between secondary
particles.  Each subsequent generation of particles has smaller energies than
the previous one.  At some cascade generation the energy of the photons drops below
the pair formation threshold and the cascade terminates.  If the energy of the parent
particle is divided roughly equally between its secondaries, i.e. the photon's
energy is roughly equally divided between electron and positron, and each pair
member emits several hard photons of approximately the same energy, then at the
last cascade step the available energy will be approximately equally split
between photons with energies somewhat above the pair formation threshold.
These photons will create the last generation of pairs.  The number of pairs in
such a cascade grows as a geometric progression at each generation and most of the
pairs will be created at the last cascade step.  In an ideal case, when both
primary and secondary particles radiate all their energies as pair producing
photons, the multiplicity $\kappa$ of such a cascade (the number of particles
produced by each primary particle) would be
\begin{equation}
  \label{eq:kappa_max}
  \SUB{\kappa}{max}\simeq{}2\, \frac{\epsilon_p}{\epsilonGesc}\,.
\end{equation}
where $\epsilonGesc$ is the maximum energy of the photons which escape the
cascade (or the minimum energy of pair producing photons) and $\epsilon_p$ is
the energy of primary particles. For convenience from here on, all particle and
photon energies will be quoted in terms of $m_ec^2$. In a real cascade both
primary and secondary particles do not radiate all their kinetic energy as pair
producing photons and $\SUB{\kappa}{max}$ can be considered as an upper limit on
the multiplicity.

In the above ideal limit, the energy of the primary particle is divided
into chunks of the size $\epsilonGesc$.  Usually $\epsilonGesc\gg2$ and even in
the ideal case, when energy is not lost at intermediate steps, the cascade
multiplicity is much smaller than $\epsilon_p$ (in terms of $m_ec^2$), which
would be the case if the whole energy of the primary particles went into the
\emph{rest} energy of pairs%
\footnote{Cascades \emph{can} operate in a different regime, when at each step
  one of the pair particles gets most of the parent photon's energy and then
  this secondary particle emits a single high energy photon carrying most of
  that particle's energy.  Such a cascade can produce
  $\sim\epsilon_p-\epsilonGesc\simeq\epsilon_p$
  pairs, what for $\epsilonGesc\gg2$ will result in a much higher multiplicity
  than that given by eq.~\eqref{eq:kappa_max}.  Photon emission and pair
  production in such cascades must happen in the extreme relativistic regime: for
  $\gamma{}B$ pair production and synchrotron radiation the parameter $\chi$
  must be large, $\chi\gg1$; for $\gamma\gamma$ pair production and Inverse
  Compton Scattering the energies of interacting particles
  $\epsilon_1$,$\epsilon_2$ must be $\epsilon_1\epsilon_2\gg1$.  For $\chi\gg1$
  photon injection must happen at large angles to the magnetic field, for
  $\epsilon_1\epsilon_2\gg1$ the interaction cross-section is much smaller than
  $\sigmaT$. In pulsar cascades particle acceleration zones are regulated by
  pair creation -- acceleration stops when pairs start being injected. This
  happens first at moderate values of $\chi$ and $\epsilon_1\epsilon_2$ thus
  preventing particles from achieving high enough energies to start cascade in
  the extreme relativistic regime.}.

In the de-facto standard pulsar model particles can be accelerated to very high
energies and produce dense pair plasma in the polar caps \citep{Sturrock71}, in
thin regions along last closed magnetic field lines \citep[the ``slot gap''
model of][]{Arons1983}, and in the ``outer gaps'', regions in the outer
magnetosphere along magnetic field lines crossing the surface where \citet{GJ}
charge density changes sign \citep{Cheng/Ruderman76}. The outer and slot gaps
occupy only a relatively small volume of the magnetosphere so that most of the
open magnetic field lines do not pass though them.  All open field lines
originate in the polar caps and a significant fraction of them pass through
polar cap particle acceleration zones.  The total number of primary particles in
the polar cap cascades is much larger than that in cascades in the outer pulsar
magnetosphere. Moreover, simulation of the outer gap cascades predict
multiplicities not higher than $10^4$ \citep[e.g.][]{Hirotani2006}. So, at least
in the standard pulsar model most of the pairs are produced in the polar cap
cascades.

It was demonstrated in \citet{TimokhinArons2013,Timokhin2010::TDC_MNRAS_I} that
pair formation in pulsars is an intermittent process. Time periods of efficient
particle acceleration and intense pair production alternate with periods of
quiet plasma flow when dense plasma screens the accelerating electric field and
no pairs are formed (more on this in \S\ref{sec:particle-acceleration}). As in
\PapI, here we consider cascades at the peak of the pair formation cycle, when their
multiplicity is the highest, postponing discussion on the effects of
intermittency to \S\ref{sec:discussion}.  Such cascades are generated by the
primary particles accelerated in well developed gaps.  Screening of accelerating
electric field in the gap happens very quickly, well before the multiplicity
reaches its maximum values.  Once primary particles have produced the first
generation of pairs which screen the accelerating electric field, they keep
moving in the regions of screened electric field radiating their energy away,
giving rise to extensive pair cascades.  So the PC cascades can be considered as
initiated by primary particles with given energies freely moving along magnetic
field lines.

\begin{figure}
  \includegraphics[clip,width=\columnwidth]{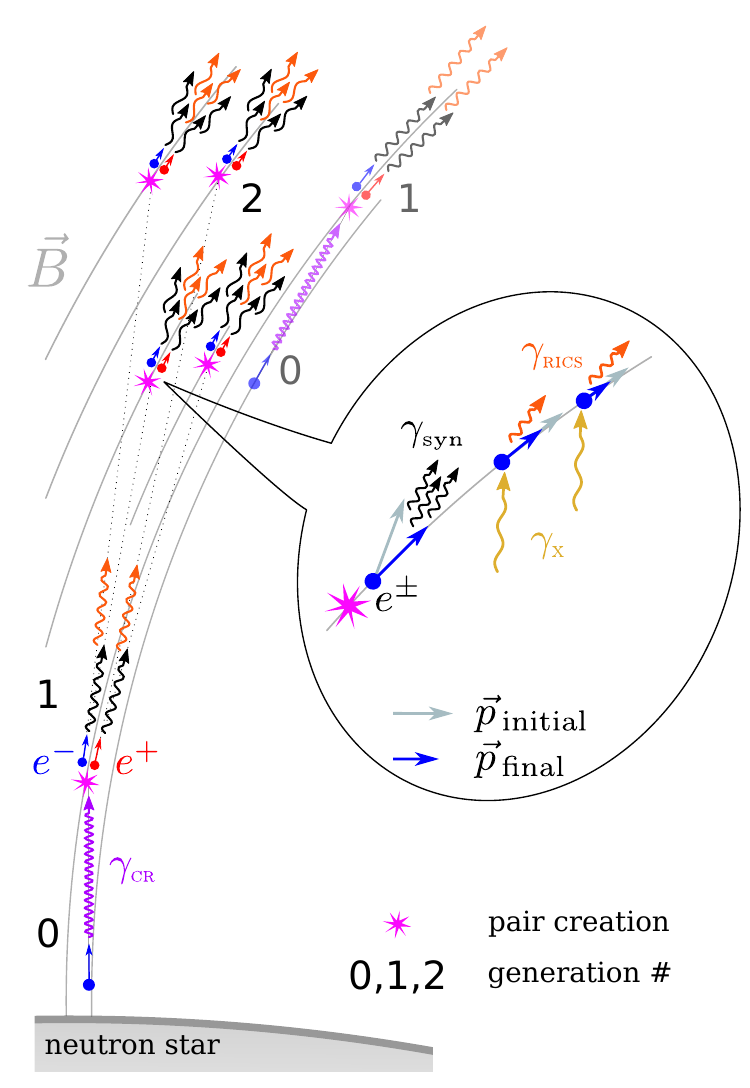}
  \caption{Schematic representation of electron-positron cascade in the polar cap
    of a young pulsar with high magnetic field, see text for description.}
  \label{fig:full_cascade}
\end{figure}

Fig.~\ref{fig:full_cascade} gives a schematic overview of processes involved in
pair plasma generation in PC cascades%
\footnote{This figure is similar to the Fig.~1 from \PapI{} but now it includes
  RICS of thermal photons by secondary pairs},
shown are the first two generations in a cascade initiated by a primary
electron.  Primary electrons emit CR photons ($\CR{\gamma}$) almost tangent to
the magnetic field lines; primary electrons and CR photons are generation 0
particles in our notation%
\footnote{Primary electrons also can emit RICS photons that produce pairs, but
  these are not shown since the numbers are too small to fully screen the
  electric field.}.
Magnetic field lines are curved and the angle between the photon momentum and
the magnetic field grows as the photon propagates further from the emission
point.  When this angle becomes large enough, photons are absorbed and each
photon creates an electron-positron pair -- generation 1 electron ($e^-$) and
positron ($e^+$).  The pair momentum is directed along the momentum of the
parent photon and at the moment of creation, the particles have non-zero
momentum perpendicular to the magnetic field.  They radiate this perpendicular
momentum almost instantaneously via SR and move along magnetic field lines. The
secondary particles can scatter thermal X-ray photons $\gamma_X$ coming from the
NS surface and lose their momenta parallel to the magnetic field.  Inverse
Compton Scattering of the thermal photons in the non-resonant regime is very
ineffective in converting the energy of parallel motion of pairs into pair
producing photons and can be neglected, see
Appendix~\ref{sec:non-resonant-inverse}. On the other hand, the RICS as it was
first pointed out by \citet{Dermer1990}, can become a very efficient emission
process in high field pulsars.
Although the secondary particles are relativistic, their energy is much lower
than that of the primaries and their curvature photons cannot create pairs.
Generation 1 photons -- synchrotron ($\SYN{\gamma}$) and RICS ($\RICS{\gamma}$)
photons produced by the generation 1 particles are also emitted (almost) tangent
to the magnetic field line -- as the secondary particles are relativistic -- and
propagate some distance before acquiring the necessary angle to the magnetic
field and creating generation 2 pairs.  These pairs in their turn radiate their
perpendicular momentum via SR, and parallel momenta via RICS emitting generation
2 photons.  The cascade initiated by a single CR photon stops at a generation
where the energy of synchrotron photons falls below $\epsilonGesc$.

Primary particles emit pair producing CR photons throughout the whole cascade
zone as they move along the field lines. Secondary particles emit all their pair
producing synchrotron photons after their creation almost instantaneously. RICS
photons are emitted by secondary particles over some distance, which is usually
much smaller than the size of the cascade zone.

In the following sections we analyze the individual factors regulating the yield
of electron-positron cascades and develop a semi-analytical technique which
models cascade development by following the general picture outline above.

\section{Photon absorption in magnetic field}
\label{sec:photon-absorption}

\subsection{Pair creation}
\label{sec:phot_pair_creation}

\begin{figure*}[t]
\includegraphics[clip,width=\textwidth]{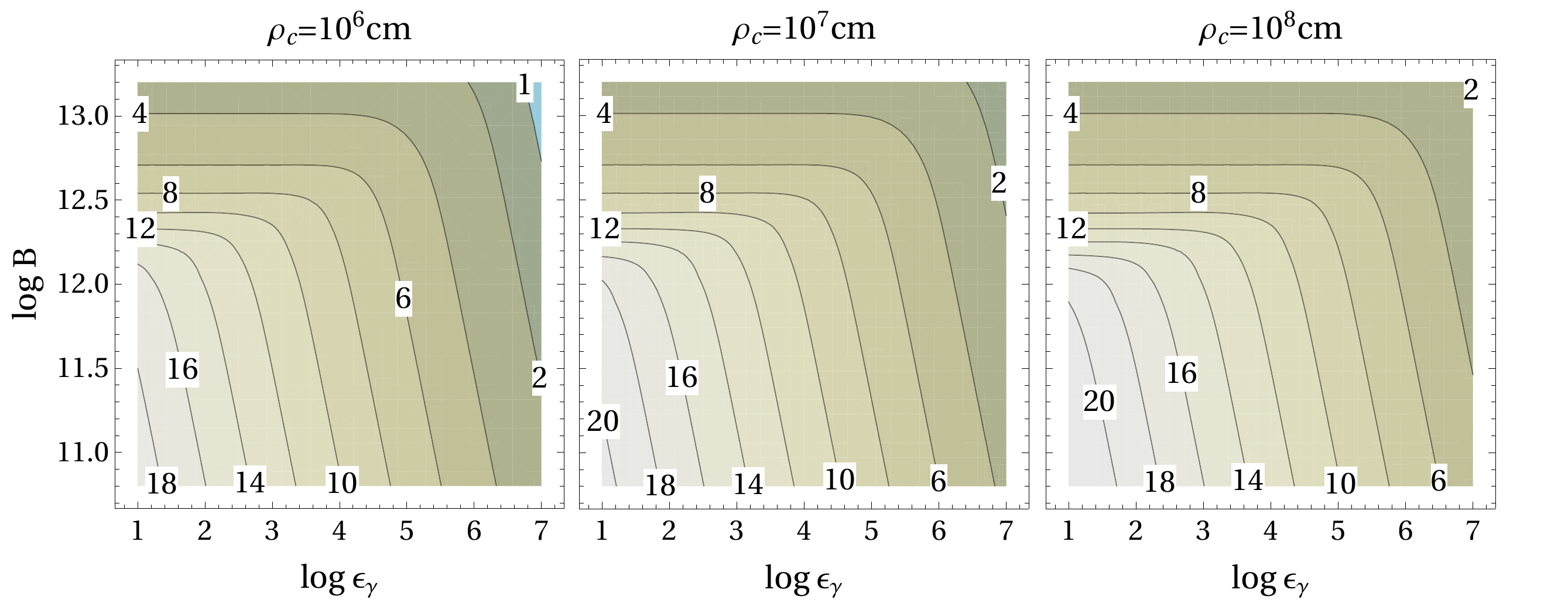}
\caption{Contour plot of $1/\chiA$ as a function of the logarithms of the
  magnetic field strength $B$ in Gauss, and photon energy $\epsilonG$ normalized
  to the electron rest energy, for three values of the radius of curvature of
  magnetic field lines $\rhoC=10^6, 10^7, 10^8\mbox{cm}$. $1\chiA$ values shown
  on this plot are calculated from eq.~(\ref{eq:chiA}).}
\label{fig:inverse_chi}
\end{figure*}

For the opacity for single photon pair production in a strong magnetic field
$\gamma\rightarrow{}e^+e^-$ we use the prescription suggested by
\citet{DaughertyHarding1983} which can be written as
\begin{equation}
  \label{eq:APP_alpha_B_DH_txt}
  \Pair{\alpha}(\epsilonG,\psi) = 0.23\,\frac{\alphaF}{\lambdaC}\:b\,\sin\psi\,
  \exp\left(-\frac{4}{3\chi}\right)\,
  f_{\alpha,1}\,.
\end{equation}
where $b\equiv{}B/B_q$ is the local magnetic field strength $B$ normalized to
the critical quantum magnetic field
$B_q=e/\alphaF\lambdaC^2=\sci{4.41}{13}$~G, 
$\psi$ is the angle between
the photon momentum and the local magnetic field, 
$\alphaF=e^2/\hbar{}c\approx{}1/137$
is the fine structure constant, and
$\lambdaC = \hbar/m c = \sci{3.86}{-11}$~cm
is the reduced Compton wavelength. The parameter $\chi$ is defined as
\begin{equation}
\label{eq:chi_def}
  \chi \equiv \frac{1}{2}\, \epsilonG b\, \sin\psi\,,
\end{equation}
where $\epsilonG$ is the photon energy in units of $m_ec^2$.
Expression~\eqref{eq:APP_alpha_B_DH_txt} differs from the usual \citet{Erber1966}
formula by the term
\begin{equation}
  \label{eq:APP_f_alpha_B_1_txt}
  f_{\alpha,1} =
  \begin{cases}
    \displaystyle
    \exp\left(-0.56\,\frac{b^{2.6962}}{\chi^{3.7}} \right), & \text{ if }\epsilonG \sin\psi\ge2\\
    0, & \text{ if } \epsilonG \sin\psi<2
  \end{cases}
\end{equation}
The function $f_{\alpha,1}$ insures that the attenuation coefficient for pair
production becomes zero below the threshold $\epsilonG \sin\psi=2$ and corrects
$\Pair{\alpha}$ for the case when absorption happens hear the threshold. The
threshold condition for $\gamma{}B$ pair production can be expressed in terms of
$\chi$ as
\begin{equation}
  \label{eq:chi_threshold}
  \chi\ge{}b \,.
\end{equation}
Expression~\eqref{eq:APP_alpha_B_DH_txt} works for high, $B>\sci{3}{12}$G, magnetic
fields, for weaker fields it reduces to the well known Erber's formula (see
Appendix~\ref{sec:optic-depths-phot}).

The optical depths for pair creation by a high energy photon in a strong
magnetic field after propagating distance $l$ is
\begin{equation}
\label{eq:tau_general}
\tau(\epsilonG,l)=\int_0^l\Pair{\alpha}(\epsilonG,\psi(x))\,dx\,, 
\end{equation}
where integration is along the photon's trajectory.  For photons emitted tangent
to the magnetic field line, $dx=\rhoC d\psi$, where $\rhoC$ is the radius of
curvature of magnetic field lines.  From eq.~(\ref{eq:chi_def}) we have
$\psi=2\chi/\epsilonG{}b$, and substituting it into eq.~(\ref{eq:tau_general})
we can express the optical depth $\tau$ to pair production as an integral over
$\chi$ as
\begin{equation}
  \label{eq:tau_chi_general}
  \tau(\epsilonG,\tilde{\chi}) = \frac{A_\tau}{\epsilonG^2}\;
  \int_0^{\tilde{\chi}} \frac{\rhoC}{b} \chi
  \exp\left(-\frac{4}{3\chi}\right)\,f_{\alpha,1}\, d\chi\,,
\end{equation}
where
$A_\tau\equiv{}0.92 \alphaF/\lambdaC\approx\sci{1.74}{8}\;\mbox{cm}^{-1}$\,.
The optical depth depends exponentially on $\chi$ and the main contribution to
the integral comes from the values of $\chi$ close to the upper boundary
$\tilde{\chi}$. For a wide range of photon energies and field strengths the
value of $\chi$ at the point where the photon is absorbed $\chiA$, changes
slowly. The mean free path of photons can be estimated from
eq.~\eqref{eq:chi_def} as
\begin{equation}
  \label{eq:MFP}
  \MFP=2\rhoC\chiA\frac{1}{b\epsilonG}\,.
\end{equation}
Both $\chiA$ and $\rhoC$ change slower than $b$ and $\epsilonG$ as the cascade
develops.  In each cascade generation the energy of particles and photons is
smaller than that in the preceding generation. The photon mean free path $\MFP$
increases because of this.  If $\MFP$ becomes comparable to the characteristic
scale of the magnetic field variation $L_B$, then the increase of $\MFP$ for the
next generation photons will be compounded by additional decrease of the
magnetic field $b$ as well, by at least $\sim$an order of magnitude (for dipolar
field). In most cases the magnetic field at the anticipated absorption point for
the next generation of photons will drop below the pair formation
threshold~\eqref{eq:chi_threshold}. Hence, the cascade generation for which
$\MFP\sim{}L_B$ should be the final one.

We consider strong cascades with large multiplicities; such cascades fully
develop before $\MFP\sim{}L_B$. For such cascades in the region where most of
the pairs are produced the magnetic field $b$ and the radius of curvature of
magnetic field lines $\rhoC$ are approximately constant. In approximation of
constant $b$ and $\rhoC$ eq.~\eqref{eq:tau_chi_general} can be written as
\begin{equation}
  \label{eq:tau_chi_general_constBRho}
  \tau(\chi) = A_\tau\frac{\rhoC}{\epsilonG^2\,b}\;
  \int_0^{\chiA} \chi
  \exp\left(-\frac{4}{3\chi}\right)\,f_{\alpha,1}\, d\chi\,
\end{equation}
and integrated analytically. The resulting expression is quite cumbersome, it is
derived in Appendix~\ref{sec:optic-depths-phot}, and given by
eq.~\eqref{eq:APP_tau_OTS_series}.

Because the opacity to pair production depends exponentially on $\chi$ it is a
reasonable approximation that all photons are absorbed when they have traveled
the distance where $\tau=1$.  We define $\chiA$ as the value of $\chi$ where the
optical depth reaches 1 through
\begin{equation}
  \label{eq:chiA}
  \chiA: \tau(\chiA)=1\,.
\end{equation}
$\chiA$ is a solution of the non-linear equation~(\ref{eq:chiA}) with $\tau$
given by eq.~\eqref{eq:APP_tau_OTS_series}.  We solved equation~(\ref{eq:chiA})
numerically for different values of $\epsilonG$, $b$, and $\rhoC$.  In
Fig.~\ref{fig:inverse_chi} we plot contours of $1/\chiA$ as functions of
$\log(\epsilonG)$ and $\log(B)$ for three different values for the radius of
curvature of the magnetic field lines $\rhoC=10^6, 10^7, 10^8\mbox{cm}$.  The
smallest value of $\rhoC$ corresponds to a strongly multipolar polar cap
magnetic field, when the radius of curvature is comparable to the NS radius.
The largest value corresponds to the radius of curvature of dipolar magnetic
field lines, which at the NS surface is given by
\begin{equation}
  \label{eq:rhoC_dipole}
  \SUB{\rho}{c, dip} = \frac{4}{3}\frac{\RNS}{\theta} =
  \sci{9.2}{7} \left(\frac{\theta}{\PC{\theta}}\right)^{-1}P^{1/2}\,\mbox{cm}\,,
\end{equation}
where $\theta$ is the colatitude of the footpoint of the magnetic field line,
$\PC{\theta}=\sci{1.45}{-2}P^{-1/2}$ -- the colatitude of the polar cap
boundary, $P$ -- the pulsar period in seconds, and $\RNS=10^{6}$cm -- the NS
radius.

$1/\chiA$ is a smooth function of $\log(\epsilonG)$, $\log(B)$, and $\log\rhoC$
-- as it is to be expected from $\Pair{\alpha}\propto\exp(-1/\chi)$ -- and can
be accurately approximated using a modest size numerical table. The change in
the behavior of $1/\chiA$ for large magnetic fields, when the contour lines
become horizontal, is due to photon being absorbed close to the pair production
threshold eq.~\eqref{eq:chi_threshold}.

\begin{figure*}[t]
\includegraphics[clip,width=\textwidth]{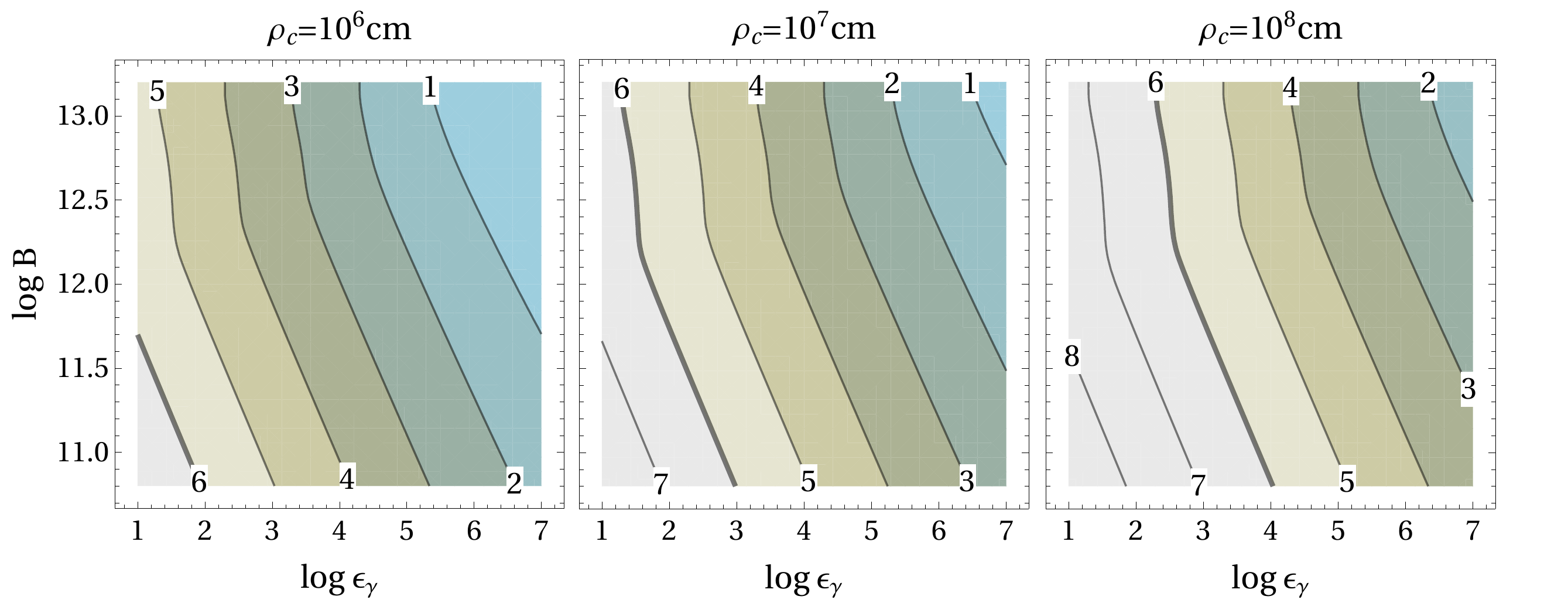}
\caption{Contour plot of the logarithm of the photon mean free path $\log\MFP$
  (in cm) as a function of the logarithms of the magnetic field strength $B$ in
  Gauss, and photon energy $\epsilonG$ normalized to the electron rest energy,
  for three values of the radius of curvature of magnetic field lines
  $\rhoC=10^6, 10^7, 10^8\mbox{cm}$.  $\log\MFP$ values shown on this plot are
  calculated from eq.~\eqref{eq:MFP}.}
\label{fig:MFP}
\end{figure*}

The values of $\chiA$ we obtained here using a more accurate expression for the 
$\gamma{}B$ opacity, eq.~\eqref{eq:APP_tau_OTS_series}, are up to 40\% higher than
that from \PapI{} where we used Erber's formula and made a simple correction for
the pair formation threshold by setting the upper limit on $\chi$ according to
eq.~\eqref{eq:chi_threshold}. This difference is larger that 10\% only for
magnetic fields with $B\gtrsim\sci{1.5}{12}$G (cf. Fig.~\ref{fig:inverse_chi}
with Fig.3 from \PapI).  Also note that the values of
$\chiA$ differ significantly from the often used value $\chiA=1/15$ first
suggested by \citet{RudermanSutherland1975}, especially for higher energy
photons.

The contour plots of the mean free path of the photons $\MFP$ emitted
tangentially to the magnetic field lines are shown on Fig.~\ref{fig:MFP}; $\MFP$
was calculated according to eq.~\eqref{eq:MFP}. As expected it scales linearly
with $1/\epsilonG$ and $1/B$, the deviation from the linear behavior is seen only
for the combination of $\epsilonG$ and $B$ when pair formation happens near the
threshold, at $B\gtrsim\sci{1.5}{12}$G.

\subsection{Photon splitting}
\label{sec:phot_split}

\begin{figure*}[t]
\includegraphics[clip,width=\textwidth]{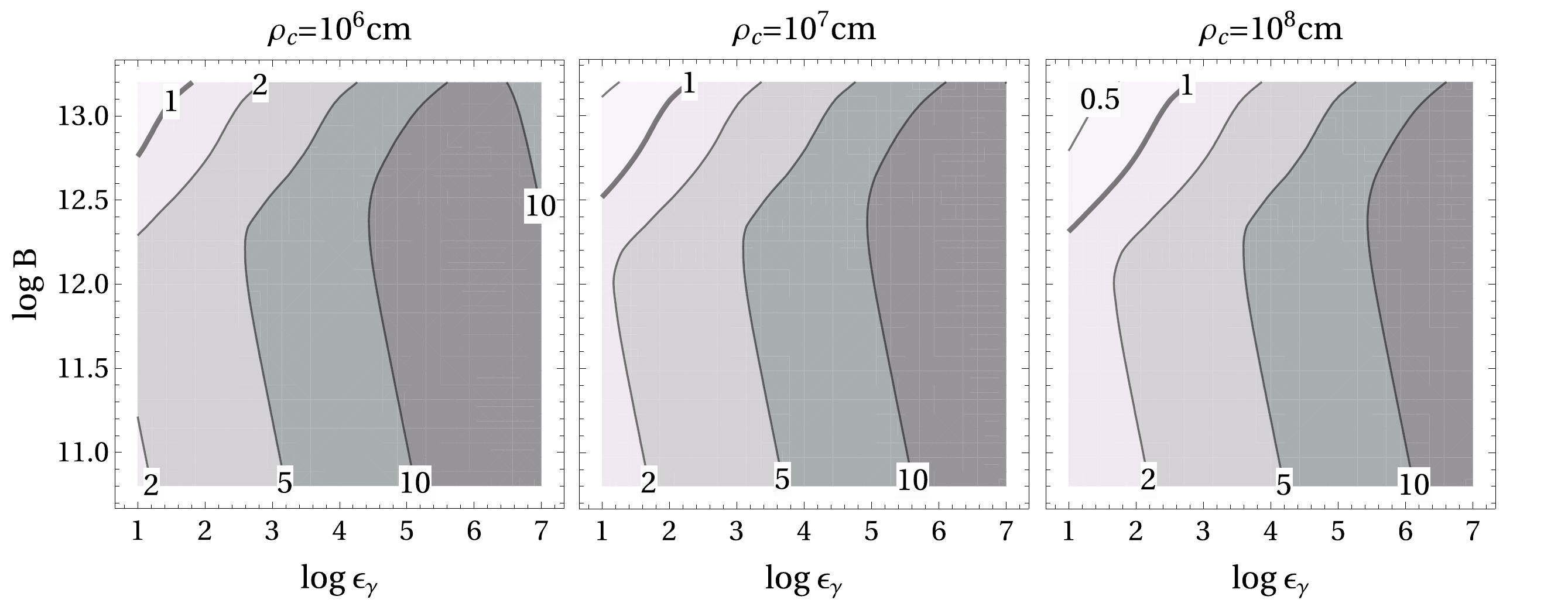}
\caption{Contour plot of the ratio of mfp for photon splitting to the mfp for
  pair production $\MFPSplit/\MFP$ (in linear scale) as a
  function of the logarithms of the magnetic field strength $B$ in Gauss, and
  photon energy $\epsilonG$ normalized to the electron rest energy, for three
  values of the radius of curvature of magnetic field lines
  $\rhoC=10^6, 10^7, 10^8\mbox{cm}$. }
\label{fig:MFP_Split2Pair}
\end{figure*}

Single photon pair creation is not the only process responsible for photon
attenuation in strong magnetic field, albeit the most significant one.  The most
important competing process to pair creation is magnetic photon splitting
$\gamma\rightarrow\gamma\gamma$.  Besides the end product, the major differences
between pair creation and photon splitting are: (i) pair creation is a first
order QED process and splitting is a third order one, therefore splitting is
weaker than pair creation by the order of $\alphaF^2$; (ii) in contrast to pair
creation, splitting has no threshold for photon's energy; (iii) while in
moderately strong magnetic fields $B\lesssim\SUB{B}{cr}$ for pair creation both
modes of photon polarization ($\parallel$ -- when photon's electric field is
parallel to the plane containing $\vec{B}$ and photon's momentum, $\perp$ --
photon's electric field is perpendicular to that plane) have similar
cross-sections and threshold conditions, below pair formation threshold photon
splitting is allowed only for the process $\perp\rightarrow\parallel\parallel$
\citep{Adler1971,Usov2002}.

Radiation processes relevant for secondary particles in polar cap cascades of
energetic pulsars (SR, RICS) produce predominantly $\perp$ polarized
photons. Despite the inherently smaller cross-section of magnetic splitting, the
absence of an energy threshold could allow photons to split before acquiring
large enough angles to the magnetic field to produce pairs, thus reducing
cascade multiplicity.  As we are interested in the most efficient cascades, a
regime where magnetic splitting becomes important is beyond the scope of this
paper.  More details on cascade kinetics in the presence of photon splitting can
be found in
\citet{HardingBaringGonthier1997,BaringHarding1997,BaringHarding2001}.  Here we
want to establish the boundary in the parameter space where photons splitting
start affecting cascade multiplicity. To do so we consider a case of photon
splitting $\perp\rightarrow\parallel\parallel$ for photons below pair formation
threshold.

The attenuation coefficient for photon splitting
$\perp\rightarrow\parallel\parallel$ is \citep{Baring2008,BaringHarding2001}
\begin{equation}
  \label{eq:alpha_gg}
  \Split{\alpha}(\epsilonG,\psi)=
  \frac{\alphaF^3 c}{60\pi^2 \lambdaC}\epsilonG^5 b^6 \sin\psi^6 \mathcal{M}_1^2
\end{equation}
at low values of the magnetic field perpendicular to the photon's trajectory  -- for
photons below the pair formation threshold -- the scattering amplitude
$\mathcal{M}_1$ is a constant independent of $b$ $\mathcal{M}_1\approx26/315$.
Integration of $\Split{\alpha}$ over the distance gives the optical depth for photon
splitting (cf. eq.~\eqref{eq:tau_general}); the mfp for splitting can then be
estimated as
\begin{equation}
  \label{eqMFP_splitting}
  \Split{\lambda}=1.8\,\epsilonG^{-5/7}b^{-6/7}\rhoC^{6/7}\,\mbox{cm}
\end{equation}

When the mfp for splitting becomes smaller than the mfp for pair formation,
$\MFPSplit<\MFP$, the photon splits before producing a pair.  In
Fig.~\ref{fig:MFP_Split2Pair} we plot the ratio $\MFPSplit/\MFP$ as a function
of magnetic field strength and photon energy, for three values of the radius of
curvature of magnetic field lines.  It is evident from these plots, splitting is
an important attenuation mechanism only for strong magnetic fields and low energy
photons.  With the increase of the magnetic field, photon splitting starts
affecting first the last cascade generation where the energy of the photons
becomes low. If these photons split, the resulting photons will be below pair
formation threshold.  In the polar cap cascades most of the pairs are produced
in the last cascade generation, and when photon splitting becomes important, the
cascade multiplicity can drop significantly. The exact fraction of
perpendicularly polarized photons in the cascade -- which are subject to
splitting -- in general depends on particle energy distributions, but it is more
than \%50 \citep[e.g.][]{BaringHarding2001}. Hence, the multiplicity of the pair
cascade will drop by at least a factor of 2 when photon splitting becomes
important.

The critical magnetic field strength $\Split{B}$ above which cascade
multiplicity becomes affected by photon splitting is the field strength when
$\MFPSplit<\MFP$ for the last generation photons, i.e. at for photons with the
escaping energy $\epsilonGesc$ (which we calculate in the next section)
\begin{equation}
  \label{eq:Bsplit_cr}
  \Split{B}: \MFPSplit(\epsilonGesc)=\MFP(\epsilonGesc)\,.
\end{equation}

\subsection{Energy of escaping photons}
\label{sec:E_esc}

As we discussed above, photons escaping the cascade are those with mfp 
larger than the characteristic scale of magnetic field variation $\MFP>L_B$.
The formal criteria we use to calculate the energy of escaping photons
$\epsilonGesc$ is $\MFP(\epsilonGesc)=\ESC{s} \RNS$; $\ESC{s}$ is a
dimensionless parameter quantifying the escaping distance in units of $\RNS$.
From the expression for mfp, eq.~\eqref{eq:MFP}, we get a (non-linear)%
\footnote{The non-linearity in this equation is because of non-linear dependency
  of $\chiA$ on $\epsilonGesc$, $b$, and $\rhoC$.}
equation for $\epsilonGesc$
\begin{equation}
  \label{eq:eps_esc_eq}
  \epsilonGesc = 2\,\frac{\rhoC}{\ESC{s}\RNS}\frac{\chiA}{b}\,.
\end{equation}

Any global NS magnetic field near the surface decays with distance as
$(r/\RNS)^{-\delta}$, $\delta\ge3$; a dipole field, $\delta=3$, is often
considered as a reasonable assumption. A pure dipole, however, seems to be too
idealized an approximation, as the NS magnetic field is slightly disturbed by
the currents flowing in the magnetosphere.  Polar cap cascade models should
consider at least near dipole magnetic fields with different curvatures of
magnetic field lines. Hence, a reasonable estimate for $L_B$ would be the
distance of the order of the NS radius $\RNS$. For our approximation of constant
$B$ and $\rhoC$ we found that the value $L_B=\RNS/2$ -- at that distance from
the NS the magnetic field decays by at least the factor of 3 -- provides a good
fit to the results of numerical simulations described in \PapI%
\footnote{in the semi-analytical cascade model of \PapI we used $L_B=\RNS$, but
  our current model works better for $L_B=\RNS/2$ when compared with numerical
  simulations.}.
Results described in this paper are obtained assuming $\ESC{s}=0.5$.

\begin{figure}
  \centering
  \includegraphics[clip,width=\columnwidth]{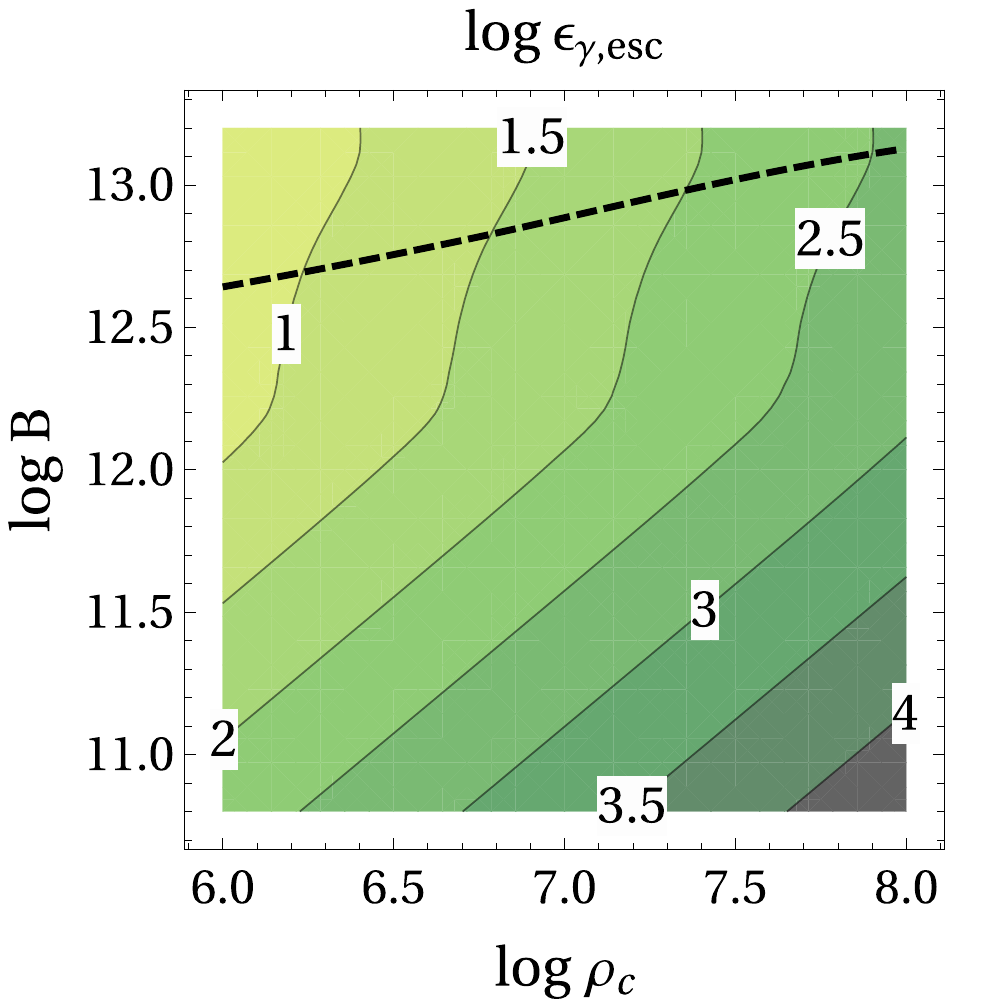}
  \caption{Energy of escaping photons: contours of $\log{\epsilonGesc}$ as a
    function of logarithms of the radius of curvature of magnetic field lines
    $\rhoC$ in cm and magnetic field strength $B$ in Gauss for $\ESC{s}=1$. The
    critical magnetic field $\Split{B}$ above which photon splitting starts
    affecting cascade multiplicity is shown by the dashed line.}
  \label{fig:Eesc}
\end{figure}

In Fig.~\ref{fig:Eesc} we plot the energy of escaping photons,
$\log{\epsilonGesc}$ as a function of the radius of curvature of magnetic field
lines $\rhoC$ and magnetic field strength $B$ for $\ESC{s}=1$.  This figure
shows (an obvious) trend that for higher magnetic field and smaller radii of
curvature, the energy of escaping photons is lower. The deviation of contours
from straight lines for $B\gtrsim\sci{3}{12}$G is due to the change of $\chiA$
behavior near the pair formation threshold (see Fig.~\ref{fig:inverse_chi} and
the next paragraph). For different values of $\ESC{s}$ the escape energy
$\epsilonGesc$ could be estimated from Fig.~\ref{fig:MFP}.

The critical magnetic field $\Split{B}$ above which photon splitting starts
affecting cascade multiplicity is shown in Fig.~\ref{fig:Eesc} by the dashed
line, $\Split{B}$ is calculated from eq.~\eqref{eq:Bsplit_cr}.  Cascade
multiplicity will drop due to photon splitting at $B\gtrsim\sci{4.4}{12}$G for
$\rhoC=10^{6}$cm, and at $B\gtrsim\sci{1.3}{13}$G for $\rhoC=10^{8}$cm.  The
increase of $\Split{B}$ for larger $\rhoC$ is due to the increase of the energy
of escaping photons -- $\MFP$ has stronger dependence on $\epsilonG$ than
$\MFPSplit$, that leads to the increase of the value $\Split{B}$ according to
eqs.~\eqref{eq:Bsplit_cr},~\eqref{eqMFP_splitting},~\eqref{eq:MFP}.

\begin{figure}
\includegraphics[clip,width=\columnwidth]{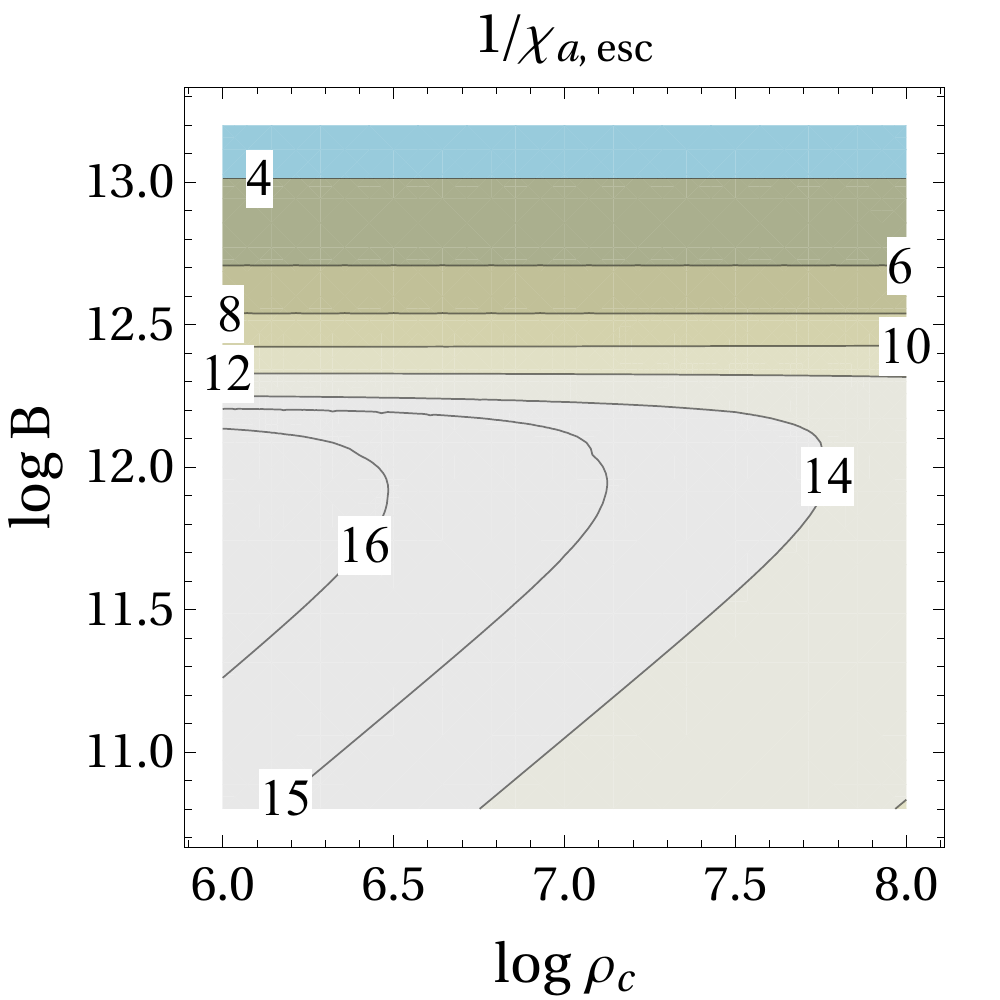}
\caption{Contour plot of $\SUB{\chi}{a,esc}$ as a function of the logarithms of
  the magnetic field strength $B$ in Gauss, and the radius of curvature of
  magnetic field lines $\rhoC$.}
\label{fig:inverse_chi_esc}
\end{figure}

It would be useful to have an analytical expression for the energy of escaping
photons; to obtain it we construct an approximation for $\chiAesc$.  We solved
eq.~\eqref{eq:eps_esc_eq} to find $\epsilonGesc$; now, using the interpolation
formula for $1/\chiA$, we can find $1/\chiAesc\equiv{}1/\chiA(\epsilonGesc)$.
In Fig.~\ref{fig:inverse_chi_esc} we show contours of $1/\chiAesc$ as a function
of $B$ and $\rhoC$. There are two distinct regions on this plot: for
$B\gtrsim\sci{3}{12}$G $\chiAesc$ changes very slowly, while for larger values
of B it changes significantly but does not depend on $\rhoC$. For
$B\gtrsim\sci{3}{12}$G the absorption of the last generation photons happens
near the pair formation threshold, when
$\epsilon_\perp=\epsilonGesc(\RNS/\rhoC)\simeq2$, and so $\chiAesc\simeq{}b$.
For weaker magnetic fields the opacity for near-threshold photons are too small
for them to be absorbed after traveling the distance $\RNS$, so the last
generation photons have energies larger than the pair formation threshold and
are absorbed at very similar values of $\chiA$. We find that the following
approximation work quite well
\begin{equation}
  \label{eq:chi_esc_Approx}
  \chiAesc =
  \begin{cases}
    b,    & \text{if $b>1/15$}   \\
    1/15, & \text{if $b\leq1/15$}
  \end{cases}
\end{equation}
The energy of the escaping photons can be expressed from
eq.~\eqref{eq:eps_esc_eq} as
\begin{equation}
  \label{eq:epsilon_esc_analythic}
  \epsilonGesc\approx{}\sci{1.8}{3}\,\frac{\SUB{\rho}{c,7}}{B_{12}}
  \left(\frac{\ESC{s}}{0.5}\right)^{-1}\chiAesc\;,
\end{equation}
where $\chiAesc$ is given by eq.~\eqref{eq:chi_esc_Approx} and
$B_{12}\equiv{}B/10^{12}$G and $\SUB{\rho}{c,7}\equiv\rhoC/10^7$cm. This simple
prescription for $\epsilonGesc$ deviates from numerical values shown in
Fig.~\ref{fig:Eesc} by no more than 20\% for $B\lesssim\sci{2}{12}$G and
$B\gtrsim\sci{8}{12}$G, the largest deviation is $\sim60\%$ at
$B\simeq\sci{3}{12}$G.

For very small values of $\chiAesc$ it is possible to get a more accurate
analytical expression for $\chiAesc$ (and $\epsilonGesc$).  Using asymptotic
expression for the optical depth in the limit of $\chi\ll1$ derived in
Appendix~\ref{sec:optic-depths-phot}, eq.~\eqref{eq:APP_tau_OTS_small_Chi} and
substituting for the photon energy the energy of escaping photons
eq.~\eqref{eq:eps_esc_eq} we get
\begin{align}
  \label{eq:inverse_chi_esc_smallB}
  \frac{1}{\chiAesc} & \approx \frac{3}{4}\ln\left(\frac{3
                       A_\tau\RNS^2}{16}\frac{b}{\rhoC}\chiAesc\,\ESC{s}^{2}\right)\\
                     & \approx 15.7+1.7\log\left(\frac{B_{12}}{\rho_7}\right)+
                       3.5\log\left(\frac{\ESC{s}}{0.5}\right)\nonumber
\end{align}

\section{Particle acceleration}
\label{sec:particle-acceleration}

Self-consistent modeling of accelerating zones in pulsar polar caps
\citep{Timokhin2010::TDC_MNRAS_I,TimokhinArons2013} demonstrated that particle
acceleration and pair formation are always non-stationary. Each period of
intense particle acceleration and pair formation is followed by a period of
quiet plasma flow when the accelerating electric field is screened and no pairs are
formed.  At the end of the quiet phase an accelerating gap begins to form -- a
region where plasma density is significantly smaller than the local GJ number
density $\GJ{\eta}/e$.  Accelerating electric field in the gap increases
linearly with distance as the gap grows.  Charged particles entering the gap
are accelerated to very high energies and emit gamma-rays which give rise to
electron-positron pair cascades.  Dense pair plasma created in the cascades
screens the electric field, stopping the growth of the gap.  The gap does not 
stay at the same place, but moves along magnetic field lines roughly
preserving its size (and the potential drop) for a while.  Most of the pair
plasma is created at or behind the trailing edge of the gap, where the high
energy particles are located%
\footnote{see e.g. Figs.~22 and 23 from \PapI{}}.
These particles move into the magnetosphere, emitting gamma-rays which convert
into electron positron pairs. Both primary and secondary particles are
relativistic and so they move together, forming a blob of pair plasma those
density increases as pair formation continues.  Some low energy particles,
however, ``leak'' from the blob, creating a tail of mildly relativistic plasma
which screens the electric field behind the blob.  When the blob with primary
particles move away from the polar cap and pair formation stops, the dense pair
plasma from the tail keeps the electric field screened for a while until most of
it has left the polar cap zones and a new cycle of pair formation begins%
\footnote{see e.g. Fig.~2 from \citet{Timokhin2010::TDC_MNRAS_I} which gives an
  overview of the entire cycle of pair formation described above -- it shows
  snapshots of the charge density distribution in the polar cap over the whole
  cycle.}.

Whether and how efficient the pair formation along given magnetic field lines
occurs depends on the ratio $\jm/\GJ{j}$ of the current density required to
support the twist of magnetic field lines in the pulsar magnetosphere
\citep[e.g.][]{Timokhin2006:MNRAS1,BaiSpitkovsky2010a},
$\jm\equiv(c/4\pi)\,|\nabla\times\mathbf{B}|$, to the local GJ current density,
$\GJ{j}\equiv\GJ{\eta}c$, where $\GJ{\eta}=B/Pc$ is the GJ charge density.
Regardless of the ability of the NS surface to supply charged particles, i.e. in both the 
space charge limited flow model of \citet{AronsScharlemann1979} and the no-particle
extraction model of \citet{RudermanSutherland1975}, particle acceleration
happens in essentially the same way.  For the \citet{RudermanSutherland1975}
regime effective particle acceleration and pair formation is possible for almost
all values of $\jm/\GJ{j}$.  In the space charge limited flow regime pair
formation is not possible if $0<\jm/\GJ{j}<1$, but is possible for all other
values of $\jm/\GJ{j}$.  A detailed description of particle acceleration in pulsar
polar caps is given in \citet{Timokhin2010::TDC_MNRAS_I} for the no-particle
extraction regime and in \citet{TimokhinArons2013} for the space charge limited flow
regime.

Although the character of plasma flow inferred from self-consistent simulation
of \citet{Timokhin2010::TDC_MNRAS_I} and \citet{TimokhinArons2013} qualitatively
differs from that assumed in both \citet{RudermanSutherland1975} and
\citet{AronsScharlemann1979} type models, the physics of particle acceleration
in the gap is similar to that of the accelerating gap in the
\citet{RudermanSutherland1975} model.  Namely, due to significant deviation of
the charge and current densities from GJ values, the electric field in the gap is
comparable to the vacuum electric field ($\sim{}h\Omega{}B/c$, $h$ is the size
of the gap) and particles are accelerated in a short gap by the strong electric
field which increases linearly with the distance. In \S6.2 of \PapI{} we
analyzed the physics of particle acceleration and derived an analytical
expression for the energy of the primary particles accelerated in non-stationary
cascades.  According to eq.~(41) in \PapI{} the final energy of particles
accelerated in the gap
\begin{eqnarray}
  \label{eq:epsilonP_final}
  \epsilonPacc & =      & \frac{49}{18} \left( \frac{\pi B_q}{\lambdaC^3 c} \right)^{1/7}
                          \chiAacc^{2/7}\; \xi_j^{1/7}  P^{-1/7} B^{-1/7} \rhoC^{4/7}\nonumber \\
               & \simeq & \sci{5}{7}\: \chiAacc^{2/7}\; \xi_j^{1/7}
                                       P^{-1/7} B_{12}^{-1/7} \SUB{\rho}{c,\,7}^{4/7}\,.
\end{eqnarray}
$\chiAacc$ is the value of the parameter $\chi$ for photons which create pairs
terminating the gap.  $\xi_j$ is a factor which shows how stronger/weaker the
electric field in the gap is compared to the field in a static vacuum gap of an
aligned rotator:
\begin{equation}
  \label{eq:xi_j_def}
  \xi_j\equiv \frac{|j-\jm|}{\GJ{j}^0} \left( 1+\frac{c}{v} \right)
  \approx{}2\,\frac{\jm}{\GJ{j}^0}\,.
\end{equation}
where $j$ is the current density in the gap.  In most cases $|j-\jm|\simeq\jm$;
$\GJ{j}^0$ is the GJ current density in an aligned rotator
\begin{equation}
  \label{eq:jGJ_0}
  \GJ{j}^0\equiv\GJ{\eta}^0c=\frac{B}{P}\,.
\end{equation}
$v$ is the velocity of the gap; in most cases the gap moves with relativistic
velocities, so $v\simeq{}c$.  Taking into account these approximations we get
the second expression for $\xi_j$ in eq.~\eqref{eq:xi_j_def}. $\jm$, and so
$\xi_j$, depend on the pulsar inclination angle and the position of the given
magnetic field line inside the polar cap (see e.g. Fig.~1 in
\citet{TimokhinArons2013}). In cascades along magnetic field lines where $\jm$
is close to the local value of $\GJ{j}$ in an aligned rotator $\xi_j\sim2$, for
the same situation in a pulsar with inclination angle of $60^{\circ}$,
$\xi_j\sim1$.  The energy of primary particles eq.~\eqref{eq:epsilonP_final} has
the same dependence on $\rhoC$, $P$ and $B$ as the expression for the potential
drop in the gap derived by \citet{RudermanSutherland1975}, their eq.~(23).  This
is to be expected as in both cases particles are accelerated by the electric
field which grows linearly with the distance and the size of the gap is
regulated by absorption on curvature photons in magnetic field.  The difference
is in the presence of factor $\xi_j$ and a different numerical factor.

\begin{figure}
  \includegraphics[clip,width=\columnwidth]{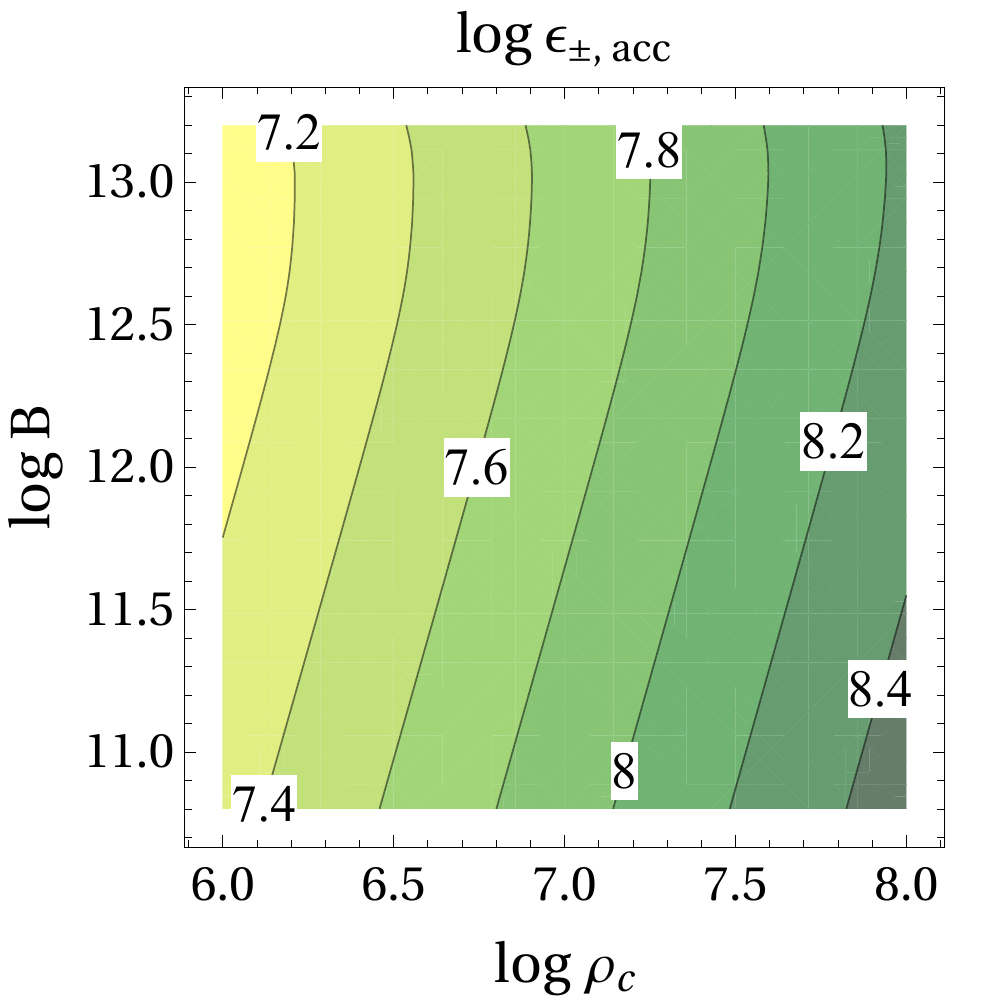} 
  \caption{Primary particle energy: contours of
    $\log\epsilonPacc$ as a function of logarithms of the
    radius of curvature of magnetic field lines $\rhoC$ in cm and
    magnetic field strength $B$ in Gauss. We used the following values
    for gap parameters $P=33$~ms, $\xi_j=2$.}
  \label{fig:e_acc}
\end{figure}

\begin{figure}
  \includegraphics[clip,width=\columnwidth]{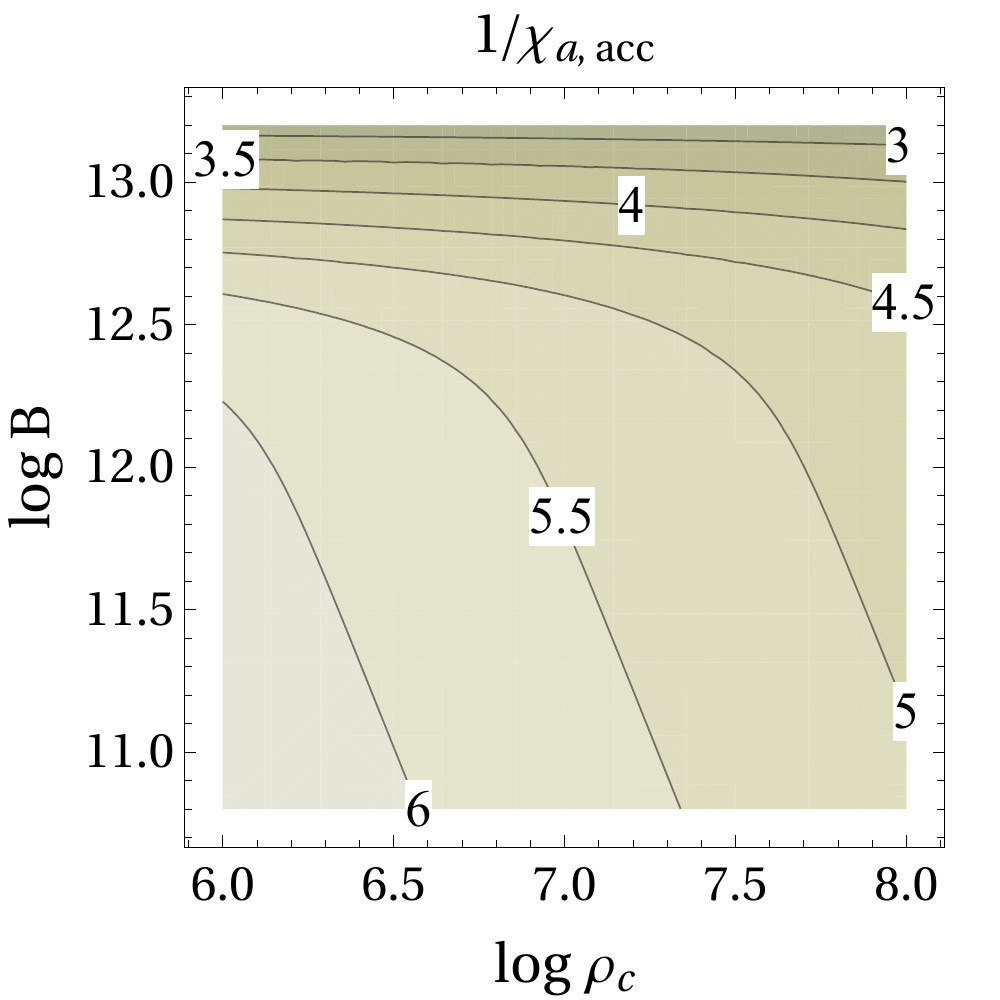} 
  \caption{Contour plot of $1/\chiAacc$ as a function of logarithms of the
    radius of curvature of magnetic field lines $\rhoC$ in cm and magnetic field
    strength $B$ in Gauss.  We used the following values for gap parameters
    $P=33$~ms and $\xi_j=2$.}
  \label{fig:chi_a_acc}
\end{figure}

Here we slightly improve the accuracy of this expression by calculating
self-consistently the value of the parameter $\chiAacc$.  Using numerical
interpolation for $\chi$ (see \S\ref{sec:phot_pair_creation}) it is easy to
obtain self-consistently the energy of primary particles accelerated in the gap
$\epsilon$ by solving numerically the equation
\begin{equation}
\label{eq:epsilonP_SC}
\epsilonPacc(\chiAacc(\epsilon))=\epsilon\,,
\end{equation}
i.e. to find the energy of particles which emit photons terminating the gap
growth taking into account the dependence of $\chiAacc$ on particle energy (via
the energy of emitted photons).  In Fig.~\ref{fig:e_acc} we show the energy of
accelerated particles for a pulsar with $P=33$~ms. The contours of constant
$\chiAacc$ deviate only slightly from straight lines corresponding to
$\propto\rhoC^{4/7}B^{-1/7}$ for higher values of $B$.  This deviation is due to
variation of $\chiAacc$ near the pair formation threshold and so the
expression~\eqref{eq:epsilonP_final} can be safely used in many cases with a
constant value of $\chiAacc$.  $\chiAacc$ itself varies slowly with gap 
parameters. In Fig.~\ref{fig:chi_a_acc} we show $1/\chiAacc$ for a pulsar with
$P=33$~ms; it was obtained together with values from
eq.~\eqref{eq:epsilonP_SC}. It is evident that the value of $\chiAacc=1/5.5$
can be used in eq.~\eqref{eq:epsilonP_final} for pulsars with $P=33$~ms.
The dependence of $\chiAacc$ on pulsar period $P$ and parameter $\xi_j$ is also
quite weak, see in Table~\ref{tab:chiAcc} where we show the variation of
$\chiAacc$ at $B=10^{12}$G, $\rhoC=10^7$cm with $P$ and $\xi_j$ and for
estimates of the primary particle energies one can use $\chiAacc=1/7$.
The dependence of the particle energy is weak and along most of magnetic field line
in the polar cap $\xi_j$ is no more than $\sim$an order of magnitude lower than
2. Assuming $\chiAacc=1/7$ and $\xi_j=2$ eq.~\eqref{eq:epsilonP_final} can be
written
\begin{equation}
  \label{eq:epsilonP_primary}
  \epsilonPacc \simeq  \sci{3.2}{7}\:  P^{-1/7} B_{12}^{-1/7} \SUB{\rho}{c,\,7}^{4/7}\,.
\end{equation}
This estimate for the primary particle energy can be used for a wide rage of
parameters of young energetic pulsars. It is $\simeq4$ times higher than the
that given by eq.~(23) in \citet{RudermanSutherland1975}.

\begin{table}[h]
  \centering
  \begin{tabular}{LLL}
    P    & \xi_j=0.25 & \xi_j=2 \\
    \hline\hline
    0.033 & 6.6        & 5.5     \\
    0.33  & 7.9        & 6.6     \\
  \end{tabular}  
  \caption{$1/\chiAacc$ for $B=10^{12}$G, $\rhoC=10^7$cm and different values of
    $P$ and $\xi_j$}
  \label{tab:chiAcc}
\end{table}

The primary particle energy has a weak dependence on pulsar period, inclination
angle (via $\xi_j$), and magnetic field strength; its strongest dependence is on
the radius of curvature of magnetic field lines. These trends are consequences
of the fact that the potential drop across the acceleration gap is regulated by
pair formation.  The gap terminates when particles reach energies high enough to
emit pair producing photons.  A gap with weak accelerating electric field due to
e.g. weaker magnetic field and/or longer period and/or smaller current density
$\jm$ will have larger height than a gap with strong accelerating field, to
accelerate particles to the energies when they emit pair producing
photons. A larger height of the gap also results in longer distances traveled by
photons; this largely alleviates the dependence of the energy of the pair
producing photons on the magnetic field strength, leaving the curvature of
magnetic field lines as the strongest factor determining the energy of primary
particles.

\section{The maximum pair multiplicity: simple estimate}
\label{sec:simple-multiplicity}

\begin{figure*}
  \includegraphics[clip,width=\textwidth]{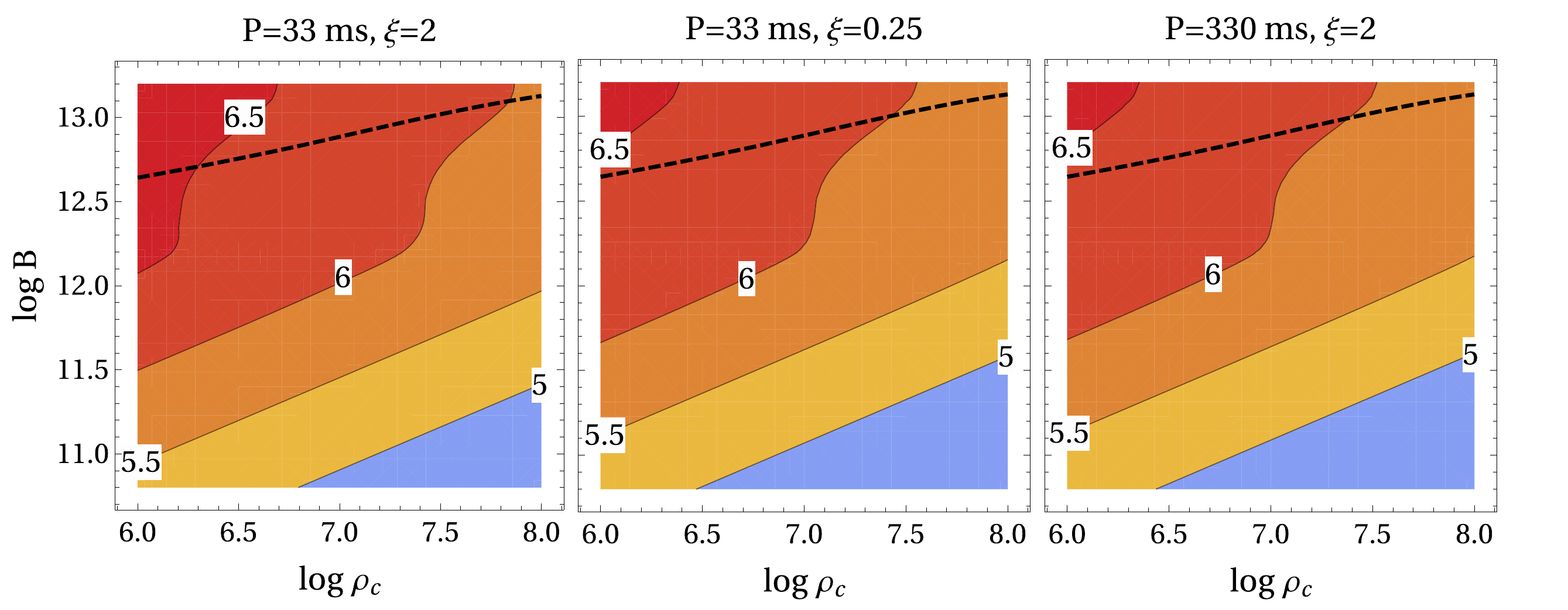}  
  \caption{Simple estimate for the maximum multiplicity of polar cap cascades:
    contours of $\log\SUB{\kappa}{max}$ as a function of logarithms of curvature
    of magnetic field lines $\rhoC$ in cm and magnetic field strength $B$ in
    Gauss for three sets of the gap parameters $(P[\mbox{s}],\xi_j)$:
    $(0.033,2)$, $(0.033,0.25)$, $(0.33,2)$.  The critical magnetic field
    $\Split{B}$ above which photon splitting starts affecting cascade
    multiplicity is shown by the dashed line.}
  \label{fig:kappa-max}
\end{figure*}

Now we can make a simple estimate of the maximum cascade multiplicity during a
burst of pair creation. As discussed in \S\ref{sec:cascades-overview}, in a
hypothetical ideal cascade the whole kinetic energy of the primary particle is
divided into energies of pairs which are produced by the photons with the
energies just above the escape energy; in such a cascade the multiplicity is given
by eq.~\eqref{eq:kappa_max} -- twice the energy of the primary particle divided
by the energy of escaping photons.  In Fig.~\ref{fig:kappa-max} we show the
estimates for the multiplicity of an ideal cascade $\log\SUB{\kappa}{max}$ as a
function of the magnetic field strength and the radius of curvature of the
magnetic field lines for three sets of the gap parameters $(P[\mbox{s}],\xi_j)$:
$(0.033,2)$, $(0.033,0.25)$, $(0.33,2)$ . Energies of primary particles and
escaping photons are calculated according to
\Ss~\ref{sec:E_esc},~\ref{sec:particle-acceleration}.  The maximum cascade
multiplicity is not very sensitive to pulsar period and inclination angle (via
$\xi_j$); the strongest dependence is on the magnetic field strength. The
maximum value of $\SUB{\kappa}{max}$ is about $\sci{3}{6}$, which is the absolute
upper limit on the polar cap cascade multiplicity.

Analytical expression for the maximum cascade multiplicity can be obtained using
expressions for $\epsilonPacc$ and $\epsilonGesc$ from
\Ss~\ref{sec:particle-acceleration},~\ref{sec:E_esc}.  Substituting
eq.~\eqref{eq:epsilonP_final} and eq.~\eqref{eq:epsilon_esc_analythic} into
eq.~\eqref{eq:kappa_max} we get an estimate on the upper limit of the cascade
multiplicity
\begin{eqnarray}
  \label{eq:kappa_max_simple}
  \SUB{\kappa}{max} & = & \sci{5.7}{4}\,P^{-1/7} \SUB{\rho}{c,\,7}^{-3/7}  B_{12}^{6/7} \nonumber \\
                    &   & \times  
  \chiAacc^{2/7}\; \xi_j^{1/7} \chiAesc^{-1} \left(\frac{\ESC{s}}{0.5}\right)\,.
\end{eqnarray}
The weak dependence of $\SUB{\kappa}{max}$ on pulsar period and inclination
angle (via $\xi_j$) is evident from this formula; this is a consequence of the
weak dependence of $\epsilonPacc$ on these parameters. The strong
dependence of $\SUB{\kappa}{max}$ on the magnetic field strength is due to the 
strong dependence of $\epsilonGesc$ on $B$.  Using values for $\chiAacc$,
$\xi_j$ assumed in \S\ref{sec:particle-acceleration} and the approximation for 
$\chiAesc$ given by eq.~\eqref{eq:chi_esc_Approx} we get two final expressions
for $\SUB{\kappa}{max}$ valid for $B\lesssim\sci{3}{12}$G
\begin{equation}
  \label{eq:kappa_max_simple_smallB}
  \SUB{\kappa}{max} =\sci{5.4}{5} \SUB{\rho}{c,\,7}^{-3/7} P^{-1/7} B_{12}^{6/7}
\end{equation}
and for $B\gtrsim\sci{3}{12}$G
\begin{equation}
  \label{eq:kappa_max_simple_largeB}
  \SUB{\kappa}{max} =\sci{1.6}{6} \SUB{\rho}{c,\,7}^{-3/7} P^{-1/7} B_{12}^{-1/7}
\end{equation}

For higher magnetic field strengths and smaller radii of curvature of magnetic
field lines the energy of the primary particles is larger and the energy of
escaping photons smaller. The energy available for the cascade and, hence, the
maximum cascade multiplicity, increases towards higher $B$ and lower $\rhoC$
values. This dependence on $B$ saturates at $B\sim\sci{3}{12}$G because photon
absorption at higher field strengths will happen near pair formation threshold,
limiting the decrease of escaping photons' energy.

In real pulsar cascades the multiplicity will be (substantially) smaller than
$\SUB{\kappa}{max}$ mainly because (i) not all of the kinetic energy of
primary and secondary particles is transferred to pair producing photons, (ii)
the last generation photons have energies above the pair formation threshold,
(iii) pair production is intermittent; no pairs are produced during the quiet
cascade phase.  The first two issues are related to the physics of the cascade
and we will address them in the next section. The last issue is directly related
to the physics of the screening of the electric field and plasma physics in the
blob of freshly formed pair plasma; it can be addressed only by means of
self-consistent high resolution simulations like
\citet{Timokhin2010::TDC_MNRAS_I,TimokhinArons2013} and will be the subject of
future research. We can provide only very rough estimates on the effect of pair
formation intermittency on the effective polar cap cascade multiplicity.

We can see from the results of this section that the absolute upper limit on the
cascade multiplicity in a single burst of pair formation is
$\SUB{\kappa}{max}\lesssim\sci{3}{6}$, with the real effective multiplicity
being significantly smaller. This already excludes the possibility of extremely
high cascade multiplicities $\sim{}10^{6}-10^{7}$ assumed in some theories of
PWNe and pulsar high energy emission
\citep[e.g.][]{BucciantiniAronsAmato2011,Lyutikov2013}.

\section{Semi-analytical cascade model}
\label{sec:cascade-devel}

\begin{figure}[t]
  \centering
  \includegraphics[clip,width=\columnwidth]{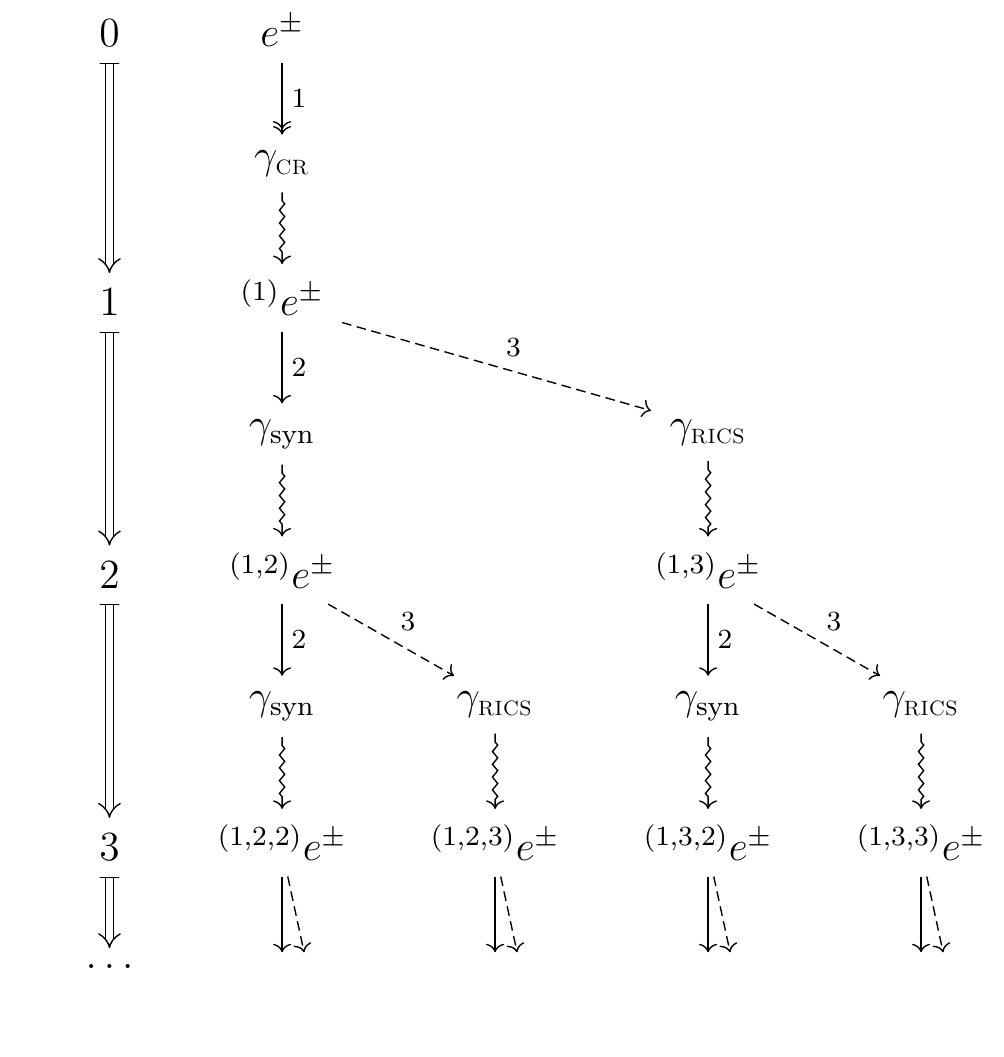}
  \caption{Diagram showing the general chain of physical processes in a strong
    polar cap cascade. Cascade generations are shown on the left -- numbers
    connected by double arrows.  Electrons and positrons $e^\pm$ produce photons
    which are turned into pairs of the next cascade generation: $\CR{\gamma}$ --
    via curvature radiation (solid line with double arrow labeled ``1''),
    $\SYN{\gamma}$ -- via synchrotron radiation (solid lines with arrow labeled
    ``2''), and $\SUB{\gamma}{RICS}$ -- via Resonant inverse Compton scattering
    (dashed lines with arrow labeled ``3'').  Numbers in parenthesis show the
    origin of each particle. }
  \label{fig:cascade-diagram}
\end{figure}

In \PapI{} we developed a simple semi-analytical cascade model which allowed us to
simulate non-branching cascades -- when only a single emission process is
involved -- and used it to explore CR-synchrotron cascades.  As we argued in
\PapI{}, such cascades develop in polar caps of moderately magnetized pulsars
(B$\lesssim10^{12}$G) where synchrotron radiation of secondary particles is the
only source of photons creating pairs in the next cascade generation.  In this
paper we are interested in extending the range of applicability of our model for
higher field pulsars as well as improving its accuracy.  Our new model differs
from the one in \PapI{} in two aspects (i) it applies to cascades with arbitrary
emission/absorption processes -- cascade branches can be arbitrarily complex and
(ii) it can account for the fact that emission mechanisms can be broadband and
not all emitted photons are able to create pairs.

\subsection{General Algorithm}
\label{sec:general-alg}

The spectral energy distribution of synchrotron and curvature radiation is
broadband with a significant amount of energy emitted well below the peak energy
$\SUB{\epsilon}{peak}$
\begin{equation}
  \label{eq:sed-synchrotron}
  F(\epsilon) \propto \frac{\epsilon}{\SUB{\epsilon}{peak}}\:
  \int_0^{\epsilon/\SUB{\epsilon}{peak}}K_{5/3}(\zeta)\,d\zeta
\end{equation}
where $K$ is the modified Bessel function of the order $5/3$.  In \PapI{} we
used a monoenergetic approximation for these processes -- all energy is
emitted as photons with energies $\SUB{\epsilon}{peak}$.  In our current model
we divide the spectrum into 3 spectral bins
$\{[0, 0.3\SUB{\epsilon}{peak}],
[0.3\SUB{\epsilon}{peak},1.5\SUB{\epsilon}{peak}],
[1.5\SUB{\epsilon}{peak},\infty]\}$%
\footnote{We experimented with larger number of spectral bins that leads only to
  a very moderate improvement in the accuracy of the results which did not
  justify the increase of computational time.}.
CR and synchrotron emission of particles is modeled as emission of photons
in each of the 3 spectral bins $j=1,2,3$ with energies
\begin{equation}
  \label{eq:epsilonSyn_bin}
  \epsilon^j\equiv{}f_\epsilon^j\:\SUB{\epsilon}{peak}\,,
\end{equation}
the number of photons emitted in each spectral bin is equal to the energy
emitted by the particle in that bin $W^i=f_w^i\,W$ ($W$ is the total emission
rate) divided by the characteristic energy of the photons
\begin{equation}
  \label{eq:N_photons_bin}
  n^j=\frac{W^j}{\epsilon^j}=\frac{f_w^j}{f_\epsilon^j}\frac{W}{\SUB{\epsilon}{peak}}\,.
\end{equation}
Coefficients $(f_\epsilon^{j}, f_w^{j})$ for energy bins we used are
$\{(0.3, 0.152)$, $(1, 0.518)$, $(1.5, 0.33)\}$; they are calculated by
integrating spectral energy distribution $F(\epsilon)$,
eq.~\eqref{eq:sed-synchrotron}, over the spectral bins.

In our algorithm both leptons and photons are macroparticles; the statistical
weight of each particle is the number of real particles it represents. We start
by calculating the energy of the primary particle accelerated in the gap
according to \S\ref{sec:particle-acceleration} and follow this particle as it
moves along magnetic field lines loosing energy emitting CR photons.  Each CR
photon initiates an electron-positron cascade with secondary particles emitting the 
next generation of pair producing photons via synchrotron radiation and Resonant
Inverse Compton scattering (RICS) of soft X-ray photons from the NS surface. We follow
every generation of photons until their energy falls below the escaping energy
according to \S \ref{sec:E_esc} and compute the number of pairs created by each
cascade generation.  The diagram in Fig.~\ref{fig:cascade-diagram} shows the
chain of physical processes initiated by a \emph{single} CR photon.  In
Fig.~\ref{fig:cascade-diagram}, ``rows'' represent different cascade generations
(particle with the same number of iterations particle/photon before their
creation starting with primary particles), while ``columns'' correspond to
branches (particles of the same generations produced by different emission
processes).  In each generation of the cascade, pairs $e^\pm$ produce photons
which are then turned into pairs of the next generation: $\CR{\gamma}$ -- via
curvature radiation (shown by solid line with double arrow labeled ``1''),
$\SYN{\gamma}$ -- via synchrotron radiation (shown by solid lines with arrow
labeled ``2''), and $\SUB{\gamma}{RICS}$ -- via Resonant inverse Compton
scattering (shown by dashed lines with arrow labeled ``3'').  Numbers in
parenthesis show the origin of each particle, for example, (1,3,2) means that
this pair was produced by synchrotron photon (2d generation) emitted by a pair
produced by RICS photon (1st generation) created by CR photon (0th generation).

Our algorithm is described in detail in Appendix~\ref{sec:algorithms}; here we
give its brief overview.  The central part of our algorithm is the recursive
function \texttt{PairCreation()}, Algorithm~\ref{alg:pair_creation_fun}.  For
each photon \texttt{PairCreation()} calculates whether and where it will be
absorbed to create a pair. The photon is counted as absorbed if its mfp is less
than the escaping distance, $\MFP\leq\ESC{s}\RNS$, and its absorption point $x$
is still inside the cascade zone, $x\leq\SUB{s}{cascade}\RNS$.  Then, for each
emission process, it calculates the energy of the next generation photons
emitted by the pair calling emission process specific function
\texttt{emissionFun()}. And, finally, it recursively calls itself for each of
the next generation photons.  We follow the primary particle as it moves along
magnetic field lines losing energy and emitting photons via curvature radiation
-- Algorithm~\ref{alg:full_cascade} in Appendix~\ref{sec:algorithms}. For each
CR photon \texttt{PairCreation()} is called and through its successive recursive
calls follow every branch of the cascade%
\footnote{In \PapI{} we considered only the CR-synchrotron cascade, i.e. we
  followed only the branch of the cascade represented by the first column in
  Fig.~\ref{fig:cascade-diagram}, particles with origins
  $\{(1), (1,2),(1,2,2),\dots\}$, cf. Fig.~(18) in \PapI; our algorithm then was
  simpler.}.
The total cascade multiplicity is calculated by integrating the number of
particles produced in cascades generated by each CR photon (computed by
recursive calls of \texttt{PairCreation()}) over the distance within the cascade
zone.  We assume that the size of the cascade zone is equal to the NS radius
$\RNS$, $\SUB{s}{cascade}=1$.

\subsection{Microscopic processes}
\label{sec:microscopic-processes}

We analyzed the microphysics of polar cap cascades of young energetic pulsars in
\PapI{} in great detail. Here we give a brief overview of how we treat the
cascade microphysics.

At the distance $s$ after exiting the acceleration
zone (hereafter all distances are normalized to $\RNS$) the energy of the
primary particle is (eq.~(19) in \PapI)
\begin{equation}
  \label{eq:CR_es}
  \epsilonP(s) = \epsilonP^0 \left[1 + 3 H \frac{(\epsilonP^0)^3}{\rhoC^2} s\right]^{-1/3}\,,
\end{equation}
where $\epsilonP^0$ is the initial particle energy,
$H=(2/3)\RNS{}r_e\approx\sci{1.88}{-7}\mbox{cm}^2$, $r_e=e^2/m_ec^2$ is the
classical electron radius. While traveling a segment of the length $ds$ the
energy emitted by the primary particle via CR (all energy quantities are
normalized to $m_ec^2$) is
\begin{equation}
  \label{eq:W_CR}
  \CR{W} = \frac{3}{2}\,\alphaF \frac{\lambdaC\RNS}{\rhoC^2}\,\epsilonP^4(s)\,ds
\end{equation}
The peak energy of CR radiation photons is
\begin{equation}
  \label{eq:Epeak_CR}
  \SUB{\epsilon}{\textsc{CR}, peak} = \frac{3}{2} \frac{\lambdaC}{\rhoC} \epsilonP{}^3
\end{equation}
The energies and statistical weights of macroparticles representing CR photons are
calculated from eq.~\eqref{eq:epsilonSyn_bin},~\eqref{eq:N_photons_bin} using
eqs.~\eqref{eq:W_CR},~\eqref{eq:Epeak_CR}

We follow the evolution of cascades initiated by CR photons by tracing the pair
producing photons as their energy degrades with each successive cascade
generation.  The energy of each pair-creating photon $\epsilonG$ is transferred
to an electron-positron pair, which is always created with a non-zero
perpendicular to the magnetic field momentum. The perpendicular energy is
emitted by the pair via synchrotron radiation shortly after pair creation. The
energy emitted as synchrotron photons is (see eq.~(13) in \PapI)
\begin{equation}
  \label{eq:W_syn}
  \SYN{W} = \epsilonG
  \left\{ 
    1 - \left[ 1+\left(\frac{\chiA}{b}\right)^2\right]^{-1/2}
  \right\}
\end{equation}
and the peak energy of the synchrotron radiation is
\begin{equation}
  \label{eq:Epeak_syn}
  \SUB{\epsilon}{syn, peak}=\frac{3}{4}\chiA \epsilonG\,,
\end{equation}
where $\chiA$ and $b$ are the value of the parameter $\chiA$ and normalized magnetic
field strength at the absorption point of the parent photon. As for CR, photons
energies and statistical weights of macroparticles representing synchrotron
photons are calculated from
eq.~\eqref{eq:epsilonSyn_bin},~\eqref{eq:N_photons_bin} using
eqs.~\eqref{eq:W_syn},~\eqref{eq:Epeak_syn}.

Pair particles can also scatter thermal photons from the NS surface. If it
happens in the cascade zone the kinetic energy associated with the motion of the particle parallel to
the magnetic field is transferred to the next generation
of pair producing photons. The maximum energy which can be emitted as RICS
photons is the pair's kinetic energy left after emission of synchrotron photons
\begin{equation}
  \label{eq:W_RICS_max}
  \RICS{W}^0 =
  \epsilonG^{i-1}\left[1+\left(\frac{\chiA}{b}\right)^2\right]^{-1/2}\,.
\end{equation}
The mean free path for RICS is given by
\citep{Zhang2000,Sturner1995,Dermer1990}:
\begin{eqnarray}
  \RICS{\lambda} & = & -0.061\, \epsilonP^2\, T_6^{-1} B_{12}^{-2}
                       \times \nonumber\\
                 && \ln^{-1}\left[1 - \exp\left( -\frac{134
                    B_{12}}{\epsilonP\, T_6 (1 -
                    \mu_s)}\right)\right] \, \mbox{cm}
  \label{eq:Lrics}
\end{eqnarray}
where $T_6$ is the temperature of the NS surface in units of $10^6$K, $B_{12}$
is the magnetic field strength in units of $10^{12}$~G, and
$\mu_s=\cos\theta_s$, where $\theta_s$ is the angle between the momenta of the
scattering photon and particle in the lab frame. Eq.~\eqref{eq:Lrics} implicitly
takes into account the condition that soft photons must be in cyclotron
resonance to be scattered, it is obtained by integration of the resonant
cross-section with a blackbody spectrum of target photons \citep{Dermer1990}.
If the mfp for RICS is larger than the size of the cascade zone,
$\SUB{\lambda}{RICS}>\RNS$, we assume that no RICS pair producing photons are
emitted. As particles move away from the NS the probability of RICS decreases
relative to that at the injection point due to the decrease on the magnetic
field strength and the number density of soft photons. To account for this
effect we assume that if $\SUB{\lambda}{RICS}<0.1\,\RNS$ all pair's kinetic
energy left after emission of synchrotron photons is transferred to RICS
photons, this fraction linearly decrease as $\SUB{\lambda}{RICS}$ is getting
bigger $\propto{}0.1\RNS/\SUB{\lambda}{RICS}$, so that the energy emitted as
RICS photons is
\begin{equation}
  \label{eq:W_RICS}
  \RICS{W} =
  \begin{cases}
    \RICS{W}^0, & \textrm{if } \RICS{\lambda}\le{}0.1\,\RNS\\  
    \frac{\RNS}{\SUB{\lambda}{RICS}}\:\RICS{W}^0, & \textrm{if } 0.1\,\RNS<\RICS{\lambda}\le\RNS\\
    0, & \textrm{if } \RICS{\lambda}>\RNS \,.\\
  \end{cases}
\end{equation}
The spectrum of RICS radiation is narrow-band and we approximate this process as
emission of monochromatic photons with the energy (see eq.~(49) in \PapI)
\begin{equation}
  \label{eq:Ephot_RICS}
  \epsilonGRICS=\epsilonG\, b\left[ 1 + \left(\frac{\chiA}{b}\right)^2 \right]^{-1/2}\,.
\end{equation}
This number of RICS photons emitted by each pair is
\begin{equation}
  \label{eq:N_RICS}
  \RICS{n}=\frac{\RICS{W}}{\epsilonGRICS}\,. 
\end{equation}

\subsection{Model applicability}
\label{sec:valid-appr}

\begin{figure}
  \includegraphics[clip,width=\columnwidth]{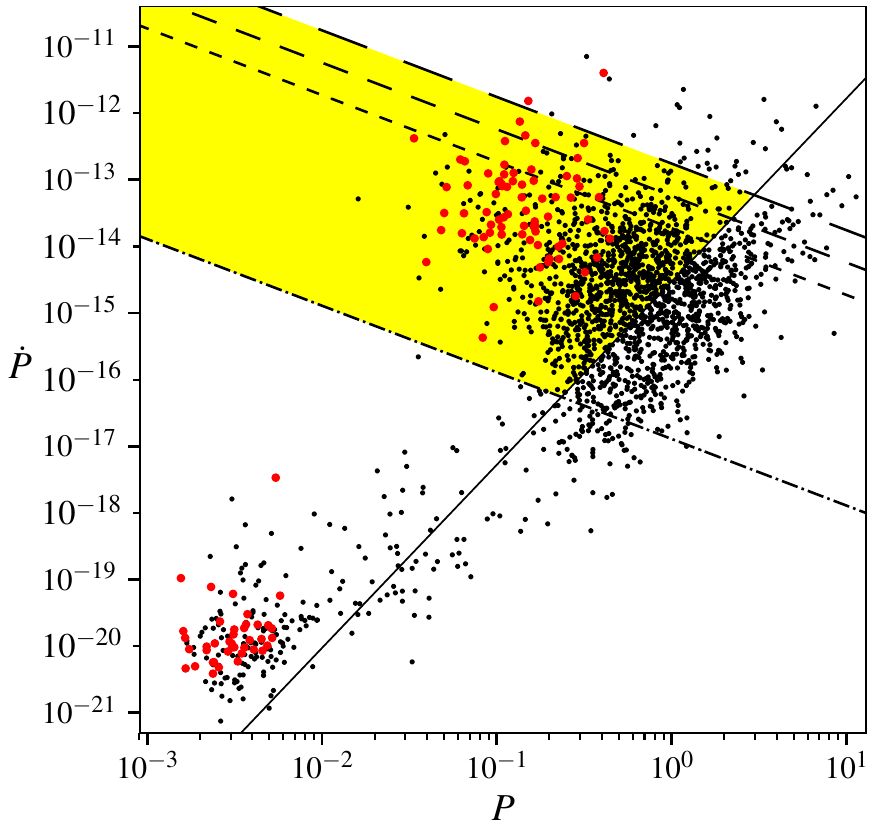}
  \caption{$P\dot{P}$ diagram with the yellow area showing the range
    of parameters where approximation for particle acceleration used
    in this paper is applicable, see text for description.  Pulsars
    from ATNF catalog (\citet{ATNF_Catalog2005},
    http://www.atnf.csiro.au/research/pulsar/psrcat) are shown by
    black dots, $\gamma$-ray pulsars from the second Fermi catalog
    \citep{Abdo2013_SecondPSRCatalog} by red dots.}
  \label{fig:ppdot}
\end{figure}

The limitations of our model come from assumptions used in derivation of the
energy of primary particles and the magnetic field strengths at which physical
processes different from the ones we consider here become important.
Eq.~\eqref{eq:epsilonP_final} for the energy of primary particles $\epsilonPacc$
was derived under the assumptions that (i) particles are accelerated freely,
i.e. radiation reaction can be neglected and (ii) the length of the gap is much
smaller that the polar cap radius, so that a one dimensional approximation can
be used.  We also do not model cascades for which photon splitting is important,
so our model is applicable (iii) for magnetic field strength below $\Split{B}$,
calculated according to eq.~\eqref{eq:Bsplit_cr}.  Constraints (i), (ii), and
(iii) together define the range of pulsar parameters where our cascade model is
applicable.  Constraints (i) and (ii) remain the same as in \PapI; they are
derived in Appendixes~B and C of \PapI{} correspondingly.  Constraint (iii) --
the magnetic field strength above which photon splitting becomes important for
photons near the threshold of pair formation -- depends on the radius of
curvature of magnetic field lines $\rhoC$.  In Fig.~\ref{fig:ppdot} we show the
range of pulsar parameters for which our model is applicable superimposed on the
$P\dot{P}$ diagram.  The one-dimensional approximation (ii) is valid to the left
of the solid line, the approximation (i) of free acceleration above the
dot-dashed line (given by eq.~54 in \PapI).  Pulsars with $B<\Split{B}$ are
below the dashed lines. The line with short dashes correspond to $\rhoC=10^6$cm,
$\Split{B}=\sci{4.4}{12}$G, the line with medium dashes -- $\rhoC=10^7$cm,
$\Split{B}=\sci{7.7}{12}$G, and the line with long dashes -- $\rhoC=10^8$cm,
$\Split{B}=\sci{1.3}{13}$. The yellow region shows the range of of pulsar
periods and period derivatives where these three assumptions are valid.  We see
that most of young normal pulsars, including gamma-ray pulsars from the Fermi
second pulsar catalog, fall in this range. Technically, the range of pulsar
parameters for which our current model is applicable is only slightly different
from that of \PapI{} (the limits on $B$ a less restrictive now, cf with Fig.~13
in \PapI), but our current model offers a considerably better treatment of
cascades for $B\gtrsim10^{12}$G.

\section{Multiplicity of the full cascade}
\label{sec:results}

\begin{figure*}
  \includegraphics[clip,width=\textwidth]{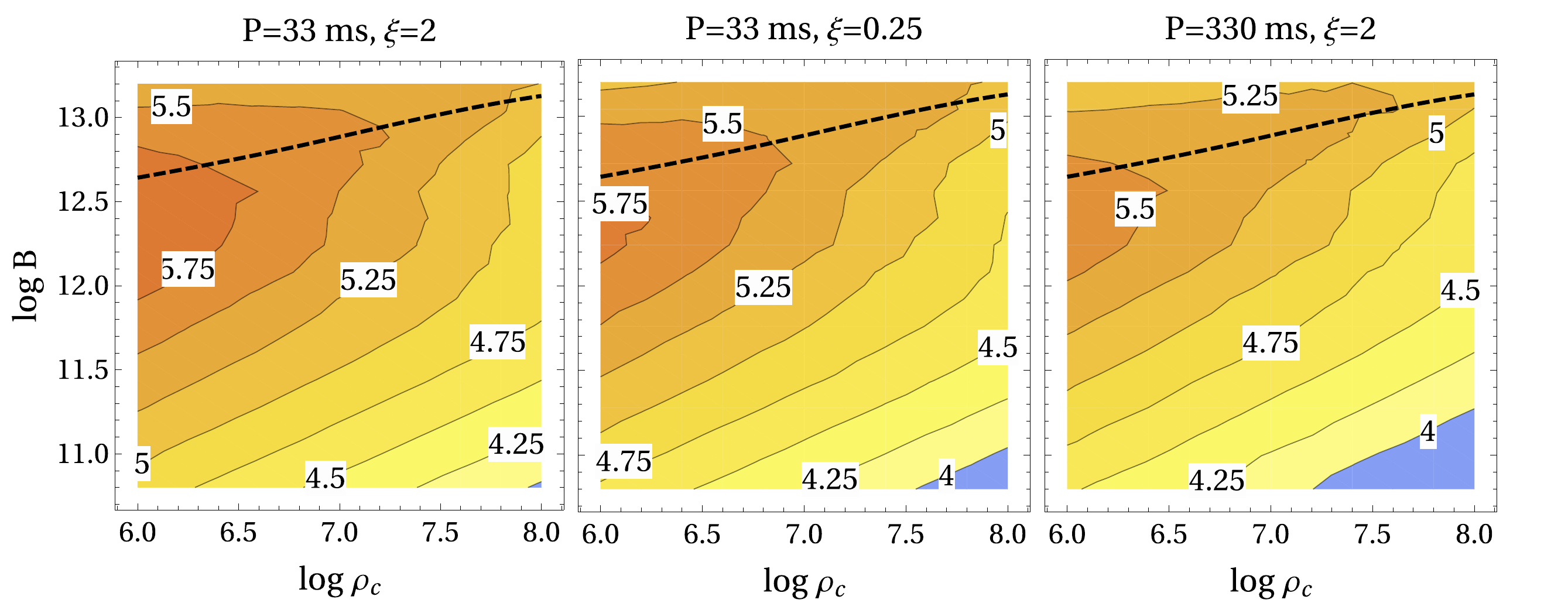}  
  \caption{Multiplicity of polar cap cascades for different pulsar parameters
    $P$ and $\xi$.  Contours of $\log\kappa$ as functions of logarithms of
    curvature of magnetic field lines $\rhoC$ in cm and magnetic field strength
    $B$ in Gauss for three sets of the gap parameters $(P[\mbox{ms}],\xi_j)$:
    $(33,2)$, $(330,0.25)$, $(33,2)$.  The temperature of NS surface was
    assumed to be $T=10^6$K. The critical magnetic field $\Split{B}$ above which
    photon splitting starts affecting cascade multiplicity is shown by the
    dashed line.}
  \label{fig:kappa-Pxi}
\end{figure*}

\begin{figure*}
  \includegraphics[clip,width=\textwidth]{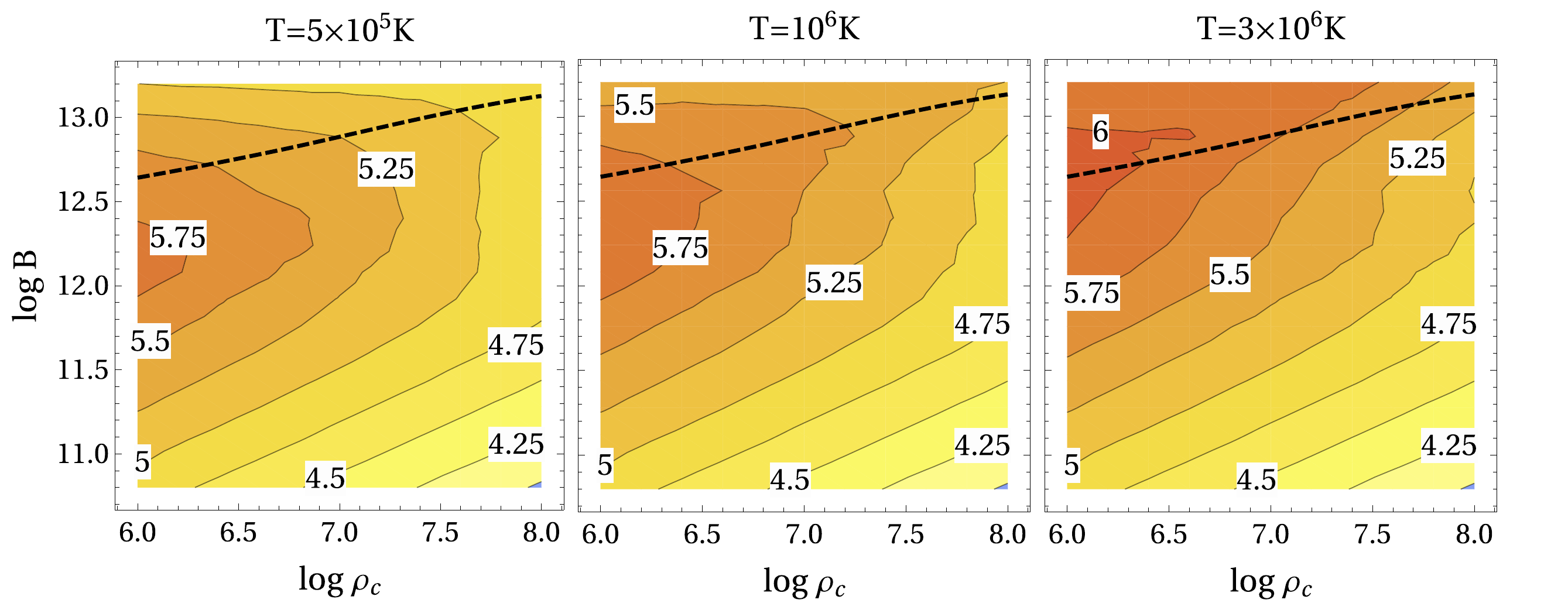}  
  \caption{Multiplicity of polar cap cascades for different temperatures of the
    NS surface. Contours of $\log\kappa$ as functions of logarithms of curvature
    of magnetic field lines $\rhoC$ in cm and magnetic field strength $B$ in
    Gauss for three temperatures of NS surface
    $T=\sci{5}{5}, 10^6, \sci{3}{6}$K.  The pulsar parameters $P$ and $\xi$ are
    assumed to be $P=33$ms, $\xi=2$.  The critical magnetic field $\Split{B}$
    above which photon splitting starts affecting cascade multiplicity is shown
    by the dashed line. Note that the middle panel is the same as the left panel
    on Fig.~\ref{fig:kappa-Pxi}.}
  \label{fig:kappa-T}
\end{figure*}

\begin{figure*}
  \includegraphics[clip,width=\textwidth]{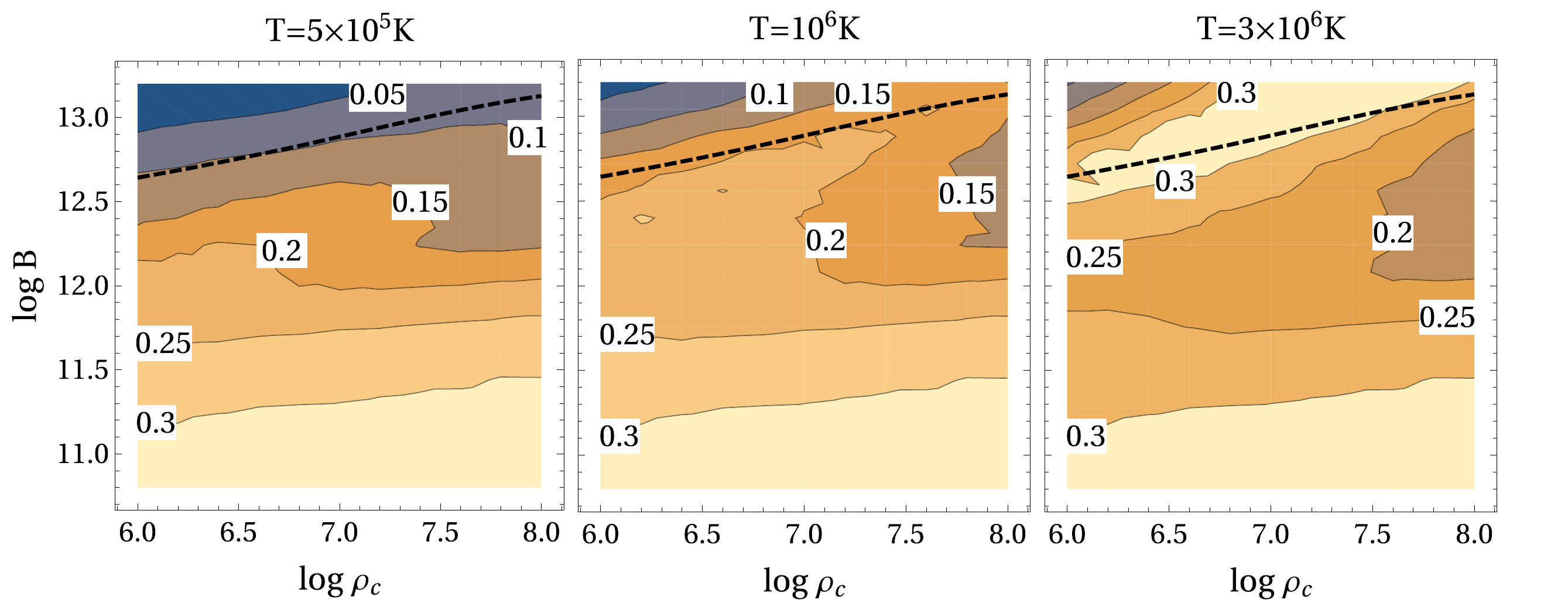}  
  \caption{Efficiency of polar cap cascades $\kappa/\SUB{\kappa}{max}$ for
    different temperatures of the NS surface.  Contours of
    $\kappa/\SUB{\kappa}{max}$ as functions of logarithms of curvature of
    magnetic field lines $\rhoC$ in cm and magnetic field strength $B$ in Gauss
    for cascade shown in Fig.~\ref{fig:kappa-T}.}
  \label{fig:cascade-efficiency}
\end{figure*}

For a wide range of pulsar parameters we computed maximum multiplicities of
polar cap cascades with $\gamma{}B$ as the pair creation mechanism and CR,
SR, and RICS of soft thermal photons from the NS surface as 
emission mechanisms according to the algorithm described in previous sections.

In Fig.~\ref{fig:kappa-Pxi} we show contours of cascade multiplicity
$\log(\kappa)$ as a function of the magnetic field strength $\log(B)$ in Gauss
and the radius of curvature of magnetic field lines $\log(\rhoC)$ in cm for a NS
with a uniform surface temperature of $T=10^6$ K. The dashed line indicates the
parameter space above which photon splitting start (negatively) to affect the
cascade multiplicity. The three plots in Fig.~\ref{fig:kappa-Pxi} are for
different pulsar periods $P=33, 330$~ms and two different filling factors
$\xi=2,0.25$. The electric field in the gap for the case ($P=33$~ms, $\xi=2$) is
an order of magnitude larger than in the other two cases, but the cascade
multiplicity is only moderately higher, it is greater by less than $\sim2$
times.  Compared to the simple estimates from \S\ref{sec:simple-multiplicity}
the total multiplicity for the same pulsar parameters is smaller by the factor
of $\sim4-5$ -- cf. Fig.~\ref{fig:kappa-Pxi} with Fig.~\ref{fig:kappa-max} where
$\SUB{\kappa}{max}$ is plotted for the same combination of parameters. The
maximum value for multiplicity reaches $\sci{6}{5}$ for smaller radii of
curvature of magnetic field lines $\rhoC\sim10^6$cm.

In the case of a pure CR-Synchrotron cascade discussed in \PapI{} (when the 
contribution of RICS is neglected), the cascade multiplicity is the highest for
magnetic fields around $B\gtrsim10^{12}$ G and drops for both higher and lower
magnetic field strengths (Fig.~14 of \PapI{}). For cascades with RICS considered
in this paper the multiplicity decrease for higher magnetic field,
$B\gtrsim10^{12}$ G, is much smaller, because the energy of pair parallel motion
is returned back to the cascade by RICS.  For lower magnetic field $B<10^{12}$G
the differences in multiplicities between Fig.~\ref{fig:kappa-Pxi} and Fig.~14
of \PapI{} vary from a few percent for the case ($P=33$~ms, $\xi=2$) to
$\sim70\%$ for the case ($P=330$~ms, $\xi=2$). These differences are because of
more accurate treatment of emission processes in this paper compared to \PapI.

In Fig.~\ref{fig:kappa-T} we show plots illustrating the dependence of the
cascade multiplicity on the NS surface temperature. Contours of $\log(\kappa)$
are plotted for three different temperatures of the NS surface
$T=\sci{5}{5}, 10^6, \sci{3}{6}$ K.  The number of soft X-ray photons available
for RICS changes dramatically with the temperature and so does the contribution of
RICS to cascade multiplicity.  For $T=\sci{5}{5}$ K the multiplicity profiles are
very similar to those of CR-synchrotron cascades.  For magnetic fields
$B\gtrsim\sci{3}{12}$ G photons are absorbed after propagating a short distance
and pairs are created with small perpendicular to $B$ momenta, which leaves a large
fraction of the parent photon's energy in the pairs' parallel motion (this was discussed
in detail in \PapI).  If there are not enough soft X-ray photons to be
up-scattered via RICS, this kinetic energy is lost from the cascade and the total
multiplicity diminish.  For higher surface temperatures the increasing number of
soft photons makes RICS more and more efficient, which leads to the decrease of
energy ``leaks'' from the cascade and the multiplicity $\kappa$ becomes similar to
the maximum multiplicity $\SUB{\kappa}{max}$.  Indeed, for low temperature
$T=\sci{5}{5}$K (left panel of Fig.~\ref{fig:kappa-T}) the plot for $\kappa$ is
similar to $\kappa$ of CR-synchrotron cascade, and for high temperature
$T=\sci{3}{6}$K (left panel of Fig.~\ref{fig:kappa-T}) the shape of $\kappa$
contours are similar to the ones of $\SUB{\kappa}{max}$ shown on
Fig.~\ref{fig:kappa-max} (left panel).

The cascade efficiency $\kappa/\SUB{\kappa}{max}$, which characterizes how the
energy available to the cascade is converted into pairs, is shown in
Fig.~\ref{fig:cascade-efficiency}, for the same parameters as $\kappa$ on
Fig.~\ref{fig:kappa-T}.  For magnetic fields below $\lesssim10^{12}$G the
efficiency for all three surface temperatures is similar and $\gtrsim20\%$; for
these magnetic fields the RICS contribution is negligible and so the cascade
behavior does not depend on the temperature.  For stronger magnetic fields, the
efficiency increases with the temperature. For high NS temperatures the cascade
efficiency can be as large as 30\%.  We should note that because of photon
splitting (which mostly affects RICS photons) the real cascade efficiency above
the dashed lines is smaller than the values shown in
Fig.~\ref{fig:cascade-efficiency}.

To get a better understanding of polar cap cascades near their peak
multiplicities we consider here several particular cascades and analyze their
properties in detail.  We consider cascades in polar caps of pulsar with
$P=33$ ms, assume $\xi=2$ and use two values of the radius of curvature of
magnetic field lines $\rhoC=10^7$ cm and $\rhoC=10^{7.9}\approx\sci{7.94}{7}$cm.
For each value of $\rhoC$ we analyze cascade properties for three values of
magnetic field strength to illustrate how the RICS contribution changes with $B$.
These examples represent cascades near their highest multiplicities in polar
caps of pulsars when (i) there is a significant non-dipolar component of the
magnetic field ($\rhoC=10^7$cm) as well as (ii) pulsars with nearly
unadulterated dipolar field ($\rhoC=10^{7.9}$cm). As we discussed earlier,
the dependence of the cascade multiplicity on pulsar parameters is weak, so these
examples should be representative for cascades in pulsars with a broad range of
periods and filling factors $\xi$.

\begin{figure*}[h]
  \begin{tabular}{l@{\hskip1in}l@{\hskip1in}l}
    \includegraphics[clip,height=2.45in]{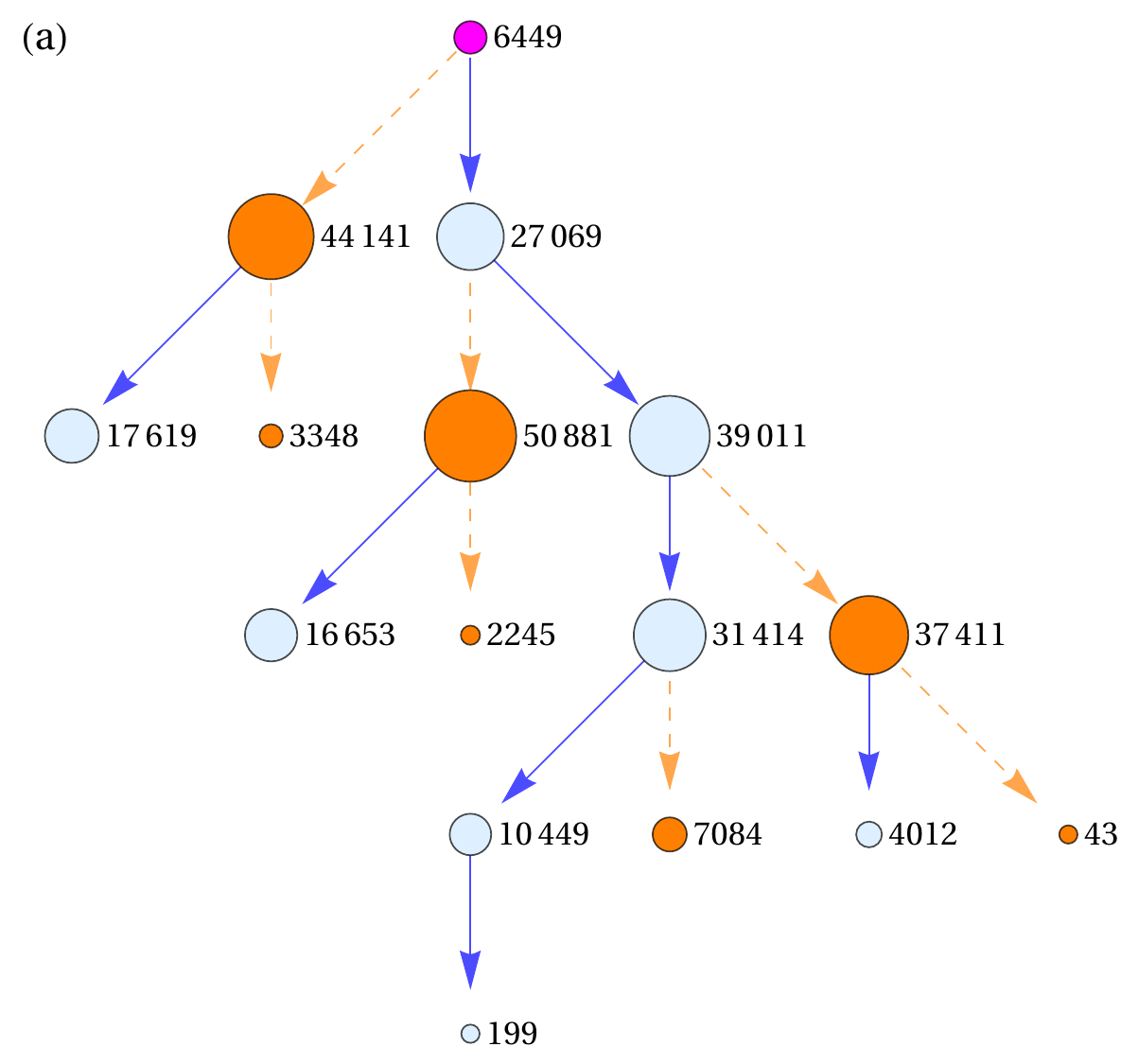} &
    \includegraphics[clip,height=2.45in]{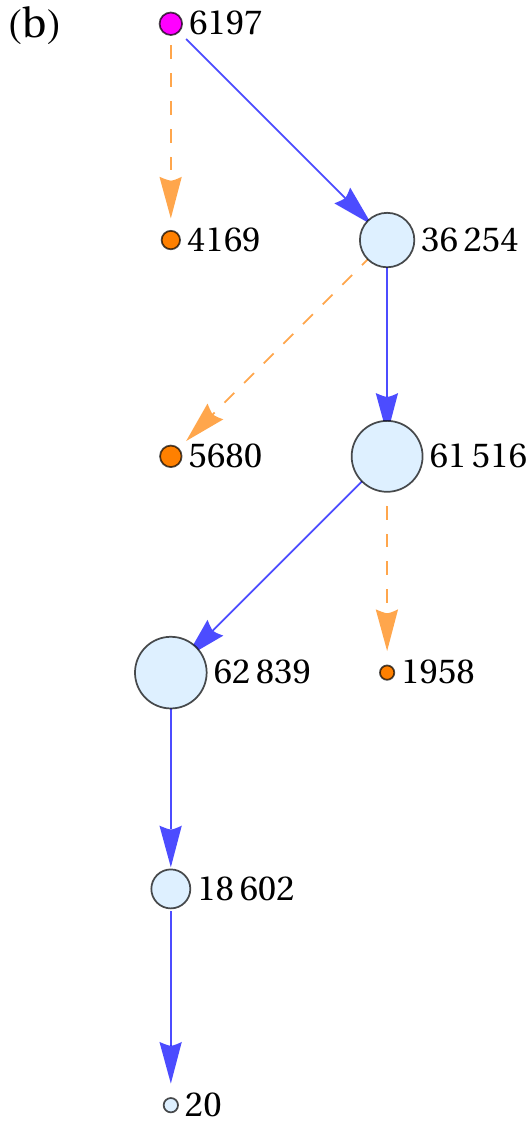} &
    \includegraphics[clip,height=2.45in]{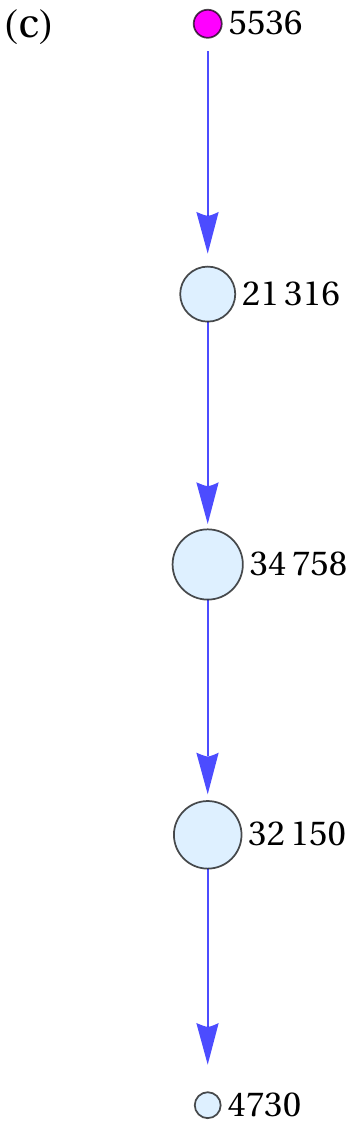}  \\
  \end{tabular}
  \caption{Cascade graphs for cascades in pulsars with $P=33$ms, $\xi=2$,
    $\rhoC=10^7$cm, $T=10^6$K, and the following magnetic field strengths: (a)
    $B=10^{12.5}$G, (b) $B=10^{12}$G, (c) $B=10^{11.5}$G.  Each graph's vertex 
    represents pairs of the same origin -- pairs produced in a certain cascade
    generation by photons emitted by pairs of the previous generation by the
    same emission mechanism.  Lines connect vertices representing parent
    particles to the vertices representing child pairs with arrows directed
    towards the child pairs.  Blue vertices and lines correspond to pairs
    created by synchrotron photons, orange vertices and dashed lines -- to
    pairs created by RICS photons, and magenta vertices -- to pairs created by
    CR photons.  The size of each vertex (its area) is proportional to the
    total number of pairs, the numbers show the number of pairs represented by
    the vertices.  The relative sizes of the vertices are consistent only within
    the same graph, sizes of vertices in different graphs are not related.}
  \label{fig:CascadeGraph_abc}
\end{figure*}

\begin{figure*}
  \begin{tabular}{l@{\hskip.2in}l@{\hskip.2in}l}
    \includegraphics[clip,height=1.3in]{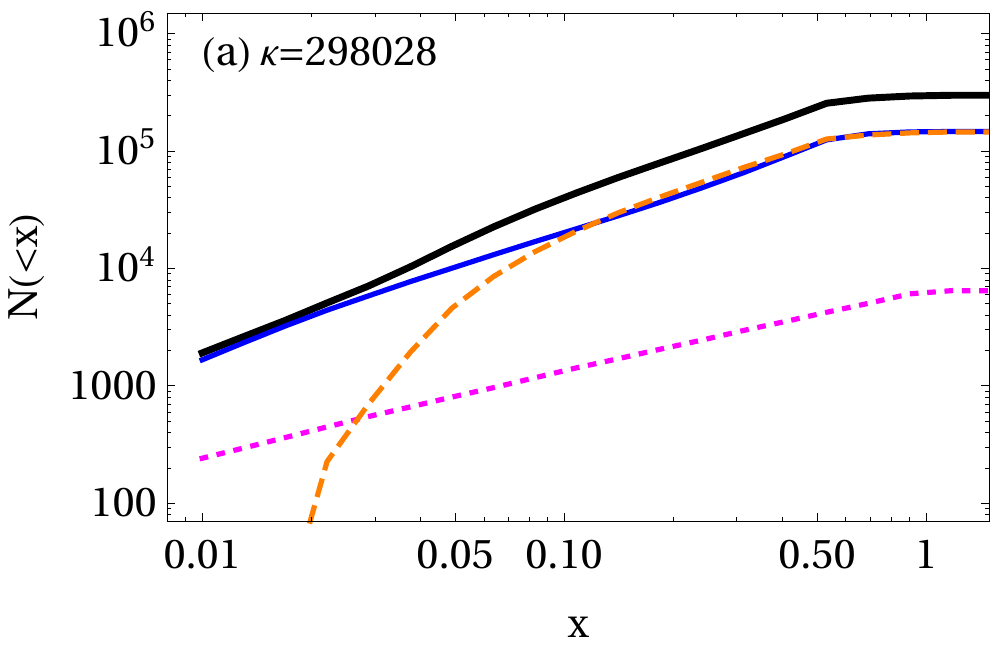} &
    \includegraphics[clip,height=1.3in]{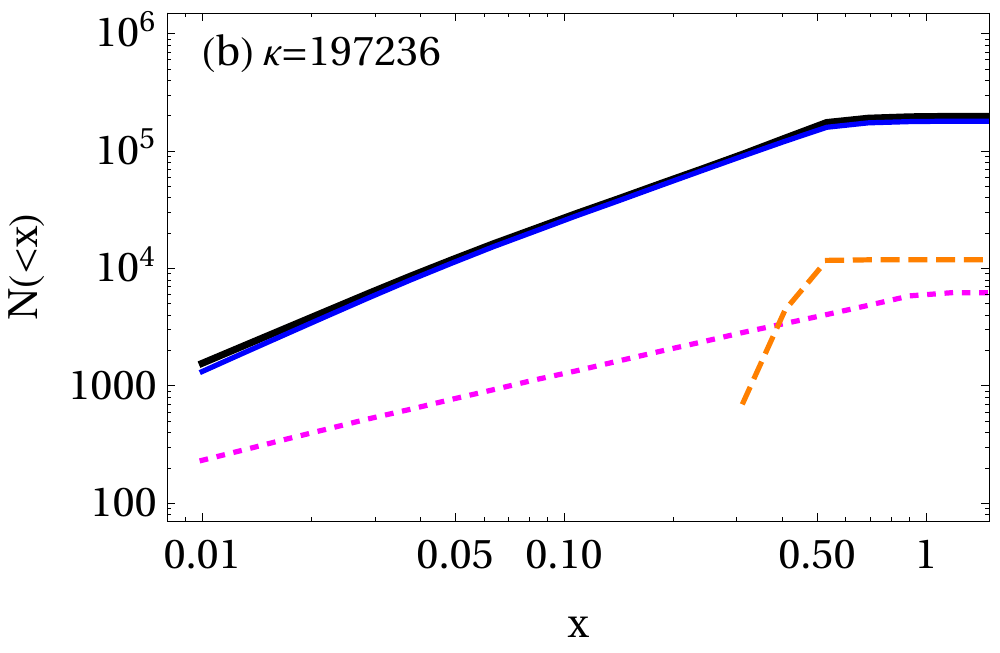} &
    \includegraphics[clip,height=1.3in]{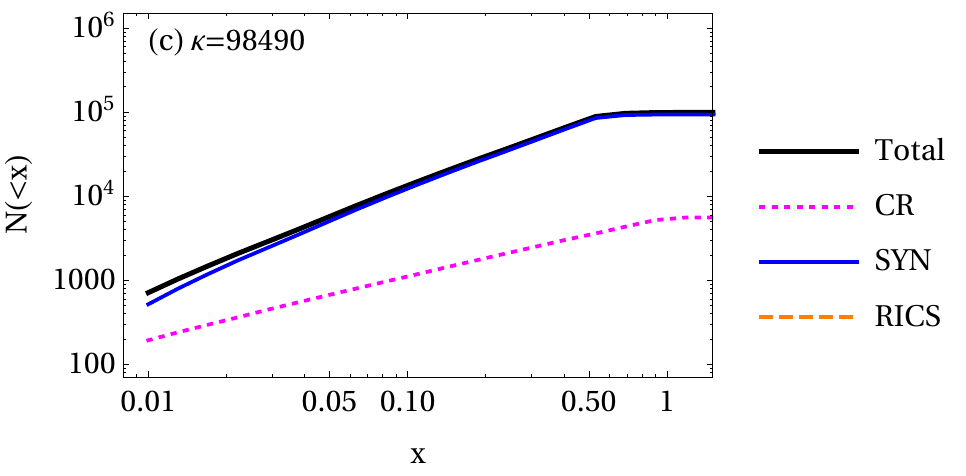}  \\
  \end{tabular}
  \caption{Cumulative pair injections rates $N(<x)$ for different emission
    mechanisms for cascades in pulsars with the same parameters as in
    Fig.~\ref{fig:CascadeGraph_abc}.  $N(<x)$ is the total number of pairs
    created at the distances less than the distance $x$.  The total pair
    injection rate is shown by the black solid line.  Dotted magenta line is the
    number of pairs created by CR photons, blue solid line -- by synchrotron
    photons, dashed orange line -- RICS photons.  The total cascade multiplicity
    is shown in the upper left corner of each plot.}
  \label{fig:CascadeMult_abc}
\end{figure*}

\begin{figure*}
  \begin{tabular}{l@{\hskip.2in}l@{\hskip.2in}l}
    \includegraphics[clip,height=1.3in]{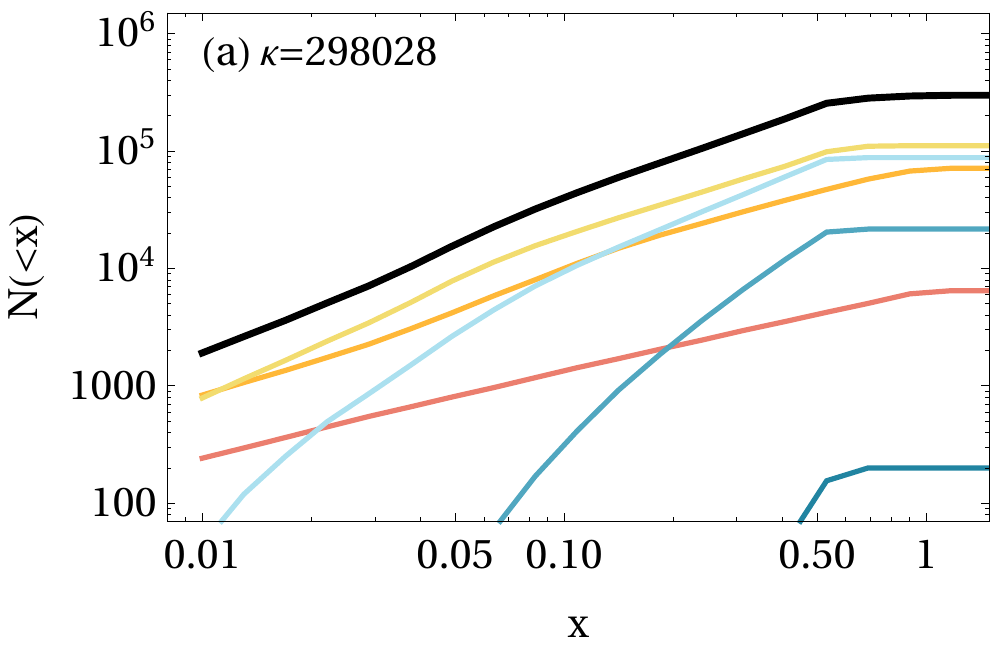} &
    \includegraphics[clip,height=1.3in]{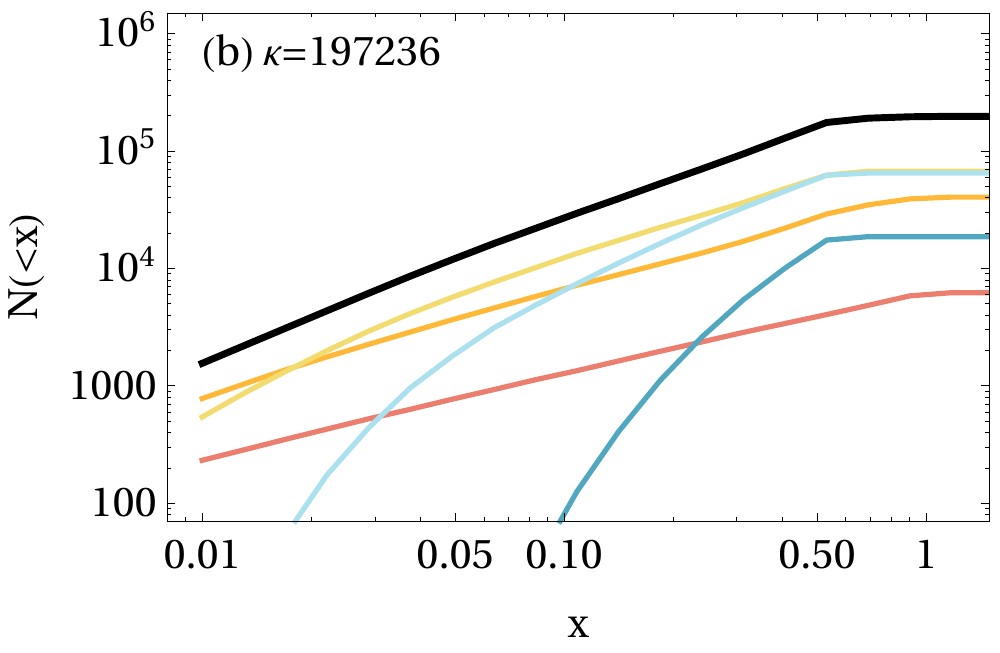} &
    \includegraphics[clip,height=1.3in]{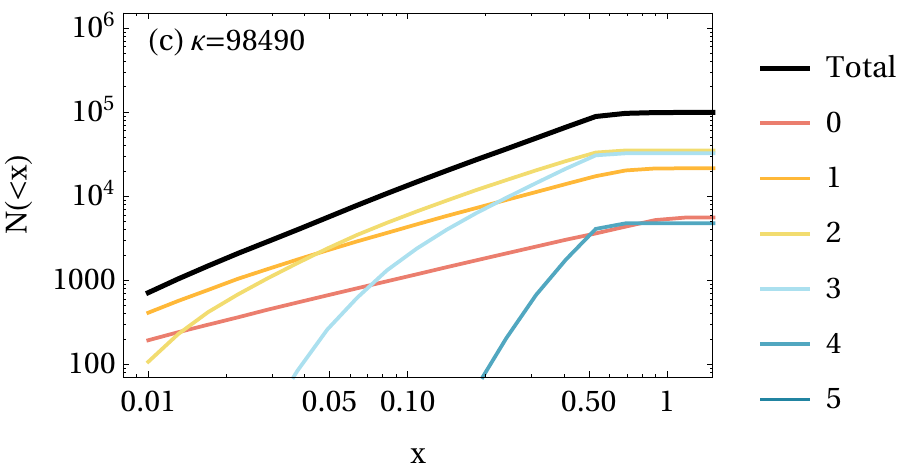}  \\
  \end{tabular}
  \caption{Cumulative pair injections rates $N(<x)$ for different cascade
    generations for cascades in pulsars with the same parameters as in
    Fig.~\ref{fig:CascadeGraph_abc}. The total pair injection rate is shown by
    the black solid line. Pair injection rates for different cascade generations
    are shown by color lines according to the legend to the right of the panel
    (c).  }
  \label{fig:CascadeMultGen_abc}
\end{figure*}

\begin{figure*}
  \begin{tabular}{l@{\hskip0.2in}l@{\hskip.2in}l}
    \includegraphics[clip,height=2.65in]{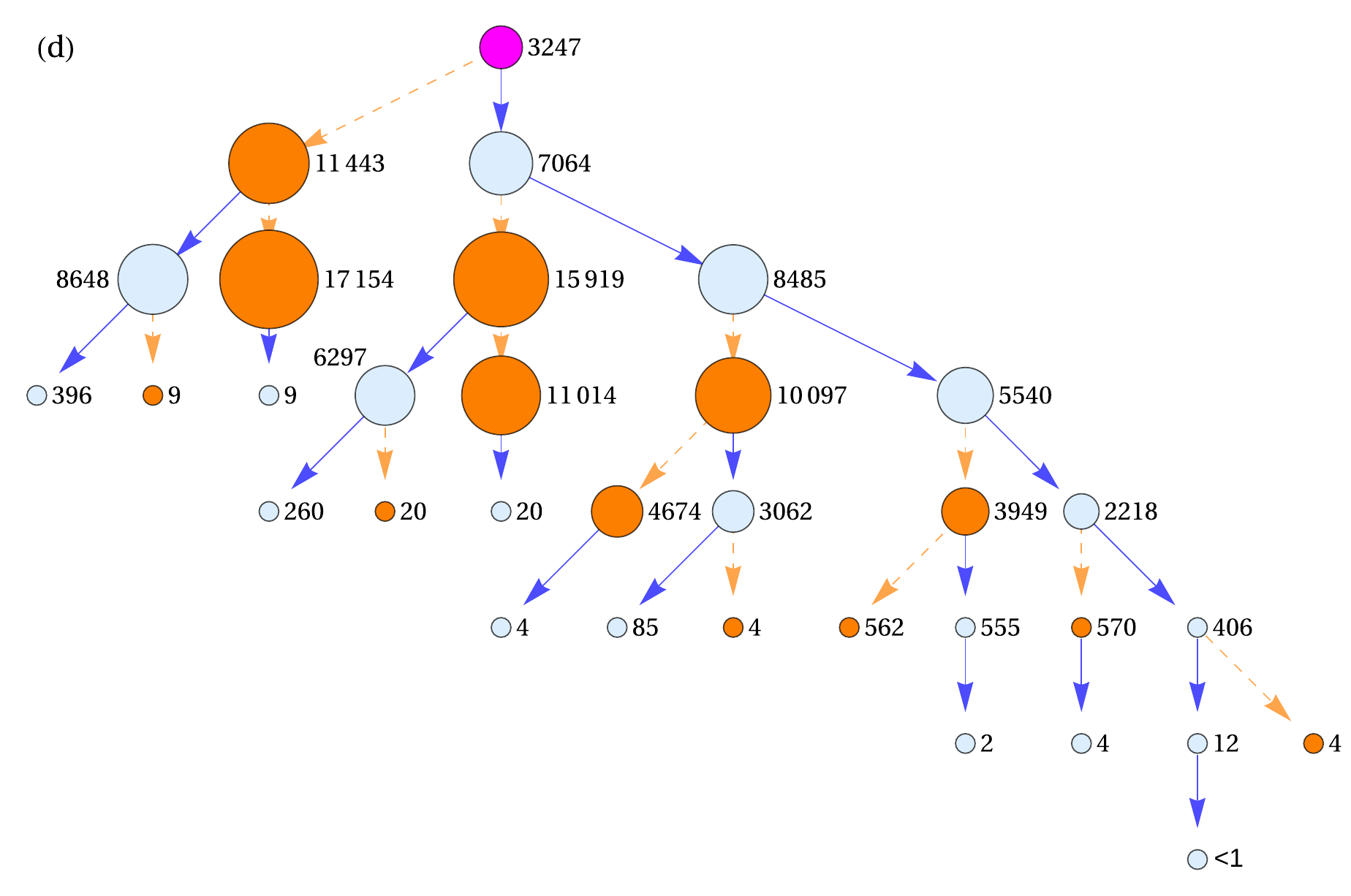} &
    \includegraphics[clip,height=2.65in]{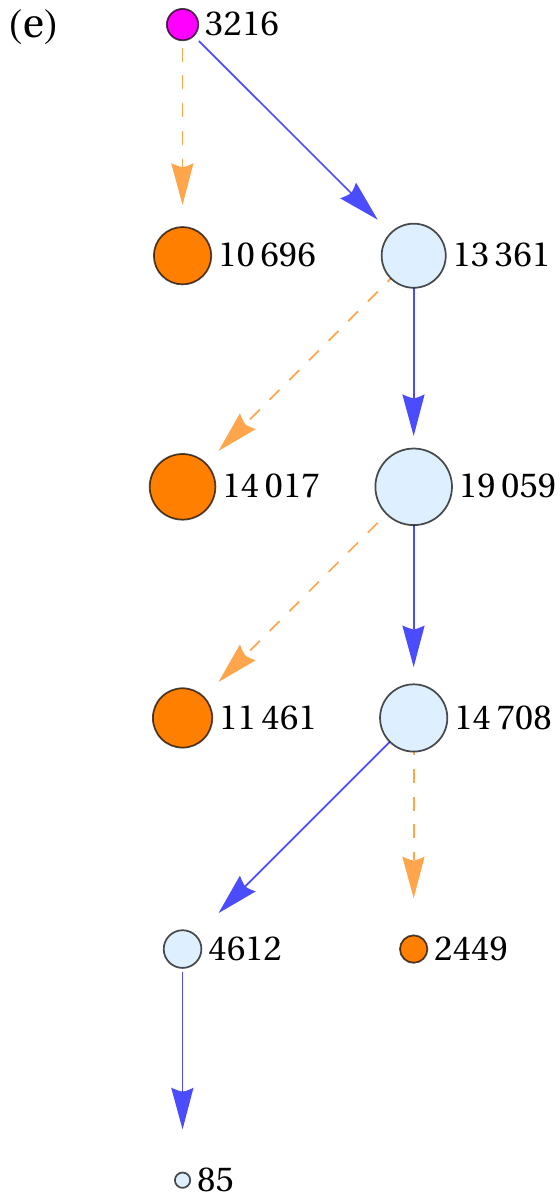} &
    \includegraphics[clip,height=2.65in]{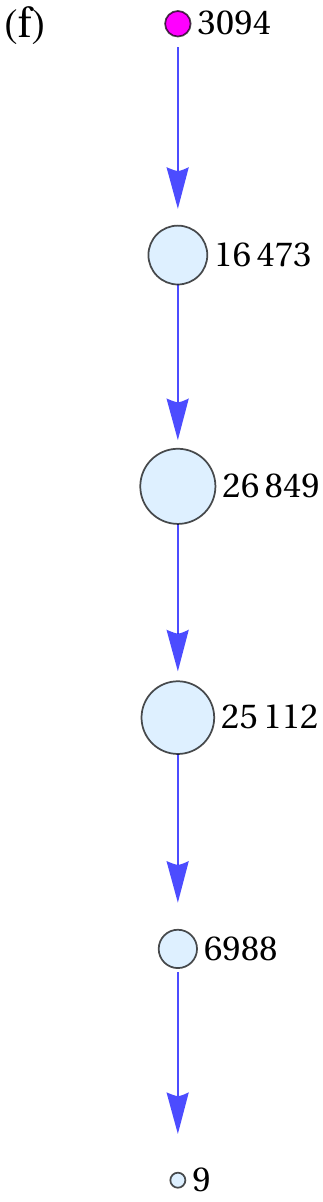}  \\
  \end{tabular}
  \caption{Cascade graphs for cascades in pulsars with $P=33$ms, $\xi=2$,
    $\rhoC=10^{7.9}\approx\sci{7.94}{7}$cm and the following magnetic field strengths: (d)
    $B=10^{12.9}$G, (e) $B=10^{12.5}$G, (f) $B=10^{12}$G. Notations are the same
    as in Fig.~\ref{fig:CascadeGraph_abc}. }
  \label{fig:CascadeGraph_def}
\end{figure*}

\begin{figure*}
  \begin{tabular}{l@{\hskip.2in}l@{\hskip.2in}l}
    \includegraphics[clip,height=1.3in]{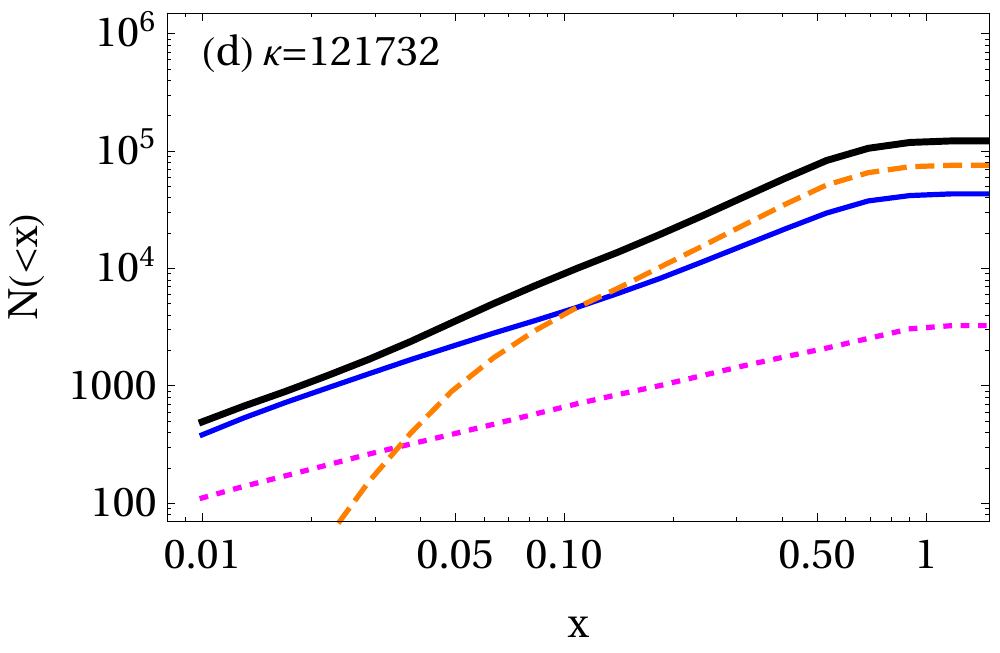} &
    \includegraphics[clip,height=1.3in]{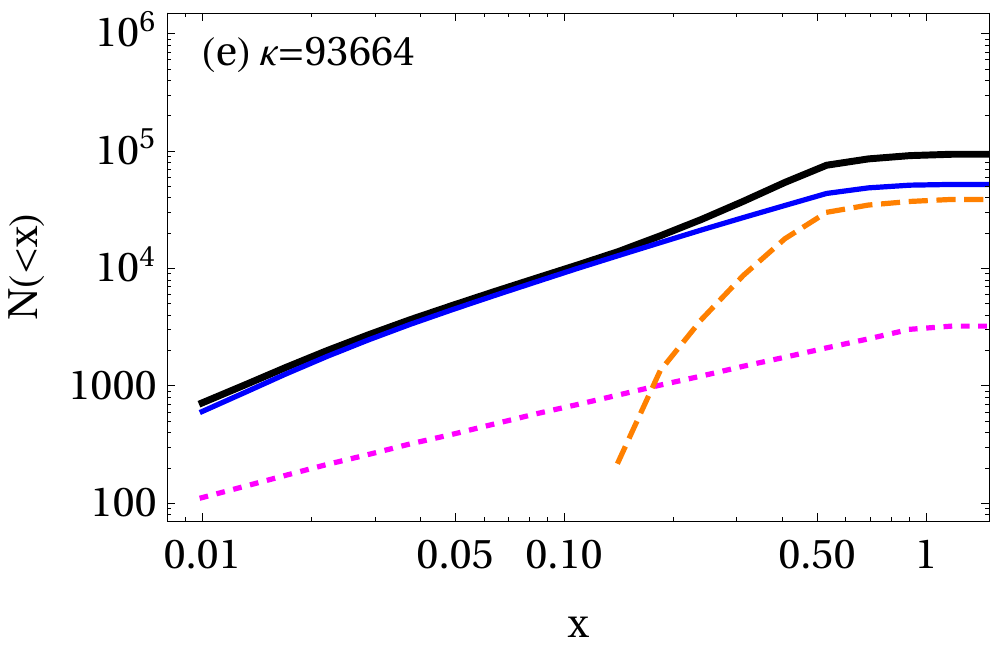} &
    \includegraphics[clip,height=1.3in]{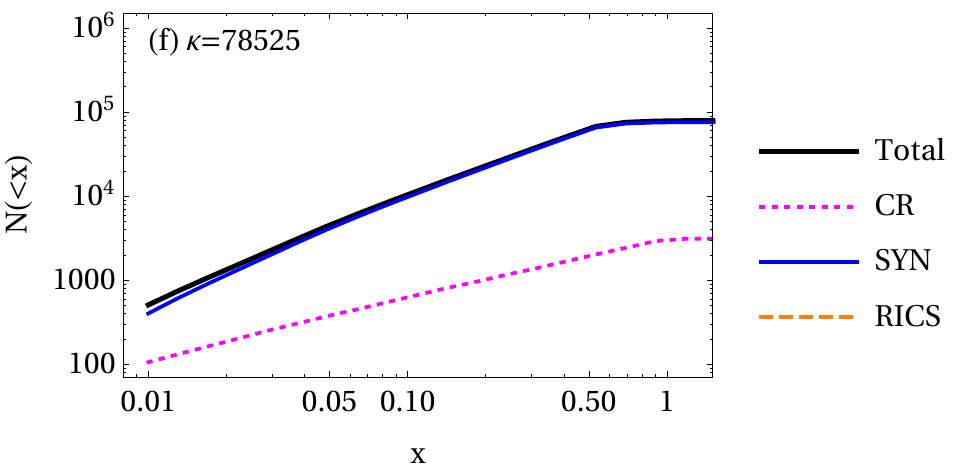}  \\
  \end{tabular}
  \caption{Cumulative pair injections rates $N(<x)$ for different emission
    mechanisms for cascades in pulsars with the same parameters as in
    Fig.~\ref{fig:CascadeGraph_def}. Notations are the same as in
    Fig.~\ref{fig:CascadeMult_abc} }
  \label{fig:CascadeMult_def}
\end{figure*}

\begin{figure*}
  \begin{tabular}{l@{\hskip.2in}l@{\hskip.2in}l}
    \includegraphics[clip,height=1.3in]{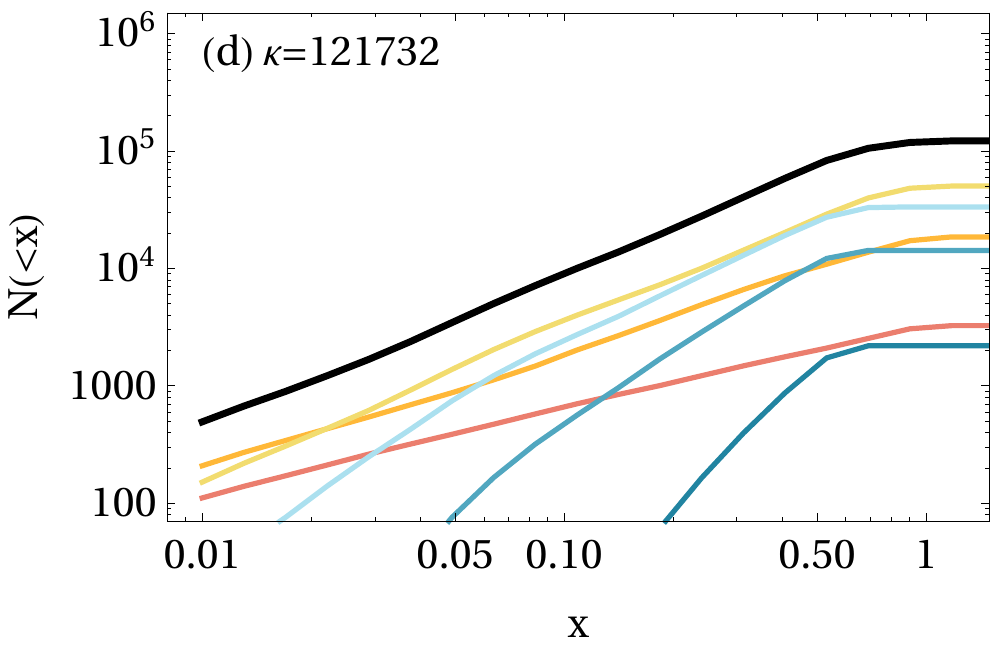} &
    \includegraphics[clip,height=1.3in]{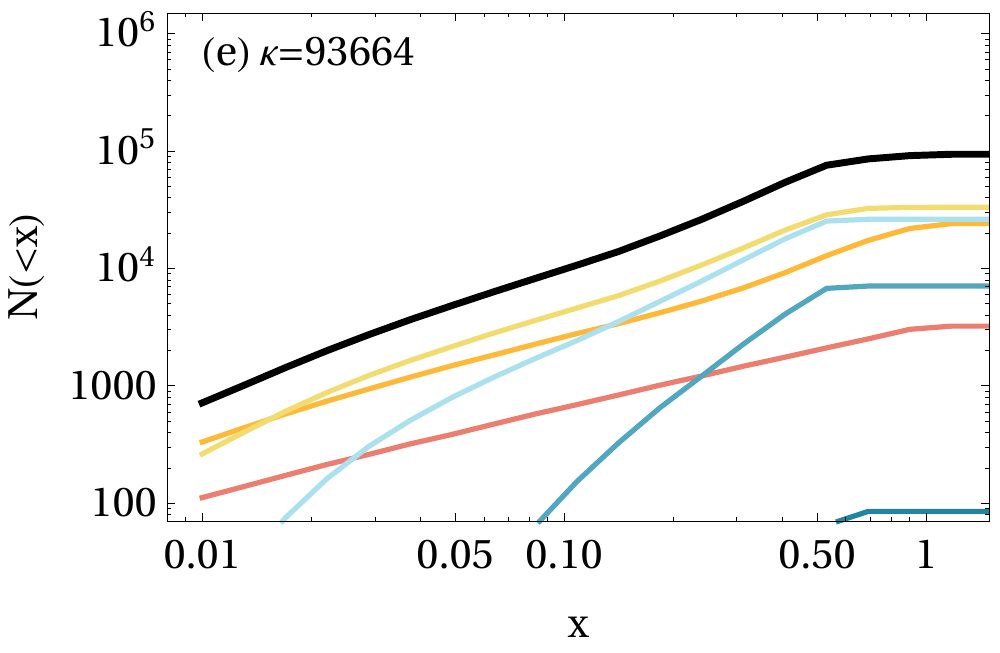} &
    \includegraphics[clip,height=1.3in]{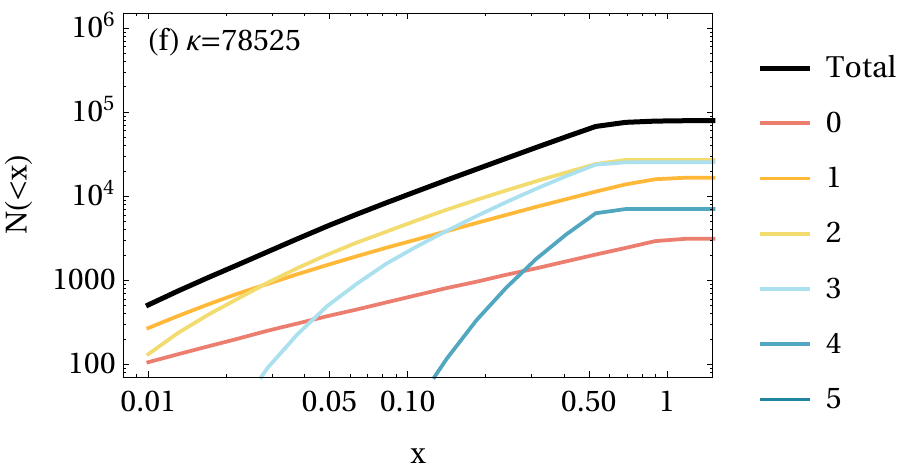}  \\
  \end{tabular}
  \caption{Cumulative pair injections rates $N(<x)$ for different cascade
    generations for cascades in pulsars with the same parameters as in
    Fig.~\ref{fig:CascadeGraph_abc}.  Notations are the same as in
    Fig.~\ref{fig:CascadeMultGen_abc}}
  \label{fig:CascadeMultGen_def}
\end{figure*}

We visualize cascade development with three types of plots: (i) the cascade
graphs (Figs.~\ref{fig:CascadeGraph_abc}, \ref{fig:CascadeGraph_def}), the
cumulative pair injections rates decomposed according to (ii) emission
mechanisms (Figs.~\ref{fig:CascadeMult_abc}, \ref{fig:CascadeMult_def}) and
(iii) cascade generations (Figs.~\ref{fig:CascadeMultGen_abc},
\ref{fig:CascadeMultGen_def}).  These plots show properties of the whole
cascade, i.e. all pairs created by a single primary particle as it moves along
magnetic field lines and emits CR photons which initiate multiple individual
cascades.  The cascade graphs in Figs.~\ref{fig:CascadeGraph_abc},
\ref{fig:CascadeGraph_def} are the quantitative representations of the cascade
diagram shown in Fig.~\ref{fig:cascade-diagram}. In these graphs vertices
represents pairs with the same origin -- the chain of emission processes which
led to emission of these pairs' parent photons is the same.  The area of the
circles at the graphs' vertices is proportional to the number of pairs, however,
their sizes are consistent only within the same graph, the sizes of the vertices
of different graphs are not related.  The cumulative pair injection rates
$N(<x)$ for a given distance $x$, shown in Figs.~\ref{fig:CascadeMult_abc},
\ref{fig:CascadeMultGen_abc}, \ref{fig:CascadeMult_def},
\ref{fig:CascadeMultGen_def}, are the total number of pairs created at distances
less than $x$.

Polar cap cascades have 5-8 of generations at most. At each cascade generation
pairs emit several photons which split the energy of the particle.  This leads
to rapid energy degradation through cascade generations and the cascade dies
after several generations.  On average, the most contribution comes from
generation 4 (see
Figs.~\ref{fig:CascadeMultGen_abc},~\ref{fig:CascadeMultGen_def})%
\footnote{There is no contradiction to the statement we made in
  \S\ref{sec:cascades-overview} that most pairs are produced at the last or
  penultimate cascade generations.  Here we consider cumulative properties of
  all cascade generated by CR photons emitted by a primary particle when it
  moves through the cascade zone.  Cascades initiated by individual CR photons
  still produce most of the pairs at the last generation, but as the energy of
  the primary particle decreases individual cascades have less
  generations. These less energetic cascades dominate the total pair output.}.
The total multiplicity does not necessarily increase with the number of cascade
branches -- despite there being more branches and generations in case (d), the
total multiplicity of these cascades is $\sim2.5$ times lower than that of
cascades for case (a).

Cases a-c and e-f represent cascades with the same values of $\rhoC$ ($10^7$cm
and $10^{7.9}$cm correspondingly) and decreasing magnetic field strength.  The
role of RICS in cascades decreases with decreasing of $B$. For case (a) RICS is
responsible for comparable, and for case (d) even slightly larger, number of
pairs, see Figs.~\ref{fig:CascadeMult_abc}(a),~\ref{fig:CascadeMult_abc}(d).  As
a result cascades (a) and (d) have more branches that their counterparts with
lower magnetic fields ((b), (c) and (e), (f) correspondingly), see
Figs.~\ref{fig:CascadeGraph_abc},~\ref{fig:CascadeGraph_def}. RICS never becomes
a dominating process, but it can contribute a comparable amount of pairs to the
cascade as synchrotron radiation for high magnetic fields.

Another illustration of the fact that the number of cascade branches is not
directly related to the total multiplicity is provided by comparison of cascades
(d)-(f). The complexity of the cascades changes significantly (due to the
diminishing role of RICS) but the total multiplicities of cascades (e), (f) are
only $\sim30\%$ and $55\%$ smaller than that of cascade (d). What matters most
is the amount of energy available for the cascade, i.e. the relation of
$\epsilonPacc$ and $\epsilonGesc$, cf. Fig.~\ref{fig:kappa-max}.

We used different values for the photon escape distance $\ESC{s}=0.5$ and the
size of the cascade zone $\SUB{s}{cascade}=1$.  The value of $\ESC{s}$ has a
direct impact on the energy of escaping photons according to
eq.~\eqref{eq:eps_esc_eq}, and the multiplicity dependence is close to
$\propto{}1/\ESC{s}$.  On the other hand, the exact value of $\SUB{s}{cascade}$,
which limits the range where photons can be emitted and absorbed, has little
impact on the total multiplicity as long as $\SUB{s}{cascade}\geq\ESC{s}$. As is
evident from Figs.~\ref{fig:CascadeMultGen_abc},\ref{fig:CascadeMultGen_def}
most pairs are produces at distances $<\ESC{s}$ from the NS which implies that
their parent photons are emitted at distances a few time smaller than
$\SUB{s}{cascade}$.  The contribution from photons emitted at distances
approaching $\SUB{s}{cascade}$ is rather small -- primary particles have lost a
large fraction of their energy and resulting cascades have only 1-2 generations.

Synchrotron photons have mostly $\perp$ polarization and RICS $\parallel$ one.
The photons susceptible to splitting are the ones with $\perp$ polarization. At
the field strength where photon splitting becomes important RICS photons provide
a comparable of larger number of pairs, hence the splitting should significantly
affect cascade multiplicity.  The highest cascade multiplicity is reached for
$B$ and $\rhoC$ values near the dashed lines in
Figs.~\ref{fig:kappa-Pxi},~\ref{fig:kappa-T}. So, the pulsars with highest
multiplicities should have $B\sim\sci{4.4}{12}-\sci{1.4}{13}$G, depending on
$\rhoC$, and surface temperature $T\gtrsim10^6$ K. On the $P\dot{P}$ diagram,
Fig.~\ref{fig:ppdot}, the pulsars with the highest maximum multiplicity are
those near the dashed lines.

The spatial distribution of pair injection rate shows that most of the pairs
produced by RICS photons are created at distances comparable to $\RNS$, which
are much larger than the polar cap size $\PC{r}\approx\sci{1.45}{4}P^{-1/2}$cm.
If soft X-ray photons are emitted from the whole surface of the NS, as assumed
in our model, their number density does not decrease dramatically throughout the
cascade zone. However, if soft photons are emitted from a hot polar cap, and the
rest of the NS surface is cold, $T<10^6$K, the number density of soft photons at
distances $\gg\PC{r}$ will be much smaller than assumed here.  In that case the
role of RICS is reduced and the polar cap cascade will operate in the
CR-synchrotron regime. In the latter case the multiplicity will reach its peak
at $B\sim10^{12}$G (for detailed analysis of CR-synchrotron cascades see \PapI).

\section{Discussion}
\label{sec:discussion}

In our previous paper, \PapI, we limited ourselves to CR-synchrotron cascades,
which was an adequate approximation for most young energetic pulsars. However
right where CR-synchrotron cascades reach their highest multiplicity, RICS
becomes an important emission mechanism and in order to get an accurate limit on
the maximum cascade multiplicity it must be taken into account.  In this study
we included in our model all three processes leading to emission of pair
producing photons in polar caps of energetic pulsars -- RICS, synchrotron and
curvature radiation -- and considered the effect on photon splitting on the
cascade multiplicity.  The treatment of the radiation is improved, dividing the
spectrum into three energy bands instead of the delta-function approximation
used in \PapI.  We used a more accurate prescription for the single photon pair
production which takes into account the decrease of the attenuation coefficient
near the pair formation threshold -- an important correction for pair formation
in magnetic fields with $B\gtrsim10^{12}$G. We also used a more consistent
treatment of the particle acceleration by finding both the energy and the
primary particles $\epsilonPacc$ and parameter $\chiAacc$, which regulates pair
injection and termination of acceleration zone, simultaneously.  We developed a
new semi-analytical algorithm which can incorporate an arbitrary number of
microphysical processes, model cascades spatial evolution, and allow fast
exploration of cascades parameter space. The improvements upon the model
presented in \PapI{} allowed us to conduct a reasonably accurate study of polar
cap cascades in the regime where they reach their highest multiplicity.  Our
current model includes the most important microphysical processes relevant for
polar cap cascades in young energetic pulsars with the highest pair yield.

The goal of our study was to find the upper limit on the multiplicity of
electron positron cascades in pulsars.  We have performed a systematic study of
pair cascades above pulsar polar caps for a variety of input parameters
including surface magnetic field, pulsar rotation period, primary particle
energy, magnetic field radius of curvature, and the temperature of NS
surface. We used the modern description for particle acceleration derived from
self-consistent models of the polar cap acceleration zone, i.e. those that are
capable of generating currents consistent with global models of the pulsar
magnetosphere.  In our model we do not address directly the non-stationary
nature of pulsar cascades.  We considered pair cascades generated by primary
particles accelerated at the peak of the pair formation burst.  The
intermittency of the pair formation process \emph{reduces} the total pair yield
and by studying cascades generated by the most energetic primary particles we
achieved our goal of finding the limit on pair cascade multiplicity.

We find that pair multiplicity is maximized for pulsars with hot $T\geq10^6$K
surfaces. These must be very young pulsars which have not yet cooled down. For
such pulsars, cascade multiplicity (almost) monotonically grows with increasing
$B$ and decreasing $\rhoC$ until photon splitting becomes more important than
pair production, which happens first in the last cascade generation.  In young
hot pulsars, pair cascades reach their highest multiplicities near magnetic
field strengths where photon splitting becomes more efficient than single photon
pair production for the last generation pair-producing photons.  The maximum
multiplicity is in the range $\kappa\sim10^6-\sci{3}{5}$ for magnetic field
strengths $\sci{3}{12}\mbox{G}\lesssim{}\,B\,\lesssim{}10^{13}\mbox{G}$
depending on the radius of curvature of magnetic field lines -- the low and
upper limits are for $\rho=10^6$cm and $10^8$cm correspondingly.  For older
pulsars, whose surfaces have cooled down below $10^6$K, the maximum cascade
multiplicity is in the range $\kappa\sim\sci{5}{5}-10^5$ and it is achieved at
$B\sim{}10^{12}$G.  Even if old pulsars have hot polar caps, the density of soft
photons at distances comparable to the NS radius will be too small to sustain
efficient RICS and so the cascade operates in the CR-synchrotron regime even for
high magnetic field strengths.

Here we ignored geometrical effects caused by curvature of magnetic field lines.
For the smallest values of $\rhoC{}$, magnetic field line at large distances
from the NS within the cascade zone can bend rather significantly. Such bending
causes the displacement of particles in the lateral direction which, however,
would have a negligible effect on the cascade multiplicity in pulsars with hot
surfaces because it does not affect CR and synchrotron radiation and the
variation of the incident angle of incoming thermal photons is washed out by the
large solid angle these photons are coming from.  It would only affect the
lateral spreading of the cascade.  In long period pulsars, where the NS surface
is cold and the only source of the soft X-rays is the hot polar cap the effects
of field line bending might potentially increase the multiplicity of the
cascade. The pairs' momenta in that case would have larger angle with soft
photons what could increase the cross-section to photon scattering. However, the
efficiency of RICS as an emission process is based in part on the wide range of
angles thermal photons are coming from. Due to the large range of photon
incident angles, pairs within a wide range of energies can still scatter thermal
photons (with relatively narrow energy distribution) in the resonant regime --
pairs of different energies scatter photons coming from different directions
\citep{Dermer1990}.  In the case of hot polar caps, the range of photon incident
angles will be small (e.g. compared to the case of the hot NS surface), and the
potential increase of scattering cross-section could benefit pairs of a single
generation at best, thus making this effect of little importance.

The multiplicity at the peak of the cascade cycle is not very sensitive to the
pulsar period, magnetic field and radius of curvature of the magnetic field
lines. The multiplicity varies by less than $\sim$ an order of magnitude for the
range of pulsar parameters spanning two or more orders of magnitude.  The reason
for this is self-regulation of the accelerator by pair creation: for pulsar
parameters resulting in more efficient pair production, the size of the
acceleration gap is smaller and the primary particle energy is lower and
vice-versa.

Even the most efficient cascades typically have only several generations.  High
multiplicity is achieved because particles of each generations emit multiple
pair producing photons. The biggest contribution comes from individual cascades
with 3-4 generations.  RICS can play an important role in polar caps cascades;
it can provide an even larger number of pairs than synchrotron radiation, but
synchrotron radiation never becomes a negligible process.  When RICS is an
important process the cascade can have many branches, but the multiplicity does
not directly depend on cascade complexity (number of branches and
generations). A more important factor is the energy available for the cascade
process; the exact way of how this energy is distributed among the final pair
population, i.e. via synchrotron or RICS branches, plays a secondary role.

The main factor determining the total yield of polar cap cascades is, however,
the variation of the flux of primary particles due to intermittency of the
particle acceleration in time-dependent cascades. Simple estimates for the
``duty cycle'' of the particle acceleration, presented in \S10 of \PapI{}, predict
that the total pair yield of polar cap cascade will be lower than the
multiplicity values of cascades at the peak of the pair formation burst by a factor
ranging from a $\sim$few for the case of space charge-limited flow (free
particle extraction from the NS) with super Goldreich-Julian current density
($j/\GJ{j}>1$) and up to $\sim$several hundred for the case of
Ruderman-Sutherland gaps (no particle extraction from the PC) and space
charge-limited flow with anti Goldreich-Julian current density ($j/\GJ{j}<0$).
The effective pair multiplicity then can not exceed a $\sim\sci{\mbox{few}}{5}$
even under the most favorable conditions -- in young hot pulsars with high magnetic
field $B\sim\sci{3}{12}$G and a significant non-dipolar component of the magnetic
field in polar caps so that $\rhoC\lesssim{}10^7$cm.

An accurate estimate of the duty cycle requires self-consistent modeling of
particle acceleration with much higher numerical resolution than was done in
\citet{TimokhinArons2013,Timokhin2010::TDC_MNRAS_I}, which will be a subject of
a future paper. The results of this paper can then be easily adopted into a
consistent model of pair supply in pulsars by scaling the multiplicity values
obtained here by the duty cycle of particle acceleration. The semi-analytical
model presented here can also guide more accurate (and time consuming) numerical
simulations of cascades.

In terms of the direct astrophysical implication of this work, our main message is
simple -- under no circumstances can the pair yield of pulsars be greater than
$\sim\sci{\mbox{several}}{5}\GJ{n}$. This should be taken into account for
examples in modeling of PWNe and lepton components of cosmic rays.

\acknowledgments

This work was supported by the NSF grant 1616632 as well as \textsl{Chandra} and
\textsl{Fermi} Guest Investigator programs.

\appendix

\section{Non-resonant inverse Compton scattering in pulsar polar cap cascades}
\label{sec:non-resonant-inverse}

  Pairs can scatter soft X-ray photons emitted by the NS surface in resonant as
  well as non-resonant regime. However, as we show below, the non-resonant
  Inverse Compton Scattering (ICS) is a very inefficient emission mechanism and
  can be neglected in comparison with scattering in resonant regime.

\begin{figure}
\includegraphics[clip]{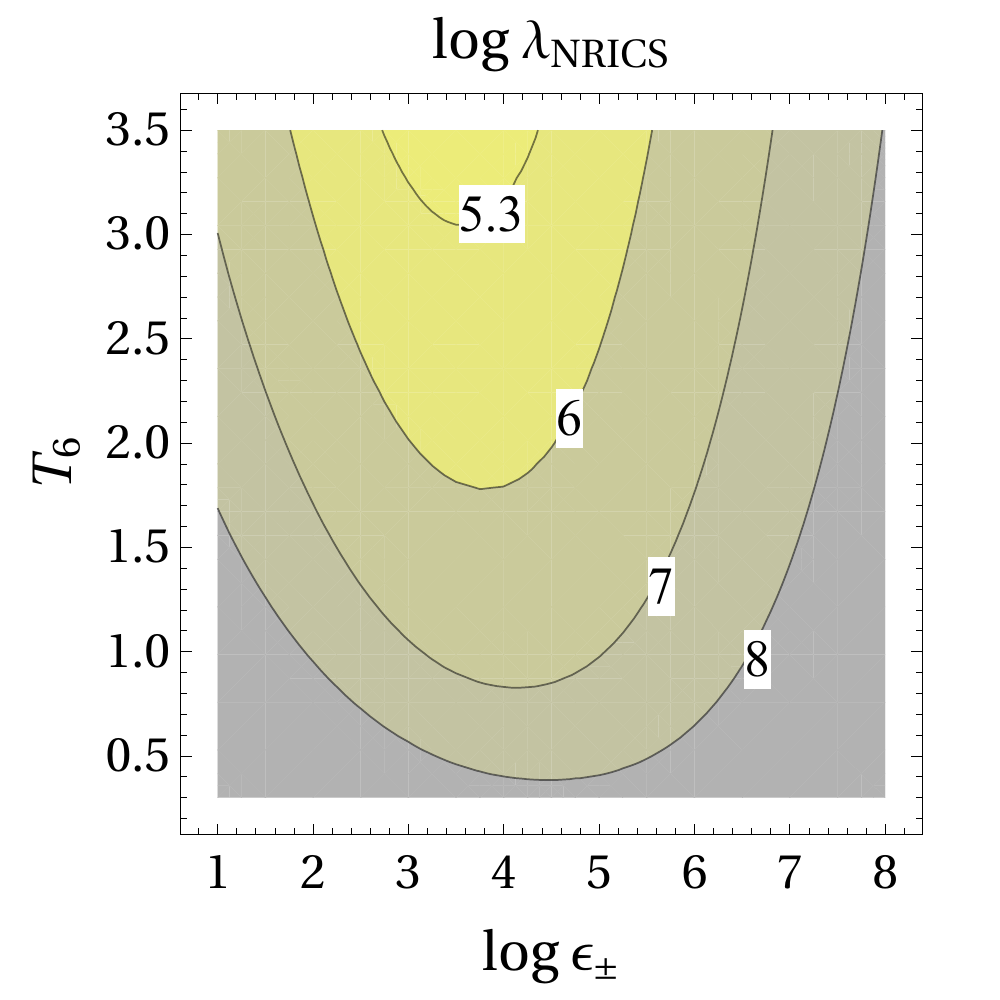}
\caption{Contour plot of the logarithm of the mean free path of a particle to
  non-resonant ICS $\SUB{\lambda}{NRICS}$ (in cm) as a function of the logarithm
  of the particle energy $\epsilonP$ and the temperature of the NS surface
  $T_{6}$ (in units of $10^{6}$K).}
\label{fig:_Lnrics}
\end{figure}

  An emission process could play a role in polar cap cascades if the distance
  over which a particle looses a substantial part of its energy to that emission
  process is less that the size of the cascade zone, which in this work is
  assumed to be $\RNS$. In Fig.~\ref{fig:_Lnrics} we plot contours of the mfp
  $\SUB{\lambda}{NRICS}$ (in cm) of an electron/positron to Non-Resonant Inverse
  Compton Scattering (NRICS) of soft photons emitted by the NS surface as a
  function of the particle's energy and the temperature of the NS. The mfp was
  calculated by integrating the full ICS cross-section over the non-isotropic
  distribution of photons emitted by the NS surface, photons are coming in the
  solid angle which is centered around particle's momentum and limited by
  $(0\leq\theta<\theta_{\max};0\leq\phi\leq{}2\pi)$, with
  $\cos\theta_{\max}=0.5$ (in this paper we use the same solid angle for
  modeling RICS). The spectral energy distribution of thermal photons was
  modeled as the Rayleigh-Jeans power law with the high energy cut-off chosen in
  such a way that the total emitted energy is consistent with the
  Stefan-Boltzmann law%
  \footnote{Our approximation is more accurate than monochromatic approximation
    used by \citet{Sturner1995} to obtain his expression for electron energy
    losses (14), which has to be integrated numerically.  We were also able to
    derive an analytical expression for electron's mfp (Timokhin 2018, in
    preparation).}
  It is easy to see from that plot that the mfp to NRICS becomes less that the
  NS radius only for very high temperatures of the NS surface,
  $T\gtrsim\sci{2}{6}$K, and even for $T\gtrsim\sci{3}{6}$K the mfp is only
  $\SUB{\lambda}{NRICS}\simeq0.2\,\RNS$. According to common models of NS
  cooling even the youngest pulsars should have surface temperatures less that
  $\sci{3}{6}$K \citep[e.g.][]{YakovlevBook2007}%
  \footnote{the \emph{polar caps} of some pulsars can be hotter than
    $T\gtrsim\sci{2}{6}$K but the solid angle of the polar cap at distances
    larger than the polar cap radius will be small, and so the efficiency of ICS
    will be significantly suppressed; also see the next paragraph.}.
  This implies that it highly
  unlikely that in polar caps of pulsars particles could loose any significant
  fraction of their energy to non-resonant ICS.

\begin{figure}
\includegraphics[clip]{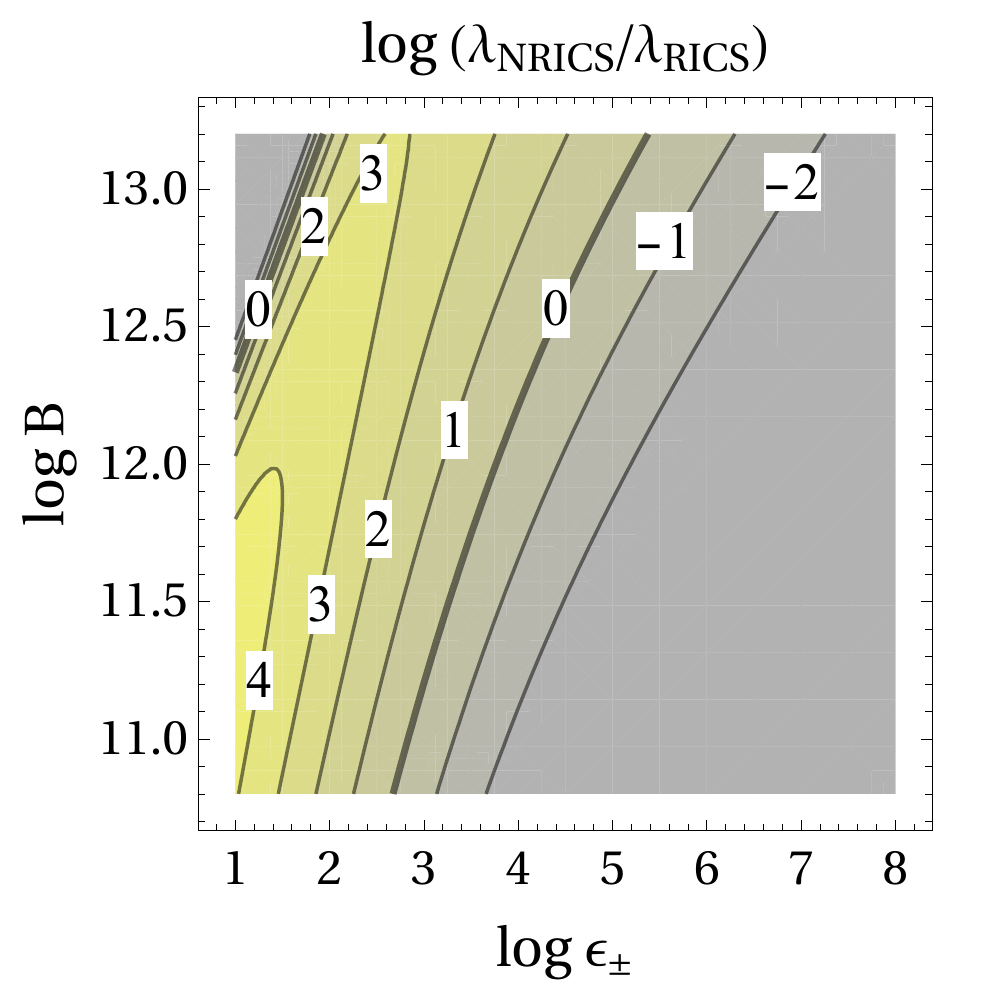}
\caption{Contour plot of the logarithm of the ratio of mean free paths of a
  particle to non-resonant and resonant ICS
  $\SUB{\lambda}{NRICS}/\SUB{\lambda}{RICS}$ as a function of the logarithm of
  the particle energy $\epsilonP$ and the logarithms of the magnetic field
  strength $B$ in Gauss.}
\label{fig:Lnrics2Lrics}
\end{figure}

  But even if the NS surface is very hot, at the upper limit of the predicted
  temperature range, $T\simeq\sci{3}{6}$K, NRICS would be still of very limited
  relevance for cascade physics. The reason is as follows. From
  Fig.~\ref{fig:_Lnrics} it is clear that NRICS might be relevant for particles
  with energies $10^{3}\lesssim{}\epsilonP\lesssim{}10^{4}$. Let us compare now
  the efficiency of resonant and non-resonant ICS.  In
  Fig.~\ref{fig:Lnrics2Lrics} we plot contours of the logarithm of the ratio of
  the mean free paths of a particle to non-resonant and resonant ICS
  $\SUB{\lambda}{NRICS}/\SUB{\lambda}{RICS}$ as a function of particle energy
  and the strength of the magnetic field for the surface temperature
  $T=\sci{3.5}{6}$K. In the energy range
  $10^{3}\lesssim{}\epsilonP\lesssim{}10^{4}$, where a significant part of
  particle's momentum could be radiated via NRICS within the cascade zone,
  $\SUB{\lambda}{NRICS}$ is less that $\SUB{\lambda}{RICS}$ only for magnetic
  fields $B\lesssim\sci{3}{12}$G. For magnetic field strengths
  $\lesssim10^{12}$G the fraction of the parent photon's energy going into the
  parallel momentum of the created pair is smaller than $\sim{}30\%$ -- most of
  the energy is emitted as synchrotron photons, see \PapI, Section~4,
  Fig.~(5). In the narrow range of magnetic field strengths
  $10^{12}\mbox{G}\lesssim{}B\lesssim\sci{3}{12}$G NRICS might become an
  important process, but only for particles of a single cascade
  generation. Indeed, because of a very steep dependence of
  $\SUB{\lambda}{NRICS}/\SUB{\lambda}{RICS}$ on particle energy for a given
  value of $B$, even if pairs of some generation with the energy in the range
  $10^{3}\lesssim{}\epsilonP\lesssim{}10^{4}$ do create photons more efficiently
  in the non-resonant regime, the next generation of pairs will scatter photons
  more efficient in the resonant regime.

\section{Optical depths for single photon pair creation in ultrastrong magnetic
  field}
\label{sec:optic-depths-phot}

In \PapI{} we adopted the widely used \citet{Erber1966} formula for the opacity
for single photon pair production in a strong magnetic field
\begin{equation}
  \label{APP_eq:alphaB_Erber}
  \Pair{\alpha}(\epsilonG,\psi) =
  0.23\,\frac{\alphaF}{\lambdaC}\:b\,\sin\psi\,\exp\left(-\frac{4}{3\chi}\right) 
\end{equation}
where $b\equiv{}B/B_q$ is the local magnetic field strength $B$
normalized to the critical quantum magnetic field 
$B_q=e/\alphaF\lambdaC^2=\sci{4.41}{13}$~G, 
$\psi$ is the angle between
the photon momentum and the local magnetic field, 
$\alphaF=e^2/\hbar{}c\approx{}1/137$
is the fine structure constant, and
$\lambdaC = \hbar/m c = \sci{3.86}{-11}$~cm
is the reduced Compton wavelength. The parameter $\chi$ is defined as
\begin{equation}
\label{eq:APP_chi_def}
  \chi \equiv \frac{1}{2}\, \epsilonG b\, \sin\psi\,,
\end{equation}
where $\epsilonG$ is the photon energy in units of $m_ec^2$.
Expression~(\ref{APP_eq:alphaB_Erber}) has been obtained in the asymptotic limit of
$\chi\ll{}1$ and $b\ll{}1$.  While $\chi\ll{}1$ is a good approximation for
gamma-rays absorbed in polar caps in all pulsars (see
\S\ref{sec:phot_pair_creation}), the approximation $b\ll{}1$ can become too
restrictive for pulsars with higher magnetic fields. For pairs created near the
kinematic threshold
\begin{equation}
  \label{eq:APP_gammaB_threshold}
  \epsilonG \sin\psi = 2
\end{equation}
eq.~(\ref{APP_eq:alphaB_Erber}) can overestimate the opacity by a factor of a few
for pulsars with magnetic fields
$\gtrsim\sci{3}{12}$G \citep{DaughertyHarding1983}.  The discrepancy becomes larger
for stronger fields.  An accurate treatment of the pair creation cross-section for
non-small
$b$ and/or near-threshold pair creation requires summation over a finite number
of cyclotron energy levels of created pairs which results in unwieldy expressions
(like eq.~6 in \citet{DaughertyHarding1983}).  For our semi-analytical model
such treatment would be an overkill, resulting in unnecessary complication of the
model.  Instead we use the numerical fit to the exact expression for the opacity
suggested by \citet{DaughertyHarding1983} (their eq.~24)
\begin{equation}
  \label{eq:APP_alpha_B_DH}
  \Pair{\alpha}(\epsilonG,\psi) =
  \begin{cases}
    \displaystyle
    0.23\,\frac{\alphaF}{\lambdaC}\:b\,\sin\psi\, \exp\left(-\frac{4
        f_\alpha(\epsilonG,b)}{3\chi}\right), & \text{ if }\epsilonG \sin\psi\ge2\\
    0, & \text{ if } \epsilonG \sin\psi<2
  \end{cases}
\end{equation}
where
\begin{equation}
  \label{eq:APP_f_alpha_B}
  f_\alpha=1 + 0.42\, \left(\frac1{2}\:\epsilonG \sin\psi\right)^{-2.7} b^{-0.0038}\,.
\end{equation}
The second term in
$f_\alpha$ is significant only for pair creation close to the threshold
\eqref{eq:APP_gammaB_threshold}.  The non-zero part of
expression~\eqref{eq:APP_alpha_B_DH} for $\Pair{\alpha}$ can be written as
\begin{equation}
  \label{eq:APP_APP_alpha_B_DH_mult}
  \Pair{\alpha}(\epsilonG,\psi) = 0.23\,\frac{\alphaF}{\lambdaC}\:b\,\sin\psi\,
  \exp\left(-\frac{4}{3\chi}\right)
  \exp\left(-0.56\,\frac{b^{2.6962}}{\chi^{3.7}} \right)\,,
\end{equation}
i.e. it differs from the usual Erber's formula~\eqref{APP_eq:alphaB_Erber} by the
exponential term
\begin{equation}
  \label{eq:APP_f_alpha_1}
  f_{\alpha,1}(b, \chi)\equiv\exp\left(-0.56\,\frac{b^{2.6962}}{\chi^{3.7}} \right)\,.
\end{equation}
This term significantly differs from 1 when pair formation occurs close to 
threshold.

The optical depths for pair creation by a photon in a strong magnetic field
after propagating a distance $l$ is
\begin{equation}
\label{eq:APP_tau_general}
\tau(\epsilonG,l)=\int_0^l\Pair{\alpha}(\epsilonG,\psi(x))\,dx\,, 
\end{equation}
where integration is along the photon's trajectory. For photons emitted tangent
to the magnetic field line, $dx=\rhoC d\psi$, where $\rhoC$ is the radius of
curvature of magnetic field lines; the angle $\psi$ is always small so the 
approximation $\sin\psi\approx\psi$ is very accurate.  In our approximation of
constant magnetic field both $b$ and $\rhoC$ are constants. From
eq.~(\ref{eq:APP_chi_def}) we have $\psi=2\chi/\epsilonG{}b$, and substituting it
into eq.~(\ref{eq:APP_tau_general}) we can express the optical depth $\tau$
to pair production as an integral over $\chi$
\begin{equation}
  \label{eq:APP_tau_OTS_general}
  \tau(\epsilonG,l) = A_\tau\, \frac{\rhoC}{\epsilonG^2 b}\;
  \int_0^{\chi(\epsilonG, \psi(l))} \chi
  \exp\left(-\frac{4}{3\chi}\right)\,
  f_{\alpha,1}(b, \chi)\:
  d\chi\,,
\end{equation}
where
$A_\tau\equiv{}0.92 \alphaF/\lambdaC\approx\sci{1.74}{8}\;\mbox{cm}^{-1}$.
If we expand the term $f_{\alpha,1}$ in a Taylor series the integral in
eq.~\eqref{eq:APP_tau_OTS_general} can be represented as a sum of integrals
$\int x^\xi \exp(-x)\,dx$; each of those integrals can be integrated
analytically.

\begin{figure}[t]
  \centering
  \includegraphics[clip]{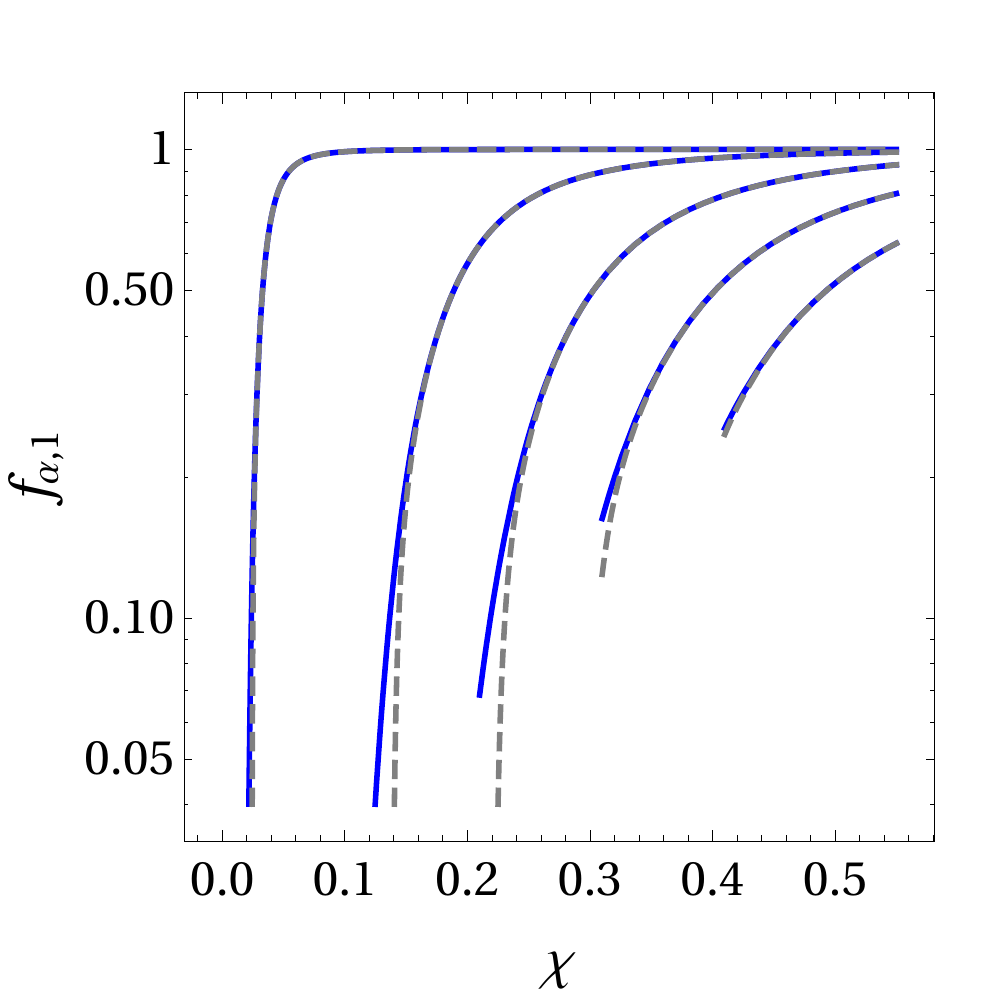}  
  \caption{Fifth order Taylor series expansion for $f_{\alpha,1}$ given by
    eq.~\eqref{eq:APP_f_alpha_B_expansion} -- dashed lines -- compared to the
    value given by eq.~\eqref{eq:APP_f_alpha_1} -- solid lines -- as a function
    of $\chi$ for 5 values of the magnetic field $b=0.01,0.11,0.21,0.31,0.41$
    (lines from left to right).}
  \label{fig:exp_Taylor_fit}
\end{figure}

We expand $f_{\alpha,1}$ in a Taylor series as
$e^{-x}\approx{}1-x+x^2/2-\dots$ and keep up to the $5^{\text{th}}$ term
\begin{align}
  \label{eq:APP_f_alpha_B_expansion}
  f_{\alpha,1}(b, \chi) \approx {}  
  & 1 - 0.56\:\frac{b^{2.6962}}{\chi^{3.7}}
    + 0.1568\:\frac{b^{5.3924}}{\chi ^{7.4}}
    - 0.02927\:\frac{b^{8.0886}}{\chi^{11.1}} \\
  & {} + \sci{4.0977}{-3}\:\frac{b^{10.7848}}{\chi^{14.8}} 
    - \sci{4.5894}{-4}\:\frac{b^{13.481}}{\chi ^{18.5}}\,. \notag
\end{align}
Only odd order term expansions are monotonic and we find that $3^{\text{rd}}$
order series is not a good enough approximation near the threshold.
Eq.~\eqref{eq:APP_f_alpha_B_expansion} provides a good fit with the minimum number
of terms -- see Fig.~\ref{fig:exp_Taylor_fit}.

Substituting expansion~\eqref{eq:APP_f_alpha_B_expansion} into
eq.~\eqref{eq:APP_tau_OTS_general} we get an analytical expression for the optical
depth as a series of special functions
\begin{multline}
  \label{eq:APP_tau_OTS_series}
  \tau(\chi) = A_\tau\, \frac{\rhoC}{\epsilonG^2 b}\;
  \left[
  \frac{16}{9}\: \Gamma \left(-2,\frac{4}{3 \chi }\right)
  -0.3434\: b^{2.6962}\, \Gamma \left(1.7,\frac{4}{3 \chi }\right) \right.\\
  + \sci{3.3165}{-2}\: b^{5.3924}\, \Gamma \left(5.4,\frac{4}{3 \chi}\right)
  - \sci{2.1354}{-3}\: b^{8.0886}\, \Gamma \left(9.1,\frac{4}{3 \chi }\right) \\
  \left. + \sci{1.0312}{-4}\: b^{10.7848}\, \Gamma \left(12.8,\frac{4}{3 \chi }\right)
  - \sci{3.9835}{-6}\: b^{13.481}\, \Gamma \left(16.5,\frac{4}{3 \chi}\right)
  \right]\,.
\end{multline}
The special function $\Gamma(a,x)$ is the so-called upper incomplete gamma
function, defined as $\Gamma(a,x)=\int_{x}^{\infty}t^{a-1}\exp(-t)\,dt$.  There
are efficient numerical algorithms for its calculation implemented in numerical
libraries and scientific software tools; using eq.~\eqref{eq:APP_tau_OTS_series} for
calculation of the optical depths result in more efficient numerical codes than
direct integration of eq.~\eqref{eq:APP_tau_OTS_general}.

For $b\ll1$, when the opacity is given by the Erber's
formula~\eqref{APP_eq:alphaB_Erber}, the optical depths to pair creation is given by
the first term in eq.~\eqref{eq:APP_tau_OTS_series}
\begin{equation}
  \label{eq:APP_tau_OTS_small_B}
  \tau(\chi) = \frac{16}{9} A_\tau\, \frac{\rhoC}{\epsilonG^2 b}\;
  \Gamma \left(-2,\frac{4}{3 \chi }\right)\,.
\end{equation}
This expression is a more compact from of the expression for $\tau(\chi)$ from
\PapI (eq.~6 in that paper) as
\begin{equation}
  \frac{16}{9}\: \Gamma \left(-2,\frac{4}{3 \chi }\right) =
  \left[
  \frac{\chi^2}{2}\left(1-\frac{4}{3\chi}\right) e^{-\frac{4}{3\chi}} - 
  \frac{8}{9}\,\mbox{Ei}\left(-\frac{4}{3 \chi }\right)
\right]\,.
\nonumber
\end{equation}
For $\chi\ll1$ eq.~\eqref{eq:APP_tau_OTS_small_B} can be simplified further by
expanding it into Taylor series around $\chi=0$ and retaining the first term
\begin{equation}
  \label{eq:APP_tau_OTS_small_Chi}
  \tau(\chi) = \frac{3}{4} A_\tau\, \frac{\rhoC}{\epsilonG^2 b}\;
  \chi^3 e^{-\frac{4}{3 \chi }}\,.
\end{equation}

\section{Algorithms for semi-analytical calculation of cascade
  multiplicity}
\label{sec:algorithms}

\begin{figure}
  \[
    \newcommand{\kxVec}{\ensuremath{\left<\kappa_0,\dots,\kappa_{nx}\right>}} 
    \newcommand{\csdMxEntry}[1]{(#1):\kxVec}
    \setlength{\kbrowsep}{1.5em} \setlength{\kbcolsep}{1.5em}
    \begin{array}{cc}
      & \textrm{\scriptsize{}Emission Process ID /Name/} \\
      \rotatebox[origin=c]{90}{\scriptsize\textrm{Cascade Generation}}
      &
      \kbordermatrix{%
      {}& 0\;\textrm{/CR/} & 1\;\textrm{/Syn/} & 2\;\textrm{/RICS/} \\
      0 & \left\{ \csdMxEntry{0} \right\}  & 0 & 0 \medskip\\
      1 & 0 & \left\{ \csdMxEntry{0,1} \right\} & \left\{ \csdMxEntry{0,2} \right\}\medskip\\
      2 & 0
      &
        \begin{Bmatrix}
          \csdMxEntry{0,1,1}\\
          \csdMxEntry{0,2,1}
        \end{Bmatrix}
      &
        \begin{Bmatrix}
          \csdMxEntry{0,2,2}\\
          \csdMxEntry{0,1,2}
        \end{Bmatrix}
      \medskip
      \\
      \vdots & \vdots & \vdots & \vdots \medskip\\
      n & 0
      &
        \begin{Bmatrix}
          \csdMxEntry{\orgn_1}\\
          \csdMxEntry{\orgn_2}\\
          \vdots
        \end{Bmatrix}
      &
        \begin{Bmatrix}
          \csdMxEntry{\orgn_1}\\
          \csdMxEntry{\orgn_2}\\
          \vdots
        \end{Bmatrix}
      \\
      }
    \end{array}
  \]
  \caption{Structure of the cascade matrix. Each row contains information about
    pairs created in the generation which number is equal to the row
    number. Each column contains information about pairs created by photons
    emitted by the emission process id which is equal to the column number; in
    this case the IDs of the emission processes are: 0-CR, 1-Synchrotron, 2-
    RICS.  Each matrix element is an associate array with entries
    $(\orgn):\kappaArr$, where the key $(\orgn)$ is the origin of the pairs --
    the label of the cascade branch where pairs have been created -- and
    $\kappaArr$ is an array of the spacial distribution of pairs created in the
    cascade branch $(\orgn)$. }
    \label{fig:cascadeMatrix}
\end{figure}

Here we show pseudo-codes of algorithms used to compute cascade multiplicity.
For calculation of $\chiA(\epsilonG,B,\rhoC)$ we computed and stored a table of
$1/\chiA$ values for a uniformly divided grid $77\times30\times20$ in
$\log\epsilonG\times\log{B}\times\log{\rhoC}$ space, and then used cubic
piece-polynomial interpolation to get $\chiA$ for parameter values required by
expressions used in the algorithms.

Our semi-numerical algorithm is build around the data structure which we call
``cascade matrix''.  Cascade matrix contains information about spatial
distribution of the pair injection rate in the cascade ordered by cascade
generations and emission processes which lead to the creation of pairs. The spatial
distribution of pair injection rates is stored as arrays \kappaArr{} of the
(fixed) length $nx$; in our simulations we usually use $nx=10$.  We divide our
calculation domain $[0,\ESC{s}]$ into $nx$ intervals, each value of $\kappa_i$
is the number of pairs injected in the interval $[s_i,s_{i+1}]$.

The structure of a cascade matrix for a cascade initiated by CR when both
Synchrotron and RICS photons create secondary pairs is shown on
Fig.~\ref{fig:cascadeMatrix}.  Row $\igen$ contains distributions for all pairs
created in cascade branches of generation $\igen$. Column $\iproc$ contains
distributions for pairs created by photons emitted via the same emission
mechanism with id=$\iproc$, i.e. pair created in cascade branches ending in
id=$\iproc$. Element $[\igen,\iproc]$ of the cascade matrix is an associative
array consisting of entries $(\orgn):\kappaArr$, where the tuple $(\orgn)$ is
pair's ``origin'' (the label of the cascade branch which lead to the injection
of the pairs) and array $\kappaArr$ is the spatial distribution of pairs created
by a given cascade branch.  Although the tuple $(\orgn{})$ is enough to identify
the position of the pair distribution regarding cascade generation and the
emission process -- tuple's length is the cascade generation and the last entry
is the id of the process producing the pair creating photon -- keeping entries
sorted according to \igen{} and \iproc{} makes interpretation and visualization
of the simulation results much easier.

Mathematical operations on the cascade matrix%
\footnote{we would need only addition and multiplication}
are defined as element-wise operations on the pairs spatial distribution arrays
$\kappaArr$ with the same $(\orgn)$, e.g. for addition of two arrays
$\kappaArr^1$ and $\kappaArr^2$ the resulting array is defined as
\begin{equation}
  \label{eq:kappa_x_op}
  \kappaArr^{\text{res}}=\kappaArr^1 + \kappaArr^2 \,:\,
  \kappa^{\text{res}}_i=\kappa_i^1 + \kappa_i^2 \text{ for } i=1\dots{}nx\,.
\end{equation}
Neither the position of the element $(\orgn):\kappaArr$, nor the value
$(\orgn{})$ change.  If an expression involving cascade matrices entry
$[i,j][(\orgn)]$ is missing in one of the matrix but not in the other(s), a
zero-filled array of the length $nx$ is inserted at the position
$[i,j][(\orgn_i)]$ of the missing element and then operation is performed as in
eq.~\eqref{eq:kappa_x_op}, e.g. for addition of two cascade matrices $M^1$ and
$M^2$ each element of the resulting matrix $M^{\text{res}}$ is given by
\begin{equation}
  \label{eq:cascade_matrix_op}
  M^{\text{res}}[i,j][(\orgn)] =
  \begin{cases}
    M^1[i,j][(\orgn)] + M^2[i,j][(\orgn)] &
    \text{for all $(\orgn)$ in both $M^1$ and $M^2$}\\
    M^1[i,j][(\orgn)] & \text{for all $(\orgn)$ only in $M^1$}\\
    M^2[i,j][(\orgn)] & \text{for all $(\orgn)$ only in $M^2$}\\
  \end{cases}
\end{equation}
Such data structure is relatively straightforward to implement in modern
scripting languages (e.g. as a list of dictionaries in Python, or a list of
associations in \textsl{MATHEMATICA}).

\begin{algorithm}[t]
  \label{alg:full_cascade}
  \SetKwFunction{KwFunPairCreation}{PairCreation}

  \SetKwData{vPhotonInput}{spctr\_bin}
  \SetKwData{vPhoton}{bin}
  \SetKwData{vPhotonList}{spectrum}
  \SetKwData{vEmissionProcessList}{emissionFun\_list}

  \SetKwData{vOrigin}{orgn}
  \SetKwData{vCsdMx}{M}

  \SetKwData{vIgen}{$\SUB{i}{gen}$}
  \SetKwData{vIproc}{$\SUB{i}{proc}$}

  \SetKwData{vParams}{params}
  \SetKwData{vX}{x}
  \SetKwData{vXX}{xx}
  \SetKwData{vJinj}{$\SUB{j}{inj}$}

  \SetKwData{vSesc}{$\ESC{s}$}
  \SetKwData{vScascade}{$\SUB{s}{cascade}$}
  
  \SetKwProg{Fn}{Function}{:}{end}
  \SetKwData{vCsdMxNew}{M\_1}
  \SetKwData{vEmissionProcessCR}{emissionFun\_CR}

  \SetKwData{vS}{$s_j$}
  \SetKwData{vSLast}{$s_{j-1}$}
  \SetKwData{vEpart}{\epsilonP}
  \SetKwData{vN}{\ensuremath{N}}
  \KwData{$\epsilonP^0$ -- energy of the primary particle, \vScascade --
    characteristic size of the cascade, \vSesc -- mfp of escaping
    photons }
  \KwResult{Cascade Matrix for all cascades initiated by CR of an primary
    particle moving in interval $[0,\vScascade]$}
  \BlankLine
  \Begin{
    divide $[0,\vScascade]$ in \vN subintervals $s_0,\dots{},s_N$\;
    \tcp{initialize empty Cascade Matrices}
    \vCsdMx$\leftarrow\left\{\right\}$\; 
    \For{$j\leftarrow 1$ \KwTo \vN}{%
      \tcp{primary particle energy at distance \vS}
      $\vEpart\leftarrow\epsilonP(\epsilonP^0, \vS)$ \tcp*{eq.~(\ref{eq:CR_es})}
      \tcp{spectrum of CR}
      \vPhotonList = \vEmissionProcessCR{$\epsilonP$,\vS,\vParams}\;
      \BlankLine
      \tcp{each CR photon starts generation 0 in a new cascade}
      $\vIgen\leftarrow 0$\;
      $\vOrigin\leftarrow (0)$\;
      \vCsdMxNew$\leftarrow\left\{\right\}$\;
      \tcp{call $\KwFunPairCreation(\,\dots)$ for each spectral bin}
      \ForEach{\vPhoton in \vPhotonList}{%
        \KwFunPairCreation{\vCsdMxNew, \vIgen, \vPhoton, \vS,
          \vOrigin, \vEmissionProcessList, \vParams }\;
      }
      \BlankLine
      \tcp{trapezoidal rule}
      $\vCsdMx \leftarrow \vCsdMx + 0.5 (\vCsdMx + \vCsdMxNew)(\vS - \vSLast)$\;
    }
    \KwRet{\vCsdMx}\;
  }
  \caption{Main function: calculates multiplicity of cascade initiated by CR
    radiation of a primary particle.}
\end{algorithm}

\begin{algorithm}[t]
  \label{alg:pair_creation_fun}
  \SetKwFunction{KwFunPairCreation}{PairCreation}

  \SetKwData{vPhotonInput}{spctr\_bin}
  \SetKwData{vPhoton}{bin}
  \SetKwData{vPhotonList}{spectrum}
  \SetKwData{vEmissionProcess}{emissionFun}
  \SetKwData{vEmissionProcessList}{emissionFun\_list}

  \SetKwData{vOrigin}{orgn}
  \SetKwData{vCsdMx}{M}

  \SetKwData{vIgen}{$\SUB{i}{gen}$}
  \SetKwData{vIproc}{$\SUB{i}{proc}$}

  \SetKwData{vParams}{params}
  \SetKwData{vX}{s}
  \SetKwData{vXX}{xx}
  \SetKwData{vJinj}{$\SUB{j}{inj}$}
  
  \SetKwData{vSesc}{$\ESC{s}$}
  \SetKwData{vScascade}{$\SUB{s}{cascade}$}
  
  \SetKwProg{Fn}{Function}{:}{end}
  \Fn(){\KwFunPairCreation{\vCsdMx, \vIgen, \vPhotonInput, $x_e$, \vOrigin,
      \vEmissionProcessList, \vParams }}%
  {
    \BlankLine
    \KwData{\\
      \vCsdMx\  -- cascade matrix, can be modified inside the function\\
      \vIgen\   -- current cascade generation\\
      \vPhotonInput\ -- spectral bin $\{\epsilonG,n_\gamma,\vIproc\}$\\
      $\quad$ $\epsilonG$: photon energy, \\
      $\quad$ $n_\gamma$: of photons in the bin,\\
      $\quad$ \vIproc: ID of the emission process which produced these photons\\
      $s_e$ -- coordinate of the photon emission point\\
      \vOrigin\ -- origin (label of the cascade branch) \\
      \vEmissionProcessList\ -- list of emission process functions\\
      \vParams\ -- parameters of the cascade zone ($\rho_c$, $T$, \vSesc,
      \vScascade, etc.)  }
    \BlankLine
    \KwResult{fills \vCsdMx with data about pairs produced in the cascade}
    \BlankLine  
    \BlankLine  
    $b = b(x_e)$\;
    $\Delta\vX=\MFP/\RNS$\tcp*[r]{eq.~\eqref{eq:MFP}}
    $\vX = s_e+ \Delta\vX$\;
    \BlankLine
    \tcp{if photons are absorbed, create new pairs}
    \If{$\Delta\vX\leq\vSesc$ and $\vX\leq\vScascade$}{
      \tcp{\vIgen and \vOrigin for newly created pairs}
      $\vIgen \leftarrow{} \vIgen+1$\;
      $(\vOrigin) \leftarrow{} (\vOrigin,\vIproc)$\;
      \BlankLine
      \tcp{index of the pair entry in array $\kappa_x$}
      $\vJinj\leftarrow{}j\,|\,\vXX[j]<\vX\le\vXX[j+1]$\;
      \tcp{multiplicity vector $\kappa_x$ for newly created pairs}
      $\left<\kappa_x\right>\leftarrow \left<0, \dots, 2n_\gamma,\dots, 0\right>\:|\:
      \kappa_x[\vJinj]=2 n_\gamma, \kappa_x[j]=0\text{ for
      }j\neq\vJinj$\;
      \BlankLine
      \tcp{add newly created pairs to Cascade Matrix}
      \eIf{$\vCsdMx[\vIgen, \vIproc][\vOrigin]$ exists}{%
        \tcp{add newly created pairs to an already existing cascade
          branch}
        $\vCsdMx[\vIgen, \vIproc][\vOrigin] \leftarrow \vCsdMx[\vIgen,
        \vIproc][\vOrigin] + \left<\kappa_x\right>$}{%
        \tcp{create new cascade branch and add newly created pairs}
        $\vCsdMx[\vIgen, \vIproc]\leftarrow \{\vCsdMx[\vIgen,
        \vIproc],\: (\vOrigin):\left<\kappa_x\right>\}$}
      \BlankLine      
      \tcp{\sc go to the next cascade generation:}
      \tcp{iterate over emission processes}
      \ForEach{\vEmissionProcess in \vEmissionProcessList}{
        \tcp{spectrum of photons emitted by newly created pairs (list
          of spectral bins)}
        \vPhotonList = \vEmissionProcess{$\epsilonG,\vX$,\vParams}\;
        \BlankLine
        \tcp{call $\KwFunPairCreation(\,\dots)$ for each spectral bin}
        \ForEach{\vPhoton in \vPhotonList}{
          \KwFunPairCreation{\vCsdMx, \vIgen, \vPhoton, $\vX$, \vOrigin,
      \vEmissionProcessList, \vParams }\;
        }
      }
    }
  }
  \caption{PairCreation function: calculates multiplicity of a photon initiated
  cascade}
\end{algorithm}

The algorithm has two structural parts. The main function shown as
Algorithm~\ref{alg:full_cascade} follows a primary particle, while it propagates
through the calculation domain and emits CR photons, integrating contributions
of cascades initiated by CR photons.  This function fills an (initially empty)
cascade matrix \texttt{M} with the data from all cascades initiated by CR
photons emitted by the primary particle.  The calculation domain is divided in
$N$ logarithmic segments%
\footnote{we used the number of points large enough to achieve numerical
  convergence, typically $N=300$}
and the energy of the primary particle $\epsilonP$ is calculated
in each segment according to eq.~\eqref{eq:CR_es} which takes into account
particle energy losses due to CR.  $\epsilonG$ is used to calculate the spectrum
of CR, which is divided in $\SUB{n}{CR}$ of spectral bins.  The recursive
function \texttt{PairCreation} is called for each spectral bin to calculate
cascades initiated by CR photons. Each CR photon belongs to a generation 0 of a
new cascade, so \texttt{PairCreation} is called with generation number
$\SUB{i}{gen}=0$, photon's origin $\orgn=(0)$, and an empty (auxiliary) cascade
matrix \texttt{M\_1}.  Matrix \texttt{M\_1} filled in recursive call(s) of
\texttt{PairCreation} is added to the matrix \texttt{M} according to the
trapezoidal integration rule.

Function \texttt{PairCreation} shown as Algorithm~\ref{alg:pair_creation_fun}
fills a cascade matrix \texttt{M}, which is passed to it as an argument, with
data about a photon initiated cascade.  Among its arguments there is a list of
functions \texttt{emissionFun\_list}; each function in that list calculates the
spectrum of the next generation cascade photons for one emission mechanism.
Emission processes relevant for the full cascade are set by the content of this
list.  In this paper the list consists of two functions, the first one
calculates spectrum of Synchrotron radiation, and the second one the spectrum of
RICS photons%
\footnote{we use monochromatic approximation for RICS spectrum},
but it can be extended to include additional processes, such as
e.g. non-resonant ICS.  Other arguments given to \texttt{PairCreation} contain
information about photon(s) initiating the next cascade generation (their energy
$\epsilonG$, number of such photons $n_\gamma$, their origin $\orgn$, number of
the current cascade generation $\SUB{i}{gen}$, coordinate of photon emission
point $x_e$), and parameters of the cascade zone.

\texttt{PairCreation} calculates the coordinate where the photon will be
absorbed.  If this point is outside of the calculation domain control is
returned to the calling program; all recursive calls of this function are
terminated in this way.  If absorption happens inside the domain, a cascade
branch is created -- cascade generation counter $\SUB{i}{gen}$ is increased and
a label for the new branch is created by adding ID of the emission process to
the tuple $\orgn$ received as a parameter.  The position of injected pairs in
array \kappaArr is computed and an array $\kappaArr$ is created for this branch
of the cascade.  This array is added to the cascade matrix either to an existing
array for the same branch, or, if the branch is not yet present in the matrix, a
new entry is created. Then, the function proceeds to the next cascade generation.
For each emission process the spectrum of the next generation photons is
computed by interaction over the function list \texttt{emissionFun\_list}.  A
new instance of \texttt{PairCreation} is called for each of the spectral bins in
the spectra of the next-generation photons. At this place the algorithm gives
control to the next cascade generation.

The general algorithms can handle a cascade process of any complexity, which
depends on the number of emission processes given to it in the list
\texttt{emissionFun\_list}, and is straightforward to parallelize. The resulting
cascade matrix is easy to analyze, i.e. to get the total multiplicity of the
cascade all entries of the matrix should be summed, for plotting the cascade
tree as shown in Figs.~\ref{fig:CascadeGraph_abc},~\ref{fig:CascadeGraph_def}
all arrays \kappaArr are summed and the matrix is traversed row-vise to create
the tree, etc.

\bibliographystyle{aasjournal} 
\bibliography{fmc_2,pulsars_theory,pulsars_obs}

\end{document}